%% file: main.tex
\documentclass[twocolumn,tighten,dvipsnames]{aastex63}
\usepackage{apjfonts}
\usepackage{textcomp}
\usepackage{color}
\usepackage{mathtools}
\usepackage{xspace}
\usepackage{comment}
\usepackage{multirow}
\usepackage{url}
\usepackage{hyperref}
\usepackage{epstopdf}
\usepackage{ragged2e}
\usepackage{tabularx}
\usepackage{url}
\usepackage{soul}
\usepackage{textcomp}
\usepackage{gensymb}
\usepackage{amssymb,amsmath,amsfonts,amsthm}
\usepackage{mathrsfs}
\usepackage{needspace}
\usepackage[T1]{fontenc}
\usepackage[english]{babel}
\usepackage[utf8]{inputenc}
\usepackage{float}
\usepackage[encapsulated]{CJK}
\usepackage{ucs}
\usepackage{longtable}
\newcommand{\cntext}[1]{\begin{CJK}{UTF8}{gbsn}#1\end{CJK}}

\def\sgra{\object{Sgr~A$^{\ast}$}\xspace}
\def\m87{\object{M87$^{\ast}$}\xspace}

\newcommand\chandra{\textsl{Chandra}\xspace}

\defcitealias{paperi}{{\m87} Paper~I}
\defcitealias{paperii}{\m87 Paper~II}
\defcitealias{paperiii}{\m87 Paper~III}
\defcitealias{paperiv}{\m87 Paper~IV}
\defcitealias{paperv}{\m87 Paper~V}
\defcitealias{papervi}{\m87 Paper~VI}
\defcitealias{SgraP1}{Paper~I}
\defcitealias{SgraP2}{Paper~II}
\defcitealias{SgraP3}{Paper~III}
\defcitealias{SgraP4}{Paper~IV}
\defcitealias{SgraP5}{Paper~V}
\defcitealias{SgraP6}{Paper~VI}

\begin{document}

\title{Millimeter light curves of Sagittarius A$^*$ observed during the 2017 Event Horizon Telescope campaign}

%
\input{authorlist.tex}
\shortauthors{Wielgus et al.}
\shorttitle{Millimeter light curves of Sagittarius A$^*$}

\begin{abstract}
The Event Horizon Telescope (EHT) observed the compact radio source, Sagittarius A$^\ast$ (\sgra), in the Galactic Center on 2017 April 5-11 in the 1.3 millimeter wavelength band. At the same time, interferometric array data from the Atacama Large Millimeter/submillimeter Array and the Submillimeter Array were collected, providing \sgra light curves simultaneous with the EHT observations. These data sets, complementing the EHT very-long-baseline interferometry, are characterized by a cadence and signal-to-noise ratio previously unattainable for \sgra at millimeter wavelengths, and they allow for the investigation of source variability on timescales as short as a minute. While most of the light curves correspond to a low variability state of \sgra, the April 11 observations follow an X-ray flare, and exhibit strongly enhanced variability. All of the light curves are consistent with a red noise process, with a power spectral density (PSD) slope measured to be between -2 and -3 on timescales between 1 min and several hours. Our results indicate a steepening of the PSD slope for timescales shorter than 0.3\,h. The spectral energy distribution is flat at 220 GHz and there are no time-lags between the 213 and 229\,GHz frequency bands, suggesting low optical depth for the event horizon scale source. We characterize \sgra's variability, highlighting the different behavior observed just after the X-ray flare, and use Gaussian process modeling to extract a decorrelation timescale and a PSD slope. We also investigate the systematic calibration uncertainties by analyzing data from independent data reduction pipelines.
\end{abstract}


%
\input{S1_Introduction.tex}


%
\input{S2_Observations.tex}

%
\input{S3_Consistency.tex}

%
\input{S4_Variability.tex}


%
\input{S5_Modeling.tex}

%
\input{S6_PSD.tex}


%
\input{S7_Summary.tex}

%
\input{acknowledgments.tex}
 \newpage
 
%
\input{S8_Appendices.tex}

\bibliographystyle{aasjournal}
\bibliography{references}

\end{document}

%% file: authorlist.tex
\author[0000-0002-8635-4242]{Maciek Wielgus}
\affiliation{Max-Planck-Institut f\"ur Radioastronomie, Auf dem H\"ugel 69, D-53121 Bonn, Germany}
\email{maciek.wielgus@gmail.com}

\author[0000-0002-5523-7588]{Nicola Marchili}
\affiliation{Italian ALMA Regional Centre, INAF-Istituto di Radioastronomia, Via P. Gobetti 101, I-40129 Bologna, Italy}
\affiliation{Max-Planck-Institut f\"ur Radioastronomie, Auf dem H\"ugel 69, D-53121 Bonn, Germany}

\author[0000-0003-3708-9611]{Iv\'an Martí-Vidal}
\affiliation{Departament d'Astronomia i Astrof\'{\i}sica, Universitat de Val\`encia, C. Dr. Moliner 50, E-46100 Burjassot, Val\`encia, Spain}
\affiliation{Observatori Astronòmic, Universitat de Val\`encia, C. Catedr\'atico Jos\'e Beltr\'an 2, E-46980 Paterna, Val\`encia, Spain}

\author[0000-0002-3490-146X]{Garrett K. Keating}
\affiliation{Center for Astrophysics $|$ Harvard \& Smithsonian, 60 Garden Street, Cambridge, MA 02138, USA}

\author[0000-0002-9248-086X]{Venkatessh Ramakrishnan}
\affiliation{Astronomy Department, Universidad de Concepci\'on, Casilla 160-C, Concepci\'on, Chile}
\affiliation{Finnish Centre for Astronomy with ESO, FI-20014 University of Turku, Finland}
\affiliation{Aalto University Mets\"ahovi Radio Observatory, Mets\"ahovintie 114, FI-02540 Kylm\"al\"a, Finland}

\author[0000-0003-3826-5648]{Paul Tiede}
\affiliation{Center for Astrophysics $|$ Harvard \& Smithsonian, 60 Garden Street, Cambridge, MA 02138, USA}
\affiliation{Black Hole Initiative at Harvard University, 20 Garden Street, Cambridge, MA 02138, USA}

\author[0000-0002-9036-2747]{Ed Fomalont}
\affiliation{National Radio Astronomy Observatory, 520 Edgemont Road, Charlottesville, 
VA 22903, USA}

\author[0000-0002-5297-921X]{Sara Issaoun}
\affiliation{Center for Astrophysics $|$ Harvard \& Smithsonian, 60 Garden Street, Cambridge, MA 02138, USA}
\affiliation{NASA Hubble Fellowship Program, Einstein Fellow}

\author[0000-0002-8247-786X]{Joey Neilsen}
\affiliation{Villanova University, Mendel Science Center Rm. 263B, 800 E Lancaster Ave, Villanova PA 19085}

\author[0000-0001-6923-1315]{Michael A. Nowak}
\affiliation{Physics Department, Washington University CB 1105, St Louis, MO 63130, USA}

% ...

%% file: S1_Introduction.tex
\section{Introduction}

Several years after its initial identification \citep{balick:74a}, the
radio source at the center of our Galaxy, now associated with the
supermassive black hole Sagittarius
A$^\ast$ (\sgra),  was discovered to be significantly
variable at radio frequencies \citep{brown:82a}. Variations of tens of percent over
year-long timescales had been recognized, with convincing evidence
for variability on timescales of $\gtrsim 1$\,day, and factor of four
variations occurring on timescales $\lesssim 10$\,days
\citep{wright:93a}. It was noted that ``flickering noise'' was
certainly possible on shorter timescales as well \citep{brown:82a}.

After \chandra's discovery of rapid X-ray flares from \sgra \citep{baganoff:01a}, however, many of the subsequent studies of its multi-wavelength variability focused on impulsive events, where the flux could grow by a factor of several tens on short timescales. The first observed X-ray flare had a duration $\approx 10$\,ks \citep{baganoff:01a}, i.e., the light crossing time for a diameter of $\approx 500\,GM/c^2$, or roughly the orbital timescale at $\approx20\,GM/c^2$ for a Schwarzschild black hole given the $\sim4\times10^6 \mathrm{M_\odot}$ mass of \sgra \citep{ghez:08a,gillessen:09a,boehle:16a,gillessen:17a,Gravity2018,Gravity2019,Do2019}. All subsequently observed X-ray flares \citep[see,e.g.,][]{Porquet2003,neilsen:13a,neilsen:15a,li:15a,ponti:15a,yuan:16a,bouffard:19a,Haggard2019} have occurred on timescales ranging from 0.4--10\,ks, with the short timescale being limited by counting statistics, and longer flares apparently being absent from the data \citep{neilsen:13a,neilsen:15a}.

Similar impulsive variability at other wavelength bands -- millimeter/submillimeter (mm/sub-mm) and infrared (IR) -- have also steered variability studies of \sgra,
especially because the first detected IR variability occurred on the short orbital timescales of the inner regions \citep{genzel:03a,ghez:04a}. The parallels to the X-ray flares have led to a strong focus on studying the flare/radiation mechanism and the relationship between the different wavebands. For cases where flares were observed simultaneously in both IR and X-ray light curves, the IR variability was not delayed from the X-ray by more than $\sim$10--15\,min \citep{eckart:04a,eckart:06a,Marrone2008}.  This suggests that the IR and X-ray emission predominantly arise from the same regions. The most recent and comprehensive analysis of X-ray to IR variability is consistent with no delay at 99.7\% confidence, but at 68\% confidence, allows for a 10--20\,min delay of the IR \citep{boyce:19a}. Multi-wavelength lags including the mm and sub-mm are more complex. The mm flux density maxima typically have a far lower relative flux density gain than flares at higher frequencies and are often delayed by 1--2\,h \citep{yusef-zadeh:08a,eckart:12a}, although \citet{Marrone2008} and \citet{Witzel2021} report delays as short as  20--30\,min and \citet{fazio:2018} report a flare with a negligible mm--IR lag. The lack of high fidelity mm light curves, and the sparse sampling compared to the IR and X-ray, have limited detailed variability and cross-correlation studies, and it has been suggested that the perceived delays between mm and IR/X-ray may in fact just be coincidental \citep{capellupo:17a}.

 Recently, the increase in quality of the \sgra IR light curves has allowed one to go beyond the studies of individual flare events and led to more detailed statistical and variability modeling over a wide range of timescales, spanning minutes to hours. Various groups have characterized the IR light curves with a red noise Fourier power spectral density (PSD; approximately $\propto f^{-2}$) on timescales longer than a few minutes, with a break to a flat, white noise PSD on timescales longer than $\approx 3$\,h \citep{meyer:09a,do:09a}. Consideration of the shortest timescales has mostly been limited by the signal-to-noise ratio (\textit{S/N}).
 Equivalently, the structure function analyses have revealed a similar result: variances consistent with the unstructured white noise on timescales longer than a few hours, and consistent with red noise on hour to minute timescales \citep{do:09a,witzel:18a,Witzel2021}. Although
 periodic signals have been searched for in the IR light curves \citep[e.g.,][]{genzel:03a}, no convincing signatures that could not instead be attributed to limited sampling of red noise have been found.

It has been only relatively recently that the quality of mm
light curves for \sgra has begun to match that in the IR, such that
a similar detailed analyses can be applied to describe the mm
behavior of \sgra on timescales from minutes to hours. In particular, \citet{Dexter2014} have shown that, similar to the IR variability, mm light curves indicate red noise characteristics, with a break to white noise on longer timescales. Additionally, detailed studies of the \sgra mm and sub-mm emission have been
enabled by high \textit{S/N} observations using the Atacama Large
Millimeter/submillimeter Array \citep[ALMA;][]{Brinkerink2015,Bower2015,Bower2018,Bower2019} and the Submillimeter Array \citep[SMA;][]{Bower2015,fazio:2018,Witzel2021}. Further, short timescale variability of \sgra mm ALMA light curves has been analyzed by \citet{Iwata2020}, based on 10 epochs of 70\,min duration.

In this work, we present the detailed analysis of ALMA and SMA light curves of \sgra obtained during the observing campaign of the Event Horizon Telescope \citep[EHT;][hereafter Papers I, II, III, IV, V, and VI]{SgraP1,SgraP2,SgraP3,SgraP4,SgraP5,SgraP6} on 2017 April 5--11. These observations consist of 5 days of SMA and 3 days of ALMA monitoring of the source for 3--10\,h per day. They constitute an uniquely long, homogeneously processed, high cadence and high \textit{S/N} mm \sgra light curve data set. We compare these observations with the historic data available at 230\,GHz. During the 2017 EHT observations, \sgra was mostly in a low variability state, with slowly varying mm flux density of 2--3\,Jy. However, on 2017 April 11, ALMA observations immediately followed a 5.5\,ks X-ray flare seen by \chandra, peaking at about 8.8 UT \citepalias{SgraP2}. The mm variability on that day was strongly enhanced, with the flux density growing by about 50\% and reaching a maximum at about 10.98 UT, 2.2\,h after the X-ray peak.

This paper is organized as follows. In Section \ref{sec:observations}, we
discuss the observations and non-standard data reduction procedures dedicated to extracting the compact source emission from the phased array data. Section \ref{sec:consistency} discusses overall data properties and consistency between individual data sets, as well as the spectral index measurements. 
In Section \ref{sec:variability}, we compare the new observations with the archival mm data sets, and characterize the variability of the light curves with correlations and structure functions. We then model the data using Gaussian process models in Section \ref{sec:modeling}. In Section \ref{sec:periodicity}, we discuss the PSDs and search for statistically significant periodicity signatures in the data. Finally, we summarize and discuss the full results in Section \ref{sec:discussion}.

%% file: S2_Observations.tex
\section{Observations and data reduction}
\label{sec:observations}

The EHT observed \sgra in April 2017 with a very-long-baseline interferometry (VLBI) array of eight stations at six distinct geographic locations \citep[][hereafter \m87 Papers I, II]{paperi,paperii}\footnote{All EHT \m87 papers \citep{paperi,paperii,paperiii,paperiv,paperv,papervi} will be hereafter referred to as \m87 Paper I, II, III, IV, V, and VI, respectively.}. A detailed analysis of the VLBI observations of \sgra on 2017 April 6 and 7 is presented in \citetalias{SgraP1,SgraP2,SgraP3,SgraP4,SgraP5,SgraP6}. Two of the participating EHT stations are connected interferometers, formed by the coherent combination of their elements; SMA located on Maunakea (Hawai`i, USA), and ALMA located on the Chajnantor plateau (Atacama Desert, Chile). An advantage of using connected-element interferometers as EHT stations, besides the enhanced sensitivity of the resulting array, is that it is possible to compute the coherency matrices among their connected elements simultaneously with the summed signals that are recorded for their later use in VLBI \citep{Goddi2019}. Therefore, as a by-product of EHT VLBI observations, we can make use of the connected-element visibilities to obtain \sgra light curves with long duration, high cadence, and high \textit{S/N}. 
Apart from their standalone scientific value, the light curve products are also employed downstream in the EHT VLBI data calibration, see Appendix \ref{appendix:VLBIfeedback}.

Utilizing observations in the VLBI mode to produce \sgra light curves allows us to access the particularly long observing windows needed for the VLBI aperture synthesis, at the cost of using a phased array in a compact configuration with relatively low resolution, and observing the source partly at an unusually low elevation. Since this is a non-standard procedure, that could be employed for similar observations in the future, in this paper we dedicate some additional effort to address the comparisons between data reduction pipelines and to recommend procedures for future VLBI observing campaigns.

The ALMA observations were carried out across four frequency sub-bands (spectral windows), each with a bandwidth of 2\,GHz, centered at 213.1 and 215.1\,GHz (B1 and B2, lower sideband), and 227.1 and 229.1\,GHz (LO and HI, upper sideband). ALMA observed \sgra on 2017 April 6, 7, and 11, typically with $\sim37$ dishes of 12\,m diameter in the phased array, with 4--10\,h tracks, see Table \ref{tab:detections}. The integration time used by the ALMA correlator was set to 4\,s. Due to the array phasing requirements, ALMA observed in a compact configuration, with the longest projected baselines reaching 160\,m on 2017 April 6, 278\,m on 2017 April 7, and 374\,m on 2017 April 11.

The use of the VLBI phased array mode at ALMA has several implications for the data properties and calibration procedures, as compared to standard ALMA observations \citep{matthews2018,Goddi2019}. In order to perform a proper VLBI polarization conversion of the ALMA signal streams \citep[using the \texttt{PolConvert} program,][]{martividal2016}, the official ALMA reduction scripts needed to be adapted in such a way that their final products are not ready for scientific use of the ALMA-only data. In particular \citep[][]{Goddi2019}:
  
\begin{itemize}

\item The ALMA phasing efficiency has to be computed at each
  integration time of the correlator, using the same subset of
  antennas that are present in the VLBI signal, regardless of the data quality of each phased element, as well as of any other factor that would imply the removal of the data under normal circumstances (e.g., shadowing among antennas). Therefore, low-quality data cannot be edited before the calibration, hence degrading the final product.

\item The system temperatures of each individual antenna are not applied to the calibration tables. Instead, a global system temperature is computed and applied to the summed signal. The effects of atmospheric opacity are computed from the overall system temperatures and then stored in the VLBI metadata. As a consequence, the opacity correction is provided in the VLBI metadata, but it is not present in the ALMA-only calibrated visibilities. 

\item In ordinary ALMA observations, amplitude calibration uses a primary flux density calibrator, e.g., a Solar System object or a monitored quasar. The calibration is then extrapolated to the secondary calibrator and bootstrapped into the target. However, when working in VLBI mode, we need to self-calibrate each ALMA subscan (16\,s segment), which implies the need to use an a priori (constant) model for the flux density of \sgra.
\end{itemize}

These limitations in the official quality assurance (QA2) calibration of the ALMA VLBI observations can be overcome with the development of independent calibration scripts, which have to handle the aforementioned peculiarities of the ALMA phasing system.
In an attempt to further limit and characterize the influence of the systematic calibration errors on the analysis, we have corrected the limitations of the QA2 calibration of the \sgra observations using two independent procedures (A1 and A2), described in more detail in the following subsections along with the SMA data reduction procedures. We consider the A1 pipeline to be the most self-consistent and reliable, given the robust assumption of a lack of structural variability of a parsec-scale image on a timescale of several days, enabling the time-dependent self-calibration of the amplitude gains. Nevertheless, comparing the two pipelines offers a valuable insight into the potential systematic errors corrupting mm light curve observations. These effects are quantified and discussed in Section \ref{sec:consistency}.

\begin{table*}[th]
     \caption{\sgra light curves presented in this paper. $\mu$ and $\sigma$ denote signal mean and standard deviation, respectively.
     }
    \begin{center}
    \tabcolsep=0.17cm
    \begin{tabularx}{0.98\linewidth}{cccccccccccc}
    \hline
    \hline
    Array &  Reduction & Band &  Day & $t_{\rm start}$ & $t_{\rm stop}$ & Duration & Samples & Flux & Modulation & max-min \\
     &   & (GHz)  &  in 2017 & UT (h) & UT (h) & (h) &  & $\mu \pm \sigma$ (Jy) & $\sigma/\mu$ &(Jy) \\
    \hline
    ALMA & A1 & B1 &  April 6 & 8.40 & 14.15 & 5.75 &2226& 2.59 $\pm$ 0.11 & 0.042 & 0.57\\
     & cadence: 4\,s & 212.1--214.1 &  April 7 & 4.39 & 14.07 &9.68 & 3549 & 2.34 $\pm$ 0.16 & 0.068 &0.68 \\
     &  min. elev.: 25$^\circ$ &  &  April 11 & 9.00 & 13.09 & 4.09 &1663& 2.44 $\pm$ 0.31 &0.127& 1.16\\
     
     \cline{3-11}
      & \textit{S/N} $\sim$ 1300  & B2 &  April 6 & 8.40 & 14.15 & 5.75 &2232& 2.59 $\pm$ 0.11 & 0.042 &0.56\\
     &  & 214.1--216.1  &  April 7 & 4.39 & 14.07 &9.68 & 3541 & 2.34 $\pm$ 0.16 & 0.068 & 0.65\\
     &    &  &  April 11 & 9.00 & 13.10 & 4.10 &1654& 2.43 $\pm$ 0.31 & 0.128& 1.20\\
     
     \cline{3-11}
      &  & LO &  April 6 & 8.40 & 14.15 & 5.75 &2222& 2.50 $\pm$ 0.09 & 0.036 &0.51\\
     &  & 226.1--228.1  &  April 7 & 4.39 & 14.07 &9.68 & 3507 & 2.26 $\pm$ 0.16 & 0.071 &0.69 \\
     &    &  &  April 11 & 9.00 & 13.09 & 4.09 &1645& 2.31 $\pm$ 0.29 & 0.126& 1.25\\
     
      \cline{3-11}
      &  & HI &  April 6 & 8.40 & 14.15 & 5.75 &2222& 2.59 $\pm$ 0.11 & 0.042 &0.57\\
     &  & 228.1--230.1  &  April 7 & 4.39 & 14.07 &9.68 & 3488 & 2.33 $\pm$ 0.16 & 0.067 &0.72 \\
     &    &  &  April 11 & 9.00 & 13.09 & 4.09 &1624& 2.40 $\pm$ 0.31 & 0.129 & 1.27\\
     
     \cline{2-11}

      & A2 & B1 &  April 6 & 8.40 & 13.75 & 5.35 &519& 2.48 $\pm$ 0.12 & 0.048 &0.49 \\
     & cadence: 18\,s & 212.1--214.1  &  April 7 & 4.93 & 13.57 &8.64 & 785 & 2.32 $\pm$ 0.12 & 0.052 &0.56\\
     &  min. elev.: 30$^\circ$  &  &  April 11 & 9.00 & 13.15 & 4.15 &461& 2.11 $\pm$ 0.25 & 0.118 &0.95\\
     
     \cline{3-11}
     
      & \textit{S/N} $\sim$ 400  & B2 &  April 6 & 8.40 & 13.75 & 5.35 &519& 2.49 $\pm$ 0.12 & 0.048 &0.49\\
     &  & 214.1--216.1  &  April 7 & 4.93 & 13.57 &8.64 & 785 & 2.33 $\pm$ 0.12 & 0.052 & 0.57\\
     &    &  &  April 11 & 9.00 & 13.15 & 4.15 &460& 2.12 $\pm$ 0.25 & 0.118& 0.95\\
     
     \cline{3-11}
     
      &  & LO &  April 6 & 8.40 & 13.75 & 5.35 &519& 2.41 $\pm$ 0.12 & 0.050 &0.49 \\
     &  & 226.1--228.1  &  April 7 & 4.93 & 13.57 &8.64 & 786 & 2.23 $\pm$ 0.12 & 0.054 &0.56 \\
     &    &  &  April 11 & 9.00 & 13.15 & 4.15 &371& 2.12 $\pm$ 0.21 & 0.099 & 0.93\\
     
       \cline{3-11}
     
      &  & HI &  April 6 & 8.40 & 13.75 & 5.35 &518& 2.51 $\pm$ 0.13 & 0.052 &0.51\\
     &  & 228.1-230.1  &  April 7 & 4.93 & 13.57 &8.64 & 786 & 2.32 $\pm$ 0.12& 0.052 &0.59\\
     &    &  &  April 11 & 9.00 & 13.15 & 4.15 &331& 2.22 $\pm$ 0.21 & 0.095 & 0.98\\
     
     \cline{1-11}
     
      SMA & SM & LO &  April 5 & 11.30 & 15.71 & 4.41 &165& 2.48 $\pm$ 0.14 & 0.056 & 0.66\\
       & cadence: 62\,s  & 226.1--228.1 &  April 6 & 11.24 & 14.54 & 3.30 &148& 2.59 $\pm$ 0.08 & 0.031 & 0.41\\
     & min. elev.: 15$^\circ$ &   &  April 7 & 11.17 & 15.59 &4.42 & 199 & 2.29 $\pm$ 0.11 &0.048 & 0.46 \\
     & \textit{S/N} $\sim$ 60 &   &  April 10 & 10.98 & 14.77 &3.79 & 157 & 2.50 $\pm$ 0.12 & 0.048 &0.54\\
     &    &  &  April 11 & 10.92 & 15.00 & 4.09 &188& 2.60 $\pm$ 0.29 & 0.112 &0.88\\
     
     \cline{3-11}
     
         &  & HI &  April 5 & 11.30 & 15.71 & 4.41 &165& 2.49 $\pm$ 0.15 & 0.060 & 0.69\\
       &   & 228.1--230.1 &  April 6 & 11.24 & 14.54 & 3.30 &148& 2.62 $\pm$ 0.08 & 0.031  & 0.43\\
     &  &   &  April 7 & 11.17 & 15.59 &4.42 & 199 & 2.30 $\pm$ 0.12 &0.052 &0.47\\
     &  &   &  April 10 & 10.98 & 14.77 &3.79 & 157 & 2.50 $\pm$ 0.12 & 0.048 &0.55 \\
     &    &  &  April 11 & 10.92 & 15.00 & 4.09 &188& 2.62 $\pm$ 0.29& 0.111 &0.90\\

    \hline 
    \hline
    FULL & ALMA+SM & LO & April 5--11 & - & - & 147.70 & 7874 & $2.36 \pm 0.22$ & 0.094 & 1.37\\
    FULL & ALMA+SM & HI & April 5--11 & - & - & 147.70 & 7834 & $2.44 \pm 0.23$ & 0.094 & 1.44\\
\hline 
\hline

    \end{tabularx}
    \label{tab:detections}
    \end{center}
\end{table*}

The SMA observations were carried out across 8 sub-bands, each with a bandwidth of 2 GHz, covering a range of frequencies between 208.1\,GHz and 232.1\,GHz. In this paper, we focus on the 227.1 and 229.1 GHz bands (LO and HI, respectively), corresponding to the two bands used for the VLBI observations with the EHT in 2017 \citepalias{paperii,paperiii}. The light curves from the other frequency sub-bands are very consistent and are summarized in Appendix \ref{appendix:SMA}. The SMA observed \sgra on 2017 April 5, 6, 7, 10, and 11, with shorter observing tracks lasting 3.3--4.4\,h, starting at a later time when compared to ALMA. The SMA observed the source with 6--7 dishes of 6\,m diameter and a correlator integration step of 10.4\,s. 

For both stations, observations were arranged into scans lasting typically 5--10\,min, interleaved with observations of calibrators \citepalias{SgraP2}. The \sgra light curve data sets analyzed in this paper are summarized in Table \ref{tab:detections}. 

\begin{figure*}
    \centering
    \includegraphics[width=0.96\textwidth]{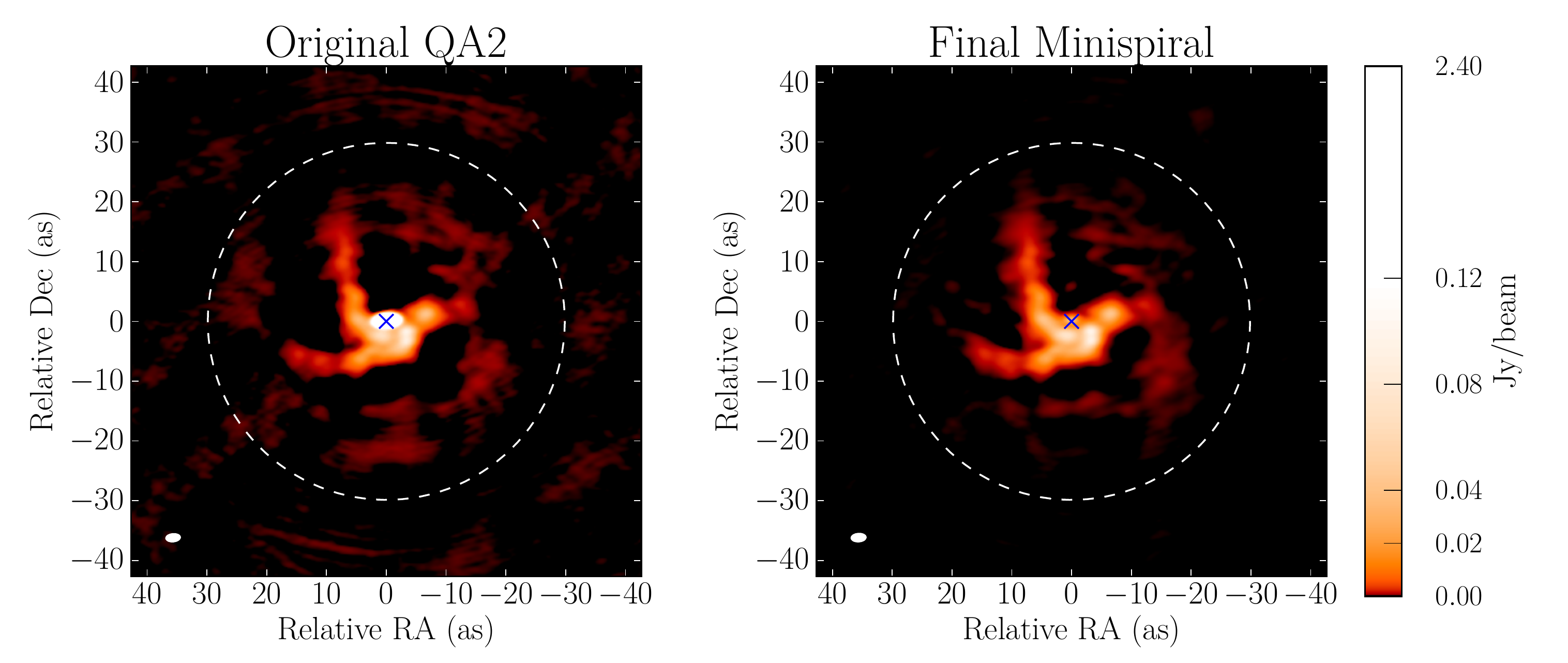}
    \caption{{\it Left:} An image of the \sgra field obtained from the
      original QA2 calibrated data, using natural weighting and a Gaussian taper in Fourier space to boost the sensitivity to the extended (minispiral) structure. {\it Right:} A final image of the minispiral, after applying the intra-field calibration and removing the signal from \sgra. In each figure, the convolving beam is shown in the bottom left and the location of \sgra is marked with a cross. The dashed line marks the region where the primary-beam response of the ALMA antennas is above 5\%.}
    \label{SgraMinispiral:fig}
\end{figure*}

\subsection{A1: intra-field flux density ALMA calibration}
\label{IntraField:subsec}

The \sgra\ image field of view can be split into two components at angular
scales of arcseconds probed by ALMA:
\begin{enumerate}
    \item An extended structure with a low brightness temperature, primarily originating from thermal emission from ionized gas and dust infalling into the central region of the
Galaxy, the so-called ``minispiral'' \citep[e.g.,][]{Lo1983,Goddi2021}. From our ALMA observations, the integrated extended flux density of the minispiral is $\sim$1.1\,Jy. Given the physical origin of this emission and the spatial scales involved (several tens of parsecs), we can assume the brightness distribution of the minispiral to remain constant during the few days of the EHT observing campaign.

\item An unresolved and highly variable component corresponding to the compact source \sgra, with a flux density typically ranging between 2--5\,Jy at 230\,GHz. 
\end{enumerate}
With the superb sensitivity of ALMA \citepalias{paperiii,SgraP2}, and the sufficiently high integrated extended flux density of the minispiral, it is possible to detect the whole field of view structure (i.e., the minispiral plus \sgra) in each ALMA 4\,s snapshot. Therefore, one can assume a two-component Fourier domain model, $V^{\rm mod}_{t}$, composed of (1) $\mathcal{F}^{e}$ -- a Fourier transform of the static extended minispiral, $F^e$, corrupted with a time-dependent amplitude gain, $G_t$, accounting for atmospheric and instrumental effects (following the QA2 calibration these effects can be modeled with a single function of time, representing the effective gain of the interferometric array), and (2) $F_t$ -- an unresolved \sgra compact component with a time-dependent flux density (still corrupted by the $G_t$ gain at this stage), so,
\begin{equation}
V^{\rm mod}_{t} = G_t \mathcal{F}^e + F_t  \ .
\label{eq:minispiral_model}
\end{equation}
If we denote the visibility observed at a time $t$ on a baseline $i$ as $V^{\rm obs}_{i,t}$, and the model sampled at the same Fourier plane location as $V^{\rm mod}_{i,t}$, the model can be fitted to the data by minimizing,

\begin{equation}
\chi^2_t(G_t,F_t) = \sum_i{ \omega_{i,t} \left| V^{\rm obs}_{i,t} - V^{\rm mod}_{i,t} \right|^2} \ 
\label{Chi2Minispiral:eq}
\end{equation}
for each time $t$, with \textit{S/N} based baseline weights, $\omega_{i,t}$, and the summation extending over all baselines available at a given time $t$ \citep{uvmultifit2014}.

Since the true integrated flux density of the minispiral is assumed to be constant, we can use the values of $G_t$ to remove the residual corruption effects in the \sgra flux density estimates, $F_t$. Hence, we produce a corrected estimate of the \sgra flux
density, $F^c_t$, using the equation,
\begin{equation}
F^c_t = \frac{F_t}{G_t}.
\label{CorrFlux:eq}
\end{equation} 

In practice, we also need to solve for the image domain minispiral model, $F^e$. We use the CLEAN algorithm \citep[e.g.,][]{Hogbom1974} implemented in the Common Astronomy Software Application (CASA) framework \citep{CASA2007} as the \texttt{tclean} task, iteratively reconstructing the image of the minispiral, recalibrating the data with $G_t$, and updating $G_t$ and $F_t$. While the minispiral structure is assumed to be constant across observed frequencies, the absolute flux density scale is allowed to vary between the sub-bands. The procedure runs until convergence. The times with unphysical or unconverged $(G_t, F_t)$ are flagged. Special attention is given to the minispiral total flux density, which is fixed per sub-band to the median of the flux densities estimated from all of the snapshots obtained throughout the EHT campaign. 

In Figure \ref{SgraMinispiral:fig}, we show two model images of the field around \sgra; the left image corresponds to the original QA2 calibrated data (the initial condition for the iterative procedure) and shows the complete source structure (minispiral and \sgra); the right image is the final minispiral model obtained after the convergence of the intra-field calibration. The high noise level seen in the QA2 image (i.e., the artifacts that are distributed across the whole field of view) is due to the effects of the time variability of \sgra. By modeling the source variability with Equation \ref{eq:minispiral_model}, the noise level in the final minispiral image (Figure \ref{SgraMinispiral:fig}, right) is reduced. We notice that the size of the ALMA primary beam at 230\,GHz is of the order of the size of the minispiral structure. This implies that different points across the image are affected by a different ALMA sensitivity. Only emission from regions with a primary-beam response $> 5$\% can be considered as detections above 5\,$\sigma$ of the ALMA sensitivity. As we approach this primary-beam threshold, marked with a dashed line in Figure \ref{SgraMinispiral:fig}, the noise effects in the brightness distribution increase. Regions far from the phase center have a negligible contribution to the visibilities, since such a contribution also scales with the primary-beam response. The effects of these regions on the calibration of the \sgra light curves will thus be small.

\begin{figure}
    \centering
    \includegraphics[width=0.49\textwidth]{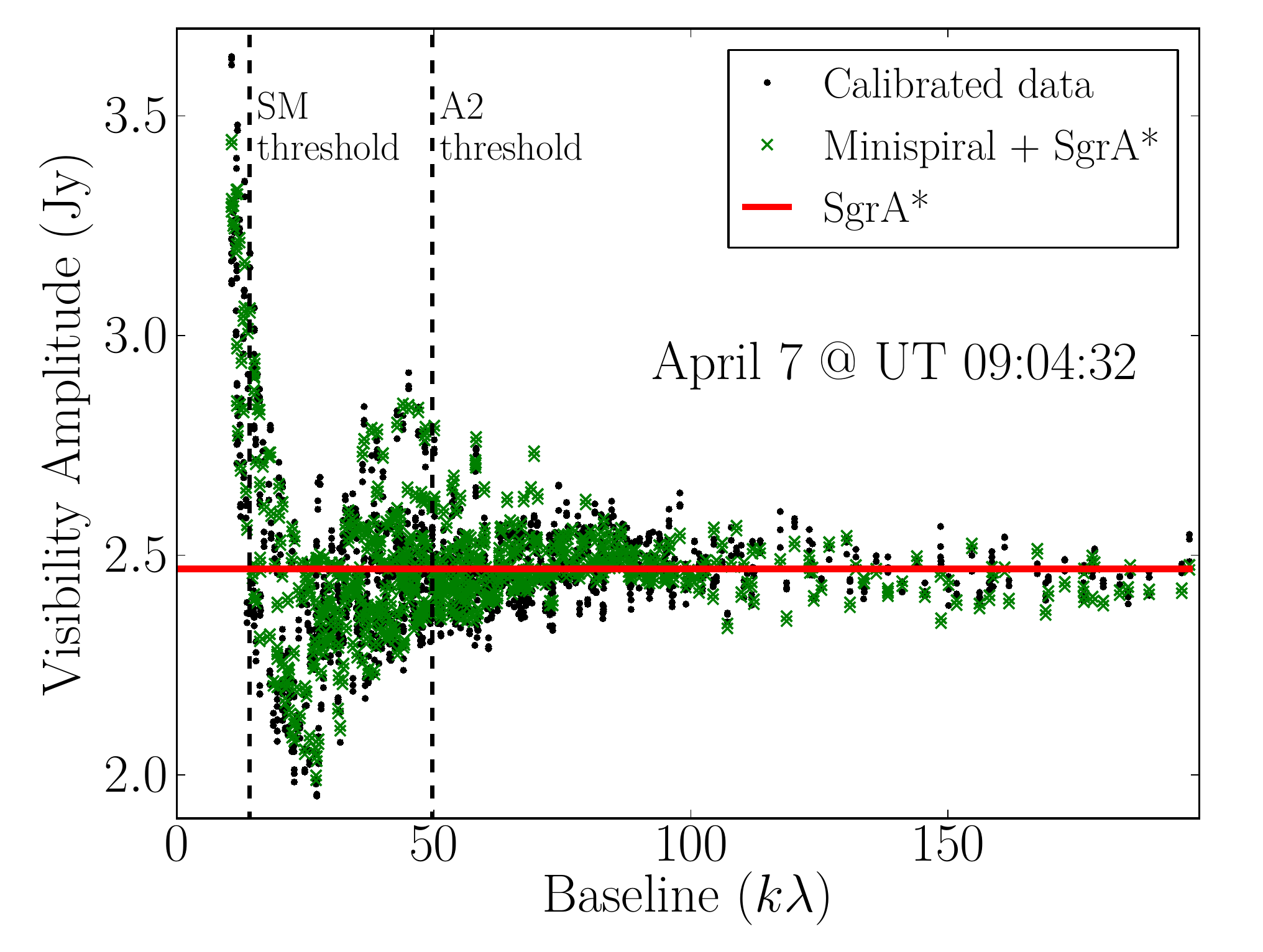}
    \caption{Calibrated visibility amplitudes of \sgra for band B1 on 2017 April 7 within a snapshot taken at 9:04:32 UT. The green crosses are the total model prediction (i.e., minispiral plus \sgra). The red line shows the instantaneous flux density of the compact unresolved \sgra. The vertical dashed lines indicate the flagging thresholds used in the A2 and SM pipelines.}
    \label{fig:uvdist}
\end{figure}

In Figure \ref{fig:uvdist}, we show the visibility amplitudes for a representative ALMA snapshot from 2017 April 7, band B1. The contribution of the extended minispiral model (green crosses) at short baselines is clearly visible. As a final product, we obtain converged light curves of \sgra with a snapshot cadence of 4\,s. Data corresponding to a source elevation below 25$^\circ$, exhibiting significant quality loss, are flagged.
The time-dependent factor, $G_t$, can also be used to correct the amplitudes of the VLBI visibilities related to the phased ALMA. Based on \cite{martividal2016}, the scaling gain factor to correct ALMA amplitudes on VLBI baselines is $\sqrt{G_t}$. This approach has been employed for a priori amplitude calibration of the EHT VLBI \sgra data \citepalias[see Appendix \ref{appendix:VLBIfeedback}, and][]{SgraP2}.

\subsection{A2: SEFD-based ALMA calibration} 

A custom script was prepared to process the non-standard array data acquired during the phased ALMA observations of \sgra using measurements of the system equivalent flux density (SEFD) of each ALMA antenna. While similar to the standard ALMA QA2 pipeline, it includes additional calibration steps necessary to produce the time-dependent light curve data.
The ALMA observations are grouped in scans, which consist of subscans of 18\,s cadence, with 16\,s of correlated data  \citep{Goddi2019}. A nearby phase calibrator, J1744--3116 (J1744-312), was observed for 30\,s every 20 min. Observations of two bright quasars, NRAO\,530 (J1733--1304, B1730--130) and J1924--2914 (B1921--293), were also included for the amplitude calibration, see Table \ref{table_sma_calibration}.
First, the phase delays associated with the atmospheric water vapor were estimated from measurements of the 183\,GHz water line, performed with high time cadence using radiometers located at each ALMA antenna. The radiometer measurements allowed us to estimate the column of water vapor above each ALMA antenna, which were then converted into a phase correction related to the atmospheric optical path.
Conversion from the relative visibility correlation amplitude to a flux density scale was performed by applying the system temperature measurements performed routinely at each antenna. The corrected data were concatenated and reduced to
produce a single CASA measurement set \citep{CASA2007} for each observing day, containing relevant data for all four sub-bands.

The second step was the bandpass calibration of all of the frequency channels in each sub-band. We used NRAO\,530 to generate the bandpass calibration tables, choosing a scan when the source was nearest to the zenith. The chosen reference antenna was located near the array center, and not shadowed by neighboring antennas.
Lower sensitivity channels near the edges of each sub-band, and data from shadowed antennas, were flagged. Following bandpass calibration, all channels within each sub-band were averaged.

In the third step, we determined the amplitude scale of the observations on
each day, and we applied the phase referencing calibration. Since all three
calibrators were observed as unresolved point sources, anomalously low amplitudes on some antennas were apparent. The few low amplitude data points, with less than about 70\% of the nominal sensitivity of the majority of antennas, were subsequently flagged. Next, the flux
density scale over the entire observation was set using NRAO\,530 as a
flux density calibrator, assuming a flux density of 1.56\,Jy at 213.1\,GHz and a spectral
index of -0.72, obtained from the ALMA calibrator catalogue\footnote{\url{https://almascience.eso.org/alma-data/calibrator-catalogue}}. The flux density calibration was applied using the average gain of the two polarizers of each antenna (X and Y), so that there was no gain bias caused by the source linear polarization. 

The flux density of the other two quasar calibrators were verified to be constant over each observation day to within a few percent. 
The last calibration step, phase-referencing between J1744--3116 and \sgra, was performed by deriving the antenna-based phase for each J1744--3116 scan with the CASA task \texttt{gaincal}, and then interpolated to each \sgra scan, which completed the calibration cycle.

The above steps provide calibrated complex visibilities from which an image containing a strong point source, \sgra, and the extended minispiral emission can be obtained. Since the phase calibrator J1744--3116 was observed only every 20\,min, the interpolated phase correction can deviate from the true values. To determine the time variability of \sgra in the presence of extended emission and calibration phase errors, we first flag baselines shorter than 70\,m (about 50 k$\lambda$, Figure \ref{fig:uvdist}).
Since virtually all of the extended emission is resolved out on longer baselines, the remaining data are reasonably consistent with a point source model, although its position may vary in time. Subsequently, given a large number of available long baselines,
we perform phase self-calibration to remove the residual phase errors remaining after the calibration with J1744--3116. The phase self-calibration algorithm determines phase corrections for each antenna and time segment, producing a set of visibilities consistent with a point source at a fixed location.

We then reconstruct CLEAN \citep{Hogbom1974} images corresponding to the phase self-calibrated long-baseline data on timescales of individual subscans. These images correspond to a near-perfect point source with a flux density equal to that of \sgra at each short time period. The relevant flux density and error estimates were obtained by fitting the CLEAN image using the CASA task \texttt{imfit}. The sequence of these short-time flux density measurements
define the time-dependent light curve of \sgra for each observing day. Finally, data corresponding to source elevations below 30$^\circ$ were found to be of poor quality and self-consistency, and were subsequently flagged in the final data set.
\begin{deluxetable}{cccc}
\tablecaption{Calibrators used in ALMA and SMA data reduction.}
\label{table_sma_calibration}
\tablehead{ \colhead{Day}     & \colhead{Bandpass}  & \colhead{Flux}  & \colhead{Gain}  }
\startdata
\hline
\multicolumn{4}{c}{ALMA A2$^{a}$}\\
\hline
April 6   & NRAO\,530 & NRAO\,530 & J1744--3116 \\
April 7   & NRAO\,530 & NRAO\,530 & J1744--3116 \\
April 11  & NRAO\,530 & NRAO\,530 & J1744--3116 \\
\hline
\multicolumn{4}{c}{SMA}\\
\hline
April 5   & 3C\,279   & Callisto  & NRAO\,530 \& J1924--2914  \\
April 6   & 3C\,273   & Ganymede  & NRAO\,530 \& J1924--2914  \\
April 7   & 3C\,454.3 & Ganymede  & NRAO\,530 \& J1924--2914  \\
April 10  & 3C\,279   & Titan     & 1749+096  \\
April 11  & 3C\,279   & Callisto  & J1924--2914 \\
\enddata
\tablenotetext{a}{J1924--2914 was also used as a flux calibration consistency check.}
\end{deluxetable}

\subsection{SM: SMA calibration and reduction} 
\label{sec:calibration:SMA}

An initial pass through the SMA data was performed with a custom MATLAB\footnote{Mathworks, Version 2019b; \url{http://www.mathworks.com/products/matlab/}} based reduction pipeline, primarily responsible for preliminary flagging and bandpass calibration. Bandpass calibration was performed using various bright calibrators, given in Table \ref{table_sma_calibration}. After these steps, the bandpass corrections were applied to the data, after which they were spectrally averaged down by a factor of 128, to a channel resolution of 17.875 MHz.

After averaging, a second round of bandpass solving was performed, and the solutions were inspected to verify that the gain corrections were consistent with unity (as the data had already been bandpass corrected). The absolute flux density scale was set by using the flux density calibrator observed closest to the time of the \sgra observations, also noted in Table \ref{table_sma_calibration}, using the Butler--JPL--Horizons 2012 models\footnote{ALMA Memo \#594}, on a spectral window by spectral window basis. 
Next, amplitude gains for individual antennas were derived using bright quasars, NRAO\,530 and J1924--2914. Analysis of the gain solutions showed that the most significant trends are correlated with the elevation of the gain calibrator, consistently with known issues with antenna pointing on the SMA at very low elevations ($\sim15^\circ$, data corresponding to lower elevations were flagged). In light of this, gain amplitudes were interpolated using a third order polynomial fitted based on the elevation of \sgra, with the median amplitude elevation-dependent correction below $25^\circ$ being approximately 5\%.

Due to the rapid fluctuations in the instrumental phase arising from the real-time phasing loop used in VLBI beamforming, the first 3 integrations ($\approx30$\,s in total) were flagged whenever the telescopes moved onto \sgra from a calibrator source.
Additionally, due to strong line absorption, presumably arising from CN foreground absorption \citep[see Appendix H.1. of][]{Goddi2021}, spectral channels between 226.6 and 227.0\,GHz were flagged.

After amplitude-only gain calibration and the aforementioned flagging, a round of phase-only gains were derived and applied using self-calibration of \sgra itself. Data for these observations were collected while the array was in a compact configuration including baselines with lengths spanning $\sim5-50\,\textrm{k}\lambda$, which at 230\,GHz are sensitive to structure up to $\sim20^{\prime\prime}$ in size, picking up extended minispiral emission surrounding \sgra. Examination of the SMA data shows a strong uptick in visibility amplitudes at $(u,v)$-distances below $15\,\textrm{k}\lambda$ (about 20\,m).
Therefore, visibilities from shorter baselines are flagged prior to self-calibration and further analysis. The remaining long baseline data are mostly sensitive to the flux density from the unresolved \sgra point source, see Figure \ref{fig:uvdist}.

Once phase self-calibration corrections were applied, a \sgra light curve was generated by taking the naturally-weighted vector average of all baselines, for each spectral window observed by the SMA.
The resultant light curves were evaluated for large fluctuations in amplitude, under the assumption that, over a 30 s interval, changes in the brightness of \sgra should be subdominant to the instrumental noise. Where fluctuations greater than 3\,$\sigma$ were seen, the measurement in question was flagged, with the total volume of data flagged in this way amounting to $\sim1\%$. Finally, to help improve the \textit{S/N} of the data, they were time-averaged over 62\,s intervals (6 integration steps).


%% file: S3_Consistency.tex
\section{Data consistency and spectral index}
\label{sec:consistency}

The light curves from all three reduction pipelines, corresponding to the HI band (229.1\,GHz) on 2017 April 6, 7 and 11, are shown in Figure \ref{fig:LC-main}. There is overall agreement of the data features between pipelines. As a preliminary step of the analysis, we quantify the data sets consistency and investigate any potential systematic discrepancies.

\subsection{Consistency between instruments and pipelines}

\begin{figure}
    \centering
    \includegraphics[width=1.0\columnwidth]{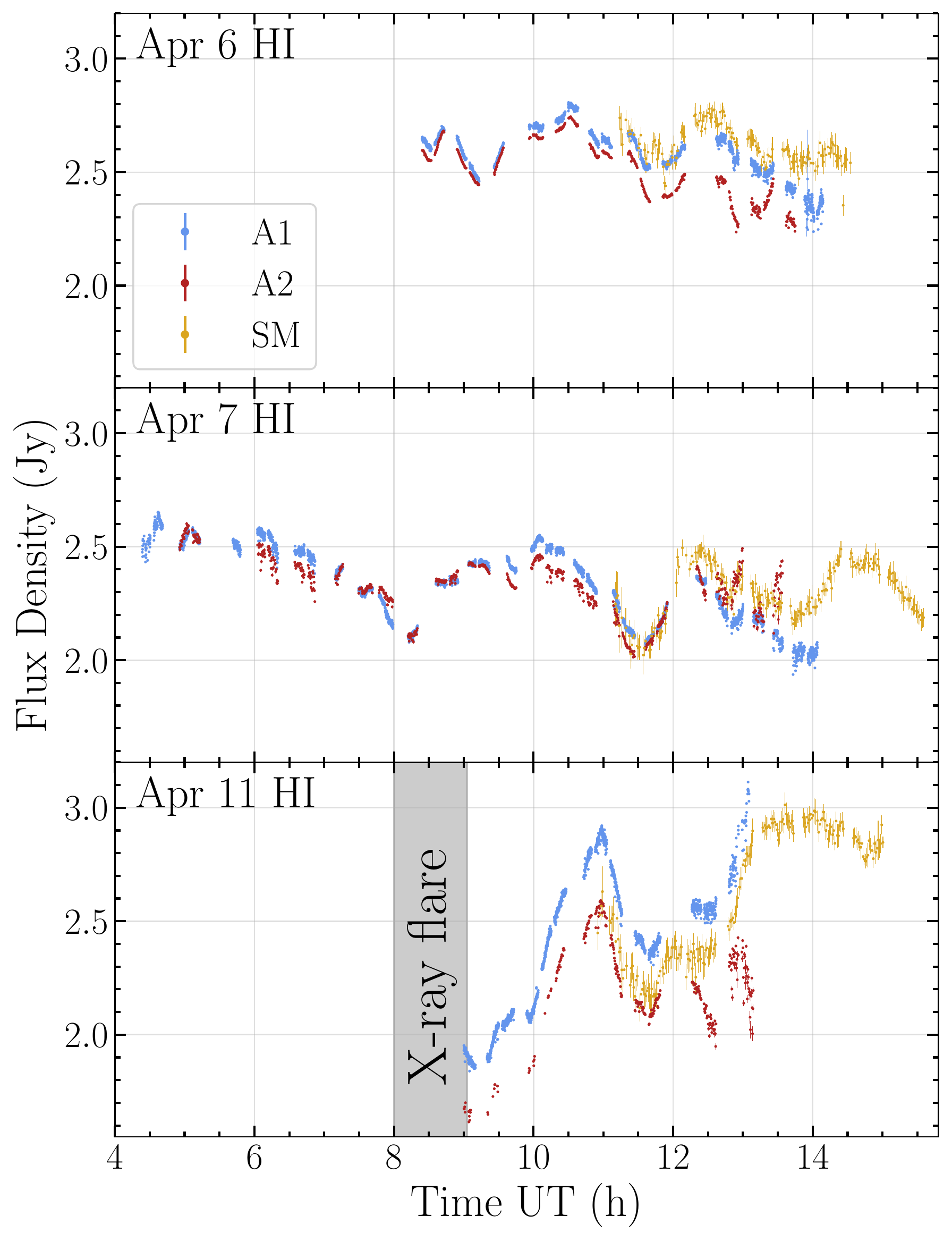}
    \caption{\sgra light curves obtained with ALMA (A1 and A2) and SMA (SM), in the HI band (229.1\,GHz) using the reduction pipelines described in Section \ref{sec:observations}. The differences between light curves originating from different pipelines are strongly dominated by the systematic calibration errors, rather than by the thermal uncertainties.}
    \label{fig:LC-main}
\end{figure}
\begin{table}[htp]
\caption{LNDCF$_0$ coefficient calculated between selected \sgra data
  sets.}
\begin{center}
\tabcolsep=0.25cm
\begin{tabularx}{0.8\linewidth}{lcccc}
\hline
Band & April 6 & April 7 & April 11 & Joined\\
\hline
\multicolumn{5}{c}{A1-A2}\\
\hline
B1    &     0.85 &  0.89 &  0.96  & 0.76\\
B2    &     0.84 &  0.87  & 0.95  & 0.76\\
LO    &     0.81 &  0.87 &  0.91  & 0.74\\
HI     &     0.87 &  0.83 &  0.92  & 0.72\\
\hline
\multicolumn{5}{c}{A1-SM}\\
\hline
LO   &      0.83 &   0.93 &  0.99  & 0.80\\
HI    &      0.87 &   0.76 & 0.99   & 0.77\\
\hline
\multicolumn{5}{c}{A2-SM}\\
\hline
LO    &     0.38 &  0.97 &  0.45  & 0.66\\
HI     &     0.59 &  0.90 &  0.44  & 0.67\\
\hline
\label{tab:lndcf0}
\end{tabularx}
\end{center}
\end{table}
The data sets were correlated through a Locally Normalised Discrete
Correlation Function (LNDCF), as defined by \citet{Lehar1992}, which
revised the standard algorithm proposed by \citet{Edelson1988},
\begin{equation}
\mathrm{LNDCF}(\Delta t)=\frac{1}{M_{\Delta t}} \sum_{i,j} \frac{ \left( a_i -\overline{a}_{\Delta t} \right) \left( b_j-\overline{b}_{\Delta t} \right) }{ \sqrt{ \left( {\sigma^2}_{a \Delta t}-e^2_{a} \right) \left( {\sigma^2}_{b \Delta t}-e^2_{b} \right) }} \ ,
\label{eq:LNDCF}
\end{equation}
where $a_{i}$ and $b_{j}$ indicate the flux density
measurements of the two compared data sets, $e_{a}$ and $e_{b}$ refer to the estimated measurement errors, and $M_{\Delta t}$ represents the
number of data pairs contributing to the lag bin, $\Delta t$. The flux
density means and standard deviations, $\overline{a}_{\Delta t},
\overline{b}_{\Delta t}, {\sigma}_{a\Delta t}, {\sigma}_{b\Delta t}$, are calculated
for each lag, $\Delta t$, using exclusively the flux density measurements that
contribute to the calculation of the $\mathrm{LNDCF}(\Delta t)$. As a comparison between the data sets, we compute the ${\rm LNDCF}(0) \equiv {\rm LNDCF}_0$, presented in Table \ref{tab:lndcf0}.

\begin{table}[htp]
\caption{Ratio of the median flux densities in the overlapping observing periods. }
\begin{center}
\tabcolsep=0.25cm
\begin{tabularx}{0.8\linewidth}{lcccc}
\hline
Band & April 6 & April 7 & April 11 & Joined\\
\hline
\multicolumn{5}{c}{A2/A1}\\
\hline
B1    &     0.96 &  1.00 &  0.86  & 0.95\\
B2    &     0.96 &  1.00  & 0.87  & 0.95\\
LO    &     0.96 &  0.99 &  0.91  & 0.95\\
HI     &     0.97 &  1.00 &  0.91  & 0.96\\
\hline
\multicolumn{5}{c}{SM/A1}\\
\hline
LO   &      1.05 &   1.08 &  0.94  & 1.03\\
HI    &      1.03 &   1.05 & 0.92   & 1.00\\
\hline
\multicolumn{5}{c}{SM/A2}\\
\hline
LO    &     1.13 &  1.06 &  1.10  & 1.09\\
HI     &     1.10 &  1.03 &  1.08  & 1.05\\
\hline
\label{tab:medians}
\end{tabularx}
\end{center}
\end{table}

\begin{figure*}
    \centering
    \includegraphics[width=1.0\columnwidth]{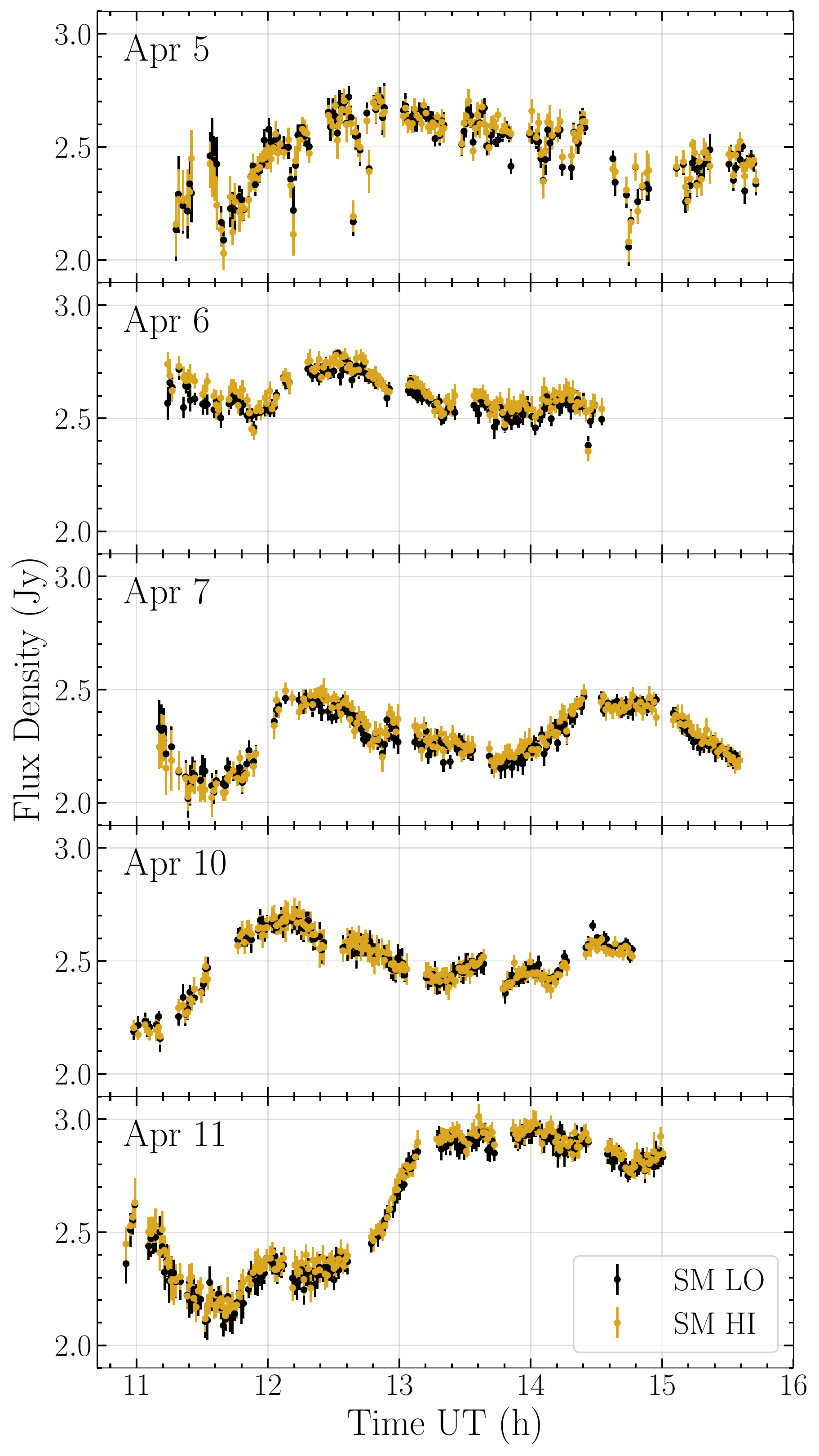}
     \includegraphics[width=1.0\columnwidth]{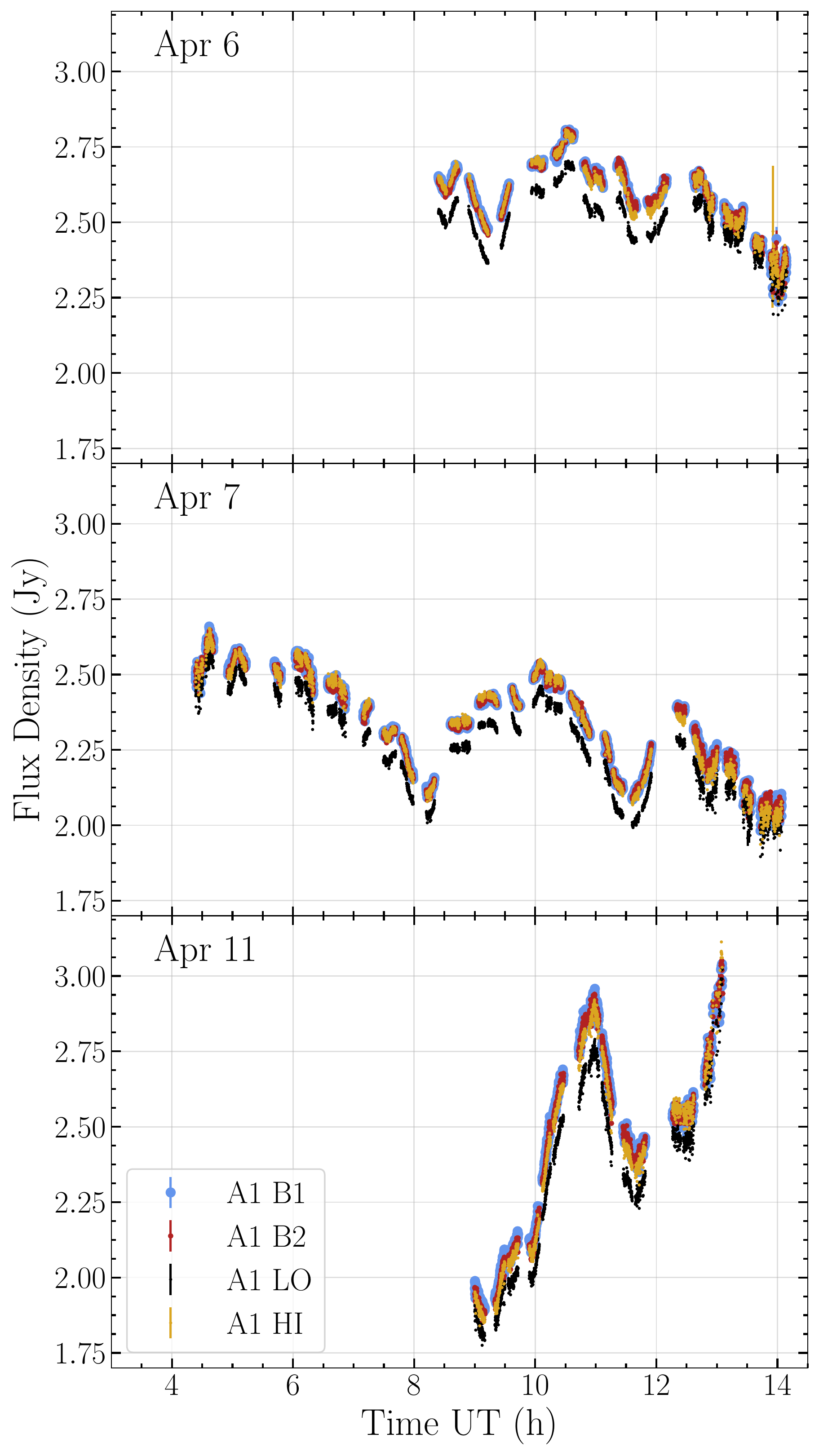}
    \caption{\sgra light curves discussed in the body of this paper. \textit{Left column:} \sgra light curves obtained with SMA in the LO and HI bands,
      for all 5 days of the EHT observations. \textit{Right column:} \sgra light curves obtained with ALMA in the B1, B2, LO, and HI
      bands, for all 3 days of the EHT observations with ALMA. Only the A1 pipeline results are shown.
        }
    \label{fig:LC-both}
\end{figure*} 

The correlation is generally high between the two ALMA pipelines, A1 and A2, with values higher than 0.8 on each individual day and band. It remains larger than 0.7 if we consider full light curves formed by joining the individual days. Similarly, there is a rather high correlation between the SMA data set and the ALMA pipeline A1, reaching above 0.75 in all cases. The correlation is less satisfactory between the A2 pipeline and the SMA, dropping below 0.5 in some cases on 2017 April 6 and 11, but remaining high for the longest and most informative light curve from 2017 April 7. Some discrepancies between A2 and SM can be directly seen in Figure \ref{fig:LC-main}. Note that the ALMA--SMA correlation is calculated only in the short overlapping time of 2--3\,h, when \sgra is seen at low elevation by both ALMA (where it is setting) and the SMA (where it is rising), contributing additional difficulty to constraining systematic errors.

Apart from the correlation, which informs us about the consistency of the variable component, we are also interested in the consistency of the absolute flux density scale. We characterize it by comparing the median flux density in the overlapping observing periods. These results are summarized in Table \ref{tab:medians}. The systematic uncertainties of the absolute flux density scaling can be as large as 10\%, and vary between the days, although the ratios are quite consistent between the bands. These uncertainties do not affect relative variability metrics such as the light curve modulation index, $\sigma/\mu$, defined as the standard deviation divided by the mean, given in Table~\ref{tab:detections} (see also Section \ref{sec:modulation_index}). Yet another way to quantify the differences between the data pipelines is through a mean flux density absolute difference between A2 and A1. In terms of this metric, the mean A1--A2 light curve consistency is 3.7\% on 2017 April 6, 2.7\% on 2017 April 7, and 16.0\% on 2017 April 11, the latter being strongly dominated by the constant offset in the flux density measurements.

We see that the overall discrepancy between light curves produced by different pipelines can be substantial. In particular, it can be significantly larger than the formal level of the thermal error in the data. Hence, we conclude that the errors are strongly dominated by the calibration systematics, which we attribute predominantly to the imperfections in the gain-calibration of the individual telescopes participating in the connected-element arrays, manifesting themselves as slowly-varying differences between the pipelines. If the pipelines were considered fundamentally equal, the consistency metrics provided in this section could serve as a proxy for quantifying systematic errors. However, since the A1 pipeline relies less heavily on a priori sensitivity estimates than the other ones, it is expected to be more robust against the relevant sources of corruption. In Figure \ref{fig:LC-main}, we observe the decorrelation between A1 and A2 to increase toward the end of the observing tracks, which suggests that these inconsistencies are related to the low source elevation, which is exactly when more severe gain-related corruptions are to be expected. In the subsequent analysis, we stress the A1 pipeline results in particular, supplementing them with the A2 and SM results, allowing us to assess our confidence in the obtained result.

\subsection{Consistency within pipelines}

The LNDCF$_0$ coefficient computed between the different frequency bands within the same pipeline is remarkably high in all cases, above 0.99. This consistency across the frequency bands can be seen in Figure\,\ref{fig:LC-both}. We also verify the ratio of medians in the overlapping observing periods, using the HI band as a reference, Table\,\ref{tab:medians_bands}. We notice that the ratio is very close to unity, which is consistent with the flat mm spectrum of \sgra \citep{Marrone2006JPhCS, Bower2015}, and expected given the narrow fractional band, $\Delta \nu /\nu \lesssim 0.1$. There is a persistent systematic effect of 4\,\% missing flux density in the LO frequency band (227.1\,GHz), seen in both of the ALMA pipelines, see Table\,\ref{tab:medians_bands} and the right panel of Figure\,\ref{fig:LC-both}. 
This could be a systematic processing/scaling error shared by both of the ALMA reduction pipelines, or an effect of absorption in the LO band. A similar effect is not seen in the SMA data, for which spectral channels possibly affected by CN absorption were flagged within the LO band (Section \ref{sec:calibration:SMA}). However, the absorption alone was estimated to be too small to be responsible for a 4\% effect \citep[Appendix H.1. of][]{Goddi2021}. As a result of this discrepancy, we refrain from using the ALMA LO band for applications such as the spectral index estimation.
\begin{table}[htp]
\caption{Ratio of the band median flux density with respect to the HI band median flux density.}
\begin{center}
\tabcolsep=0.25cm
\begin{tabularx}{0.8\linewidth}{lcccc}
\hline
Band & April 6 & April 7 & April 11 & Joined\\
\hline
\multicolumn{5}{c}{A1}\\
\hline
B1    &     1.00 &  1.00 &  1.02  & 1.00\\
B2    &     1.00 &  1.00  & 1.01  & 1.00\\
LO    &     0.96 &  0.96 &  0.96  & 0.96\\

\hline
\multicolumn{5}{c}{A2}\\
\hline
B1    &     0.99 &  1.00 &  0.97  & 0.99\\
B2    &     0.99 &  1.00  & 0.97  & 0.99\\
LO    &     0.96 &  0.96 &  0.96  & 0.96\\

\hline
\multicolumn{5}{c}{SM}\\
\hline
LO    &     0.99 &  0.99 &  0.99  & 0.99\\

\hline
\label{tab:medians_bands}
\end{tabularx}
\end{center}
\end{table}

\subsection{Spectral index}
\label{sec:spectral_index}

\begin{figure}[h!]
    \centering
    \includegraphics[width=1.0\columnwidth]{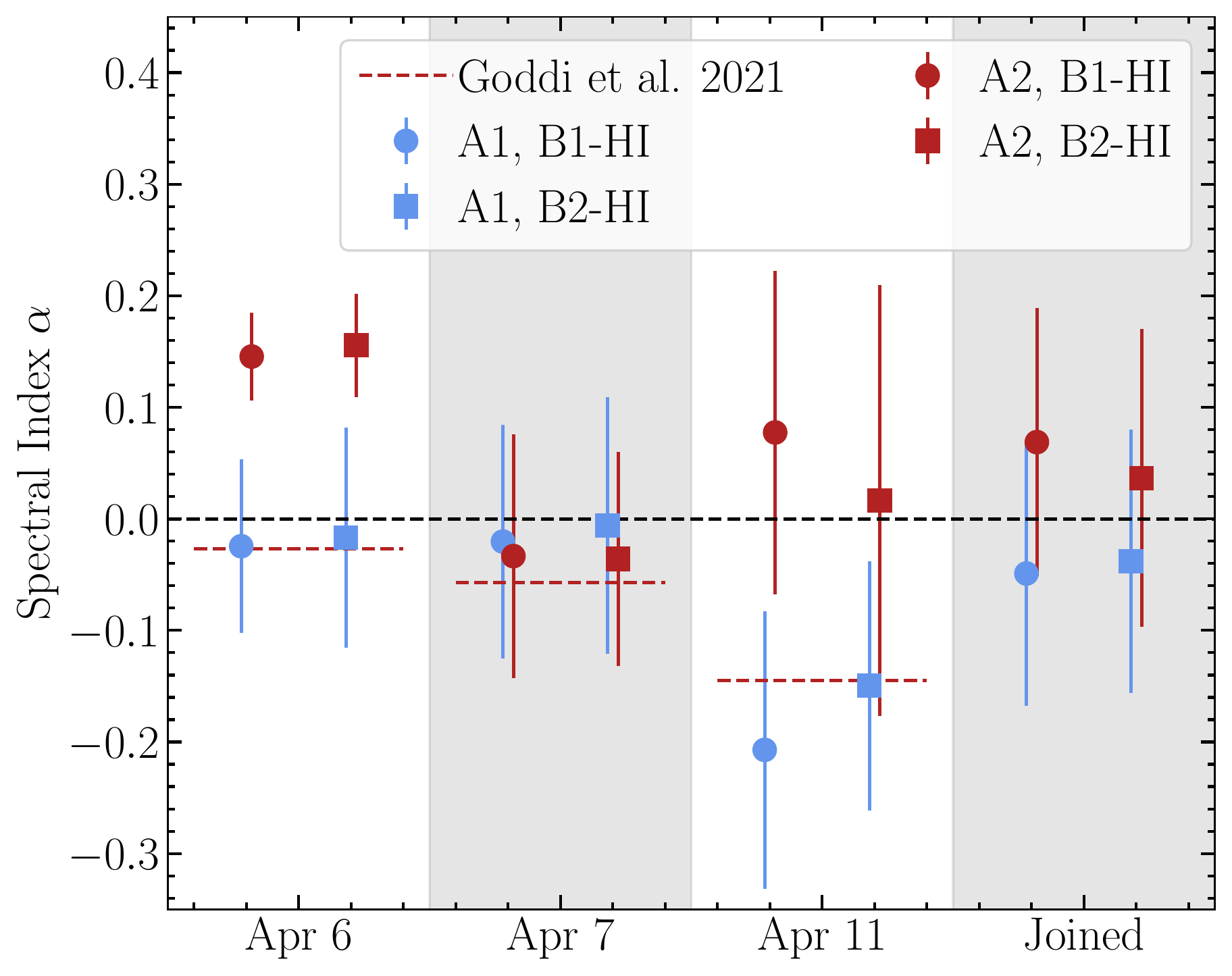}
    \caption{The spectral index at $220$\,GHz estimated from the presented
      light curves. We consider ratios between the B1 and HI bands (round
      markers) and the B2 and HI bands (square markers). Blue markers
      correspond to the A1 pipeline and red markers to the A2 pipeline. The red dashed line indicates the values reported by \citet{Goddi2021}, based on the raw QA2 ALMA data (see Section \ref{sec:observations}), consistent with the A1 pipeline measurements. }
    \label{fig:spectral_index}
\end{figure}

We model the frequency dependence of the flux density with a power law, $F_\nu \propto \nu^\alpha$, thus defining the spectral index as $\alpha$. Subsequently,  we compute $\alpha$ for each pair of simultaneous flux density measurements in the bands (B1, HI) and (B2, HI). We show the results with sample standard deviation error bars in Figure \ref{fig:spectral_index}. We conclude that the spectral index measured between 213.1\,GHz and 229.1\,GHz is consistent with zero, $\alpha_{220}= 0.0\pm0.1$. Figure \ref{fig:spectral_index} implies that the calibration-related systematic uncertainties and short timescale fluctuations of the spectral index dominate the associated error budget. In Appendix \ref{appendix:SMA}, we confirm these findings with the full-bandwidth SMA data analysis.

\begin{figure}[t]
    \centering
    \includegraphics[width=1.0\columnwidth]{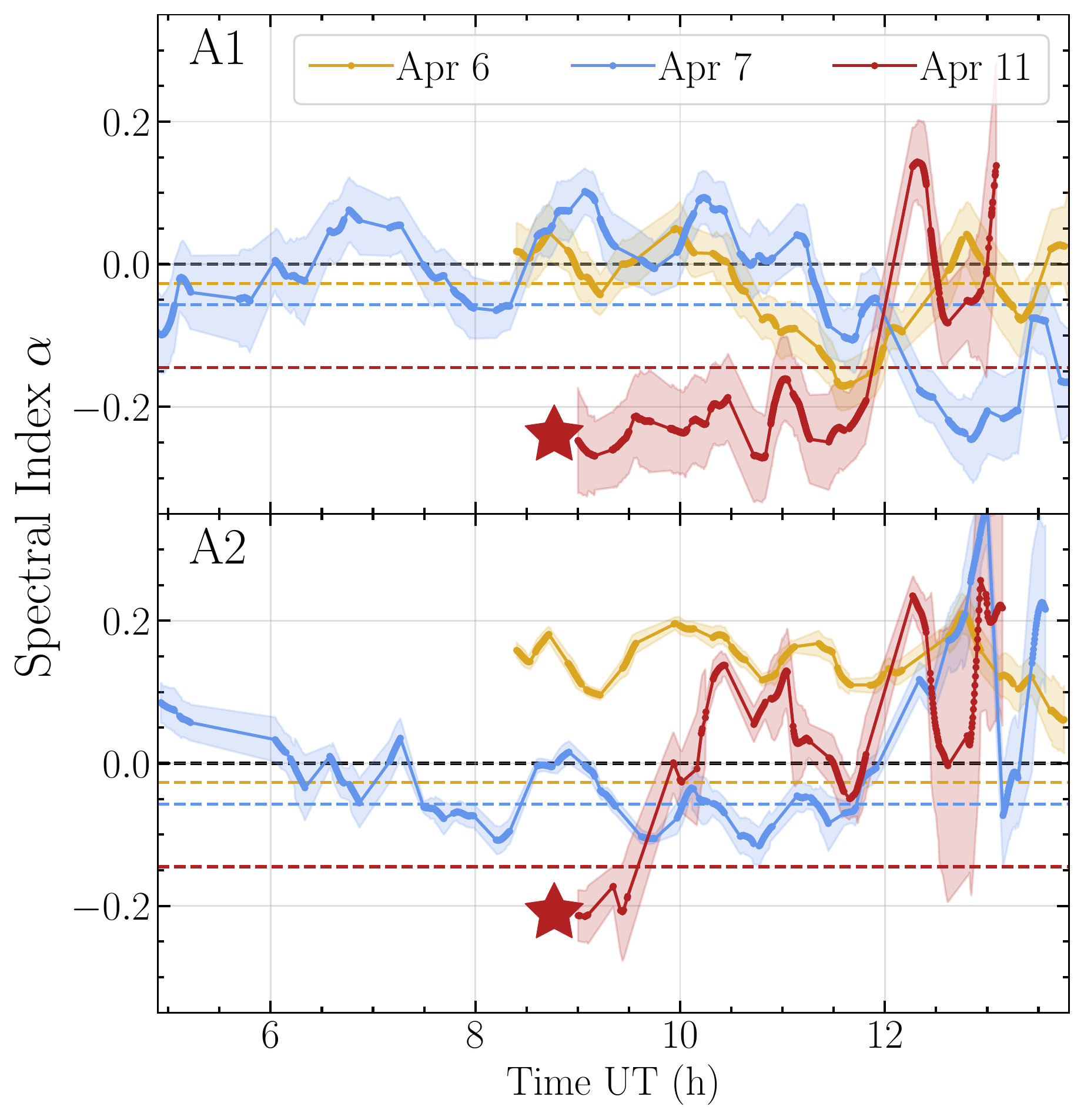}
    \caption{The time dependence of the \sgra spectral index between the HI and B1 bands for the two pipelines, A1 (top) and A2 (bottom). The lines and color bands for each day represent a mean and standard deviation calculated in a running window with 10\,min width. Dashed lines indicate values reported by \citet{Goddi2021} for each corresponding day. The red star marks the peak of the X-ray flare on 2017 April 11. While there are overall significant discrepancies between the pipelines, they both indicate a negative spectral index of \sgra immediately after the X-ray flare. }
    \label{fig:SIdynamics}
\end{figure}

Combining our flux density measurements at 220 GHz of 2.4$\pm$0.2 Jy, with the compact flux of 2.0$\pm$0.2 Jy at 86 GHz reported by \citet{Issaoun2019} based on semi-simultaneous observations on 2017 April 3, we find a spectral index
of $\alpha_{150} = 0.19\pm0.13$ at $\nu_0 = 0.5\times (86+220) \approx 150$\,GHz. Hence, we find a small positive spectral index at about 150\,GHz that becomes consistent with zero at about 220\,GHz. These findings are generally consistent with a flat spectral index at mm wavelengths reported by \citet{Bower2015} and \citet{Iwata2020}, as well as with our broad understanding of the \sgra spectral energy distribution, with a flattening spectrum in the mm approaching a peak in the sub-mm \citep[``sub-mm bump'',][]{Zylka1995, 2001Melia}. The mean light curve spectral index may be an important discriminant of the theoretical models of \sgra \citep{Ricarte2022}, as this quantity is sensitive to physical properties such as temperature, magnetic field strength, optical depth, and the electron distribution function.

One can also resolve the measured spectral index as a function of time, obtaining the results presented in Figure \ref{fig:SIdynamics}. These measurements show large fluctuations of the spectral index and swings in a range between -0.2 and 0.1 on a timescale of $\sim$1\,h. This can be interpreted as rapid fluctuations of the effective optical depth of the compact system, possibly related to the turbulent character of the accretion flow. Interestingly, both pipelines indicate that $\alpha$ was more negative immediately after the 2017 April 11 X-ray flare (ALMA observations begin at 9.0 UT, about 10--15\,min after the peak of the X-ray flare reported by \chandra in \citetalias{SgraP2}, see also Figure \ref{fig:LC-main}), reaching $-0.23\pm0.05$ and subsequently recovering to values consistent with zero on a timescale of 1--2\,h. This suggests an increased contribution of the optically thin component to the total intensity immediately after the X-ray flare. Indeed, since the synchrotron self-absorption decreases with decreasing magnetic field, $B$, and increasing plasma temperature, $T_{\rm e}$ \citep{Rybicki1979}, a flaring event injecting energy of the magnetic field into electrons through magnetic reconnection \citep{Yuan2003} is expected to reduce the effective optical depth of the system. 
Additionally, we note that the first scan by ALMA, which marginally overlaps with the X-ray flare, indicates a decrease in the 1.3\,mm emission, while all subsequent scans for the next 2 h show a growing flux density, in total by about 60\%. Such an evolution of the flux density and spectral index suggests a particle acceleration event where magnetic reconnection heats up electrons to a power-law distribution (\citealt{sironi2014,Guo_2014,Werner_2015}), thus shifting the emission to near-IR and X-ray wavelengths, and causing an inverted spectrum (i.e., $\alpha<0$). As the electrons cool down radiatively, and subsequently the reconnection layer powering the flare depletes (\citealt{RipperdaReconnection}), the optically thin emission shifts back to mm and radio wavelengths \citep[e.g.,][]{Brinkerink2015}, and the source eventually settles back to the state before the flare. 

An elevated X-ray activity was also reported in \chandra observations on 2017 April 7 at 11--13 UT \citepalias{SgraP2}. Here we see that this X-ray event was accompanied by a decreased spectral index period in our A1 pipeline data (as seen in Figure~\ref{fig:SIdynamics}), and a total flux density decrease at 11--13 UT in the A1 pipeline and the SMA observations. This was then followed up by a flux density recovery seen in the SMA data around 14 UT (as seen in Figure~\ref{fig:LC-main}). All of these observations further strengthen the presented interpretation.

%% file: S4_Variability.tex
\begin{figure*}[h]
    \centering
        \includegraphics[width=0.99\textwidth,trim={0cm 0 0 0},clip]{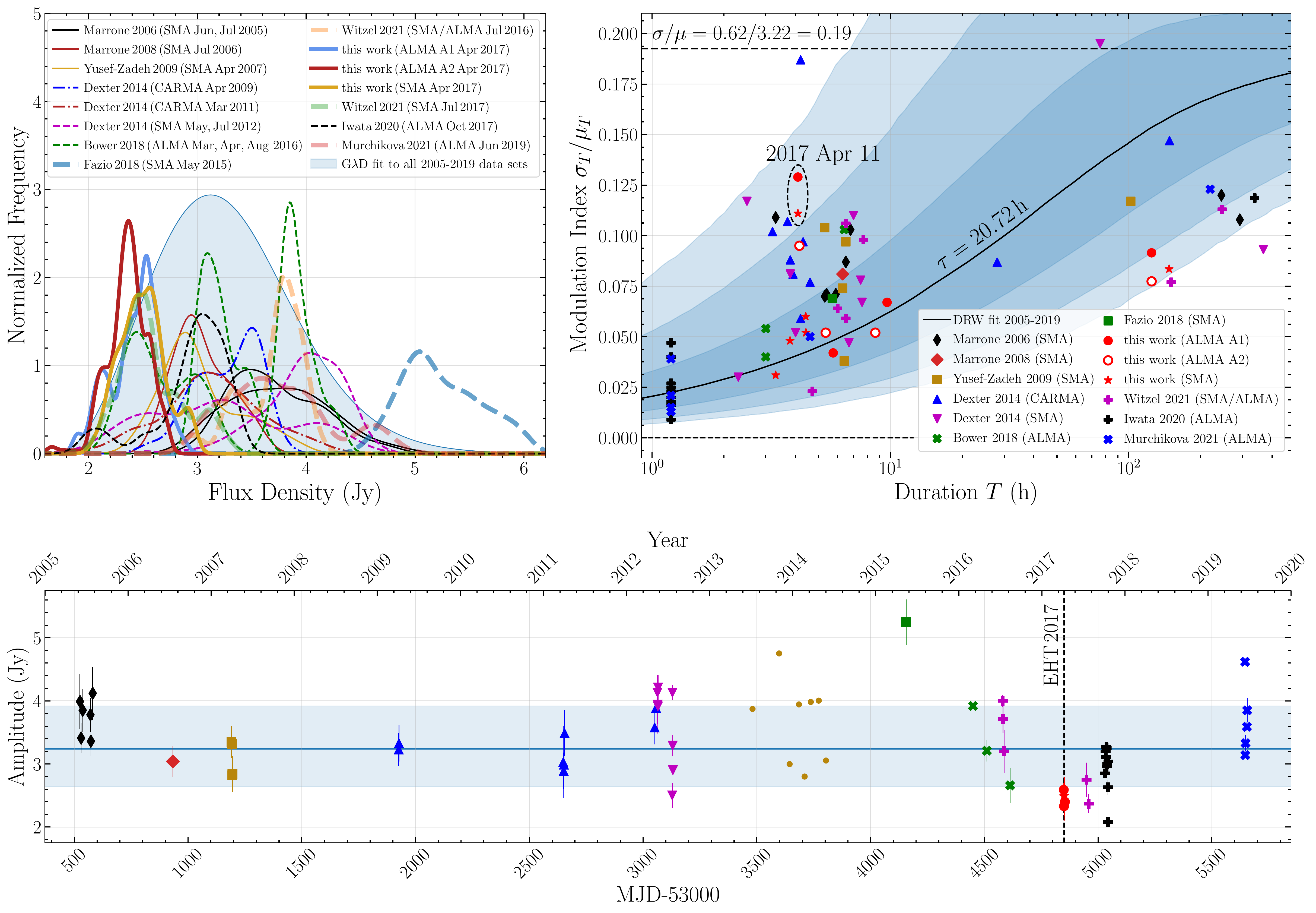}
    \caption{\textit{Top left:} Distributions of the published flux density measurements of \sgra at 230\,GHz. Each distribution represents an epoch no longer than 16 days. Across all of the epochs, the flux density remained roughly within $4\pm2$\,Jy range. \textit{Top right:} Modulation index, $\sigma_T/\mu_T$, measured in different observations as a function of the observation duration, $T$. The black median line and the 1\,$\sigma$, 2\,$\sigma$, and 3\,$\sigma$ uncertainty bands, calculated with a Monte Carlo scheme, correspond to expectations from a damped random walk model fitted to the combined 2005--2019 data set (with the timescale $\tau = 20.72$\,h, and the asymptotic modulation index of $\sigma/\mu = 0.19$ indicated with a dashed line, see also Section \ref{sec:modeling}). \textit{Bottom:} All data sets presented in Tables \ref{tab:detections} and \ref{tab:detections_other_papers} with mean values and standard deviations indicated. The markers follow the convention of the top right panel. Additional points between 2013 and 2015 correspond to ALMA measurements reported in \citet{Bower2015}. The horizontal line and blue bands correspond to the median value and 68\,\% confidence interval of a G$\lambda$D fit to combined 2005--2019 data sets.}
    \label{fig:distributions}
\end{figure*}

\begin{deluxetable*}{lclcccccc}
\tabletypesize{\footnotesize}
\tablewidth{0pt}
 \tablecaption{ Archival \sgra light curves at frequency $\sim$\,230\,GHz from the literature. \label{tab:detections_other_papers}}
 \tablehead{
 \colhead{Reference} & \colhead{Array} & \colhead{Date} & \colhead{Duration (h)}& \colhead{Samples} & \colhead{Flux density (Jy)} &\colhead{$\sigma/\mu$}& \colhead{max-min (Jy)}
 }
 \startdata 
 \hline
     \citet{Marrone2006} & SMA & 2005 June 4 & 3.3 & 15 & $3.99\pm 0.44$ & 0.109 & 1.31 \\
               &  & 2005 June 9 & 5.9 & 32 & $3.41\pm 0.24$ & 0.071 & 1.01 \\
               & & 2005 June 16 & 6.5 & 45 & $3.85\pm0.34$ & 0.087 & 1.07 \\
               & & 2005 July 20 & 5.3 & 32 & $3.78\pm0.27$ & 0.070 & 0.86 \\
               & & 2005 July 22 & 5.4 & 33 & $3.36\pm0.24$ & 0.071 & 0.87 \\
               & & 2005 July 30 & 6.8 & 33 & $4.12\pm0.42$ & 0.103 & 1.95 \\
    \hline
    \citet{Marrone2008} & SMA  & 2006 July 17 &  6.3 & 45 & $3.04\pm0.25$ &0.081 & 0.97 \\
    \hline
     \citet{Yusef2009} & SMA  & 2007 April 1 &  6.3 & 13 & $3.35\pm0.25$ &0.074 & 0.73  \\
       &   & 2007 April 3 &  5.3 & 35 & $3.32\pm0.35$ &0.104 &1.14 \\
        &   & 2007 April 4 &  6.4 & 41 & $2.82\pm0.11$ &0.038 &0.45 \\
         &   & 2007 April 5 &  6.5 & 32 & $2.84\pm0.28$ &0.097 &0.91 \\
    \hline 
      \citet{Dexter2014} & CARMA$^a$  & 2009 April 5 &  3.9 & 46 & $3.23\pm0.26$ &0.081 & 0.98\\
      &   & 2009 April 6 &  3.8 & 42 & $3.33\pm0.29$ &0.088& 1.13 \\
       &   & 2011 March 29 &  4.2 & 36 & $3.03\pm0.57$ &0.187 &2.14 \\
       &   & 2011 March 31 &  3.2 & 25 & $2.89\pm0.29$ &0.102 &0.84 \\
        &   & 2011 April 1 &  4.2 & 32 & $2.99\pm0.17$ &0.059 &0.73 \\
         &   & 2011 April 4 &  3.7 & 39 & $3.49\pm0.37$ &0.107& 1.72 \\
        &   & 2012 May 4 &  4.6 & 50 & $3.58\pm0.27$ &0.077 &1.24 \\
        &   & 2012 May 10 &  4.3 & 48 & $3.89\pm0.38$ &0.097 &1.71 \\
     \cline{2-8}
     & SMA  & 2012 May 16 &  7.6 & 82 & $4.13\pm0.28$ &0.067 & 1.35\\
      &   & 2012 May 17 &  7.5 & 76 & $3.94\pm0.31$ &0.078& 2.02 \\
       &   & 2012 May 18 &  7.0 & 60 & $3.91\pm0.43$ &0.110 &2.23 \\
       &   & 2012 May 19 &  6.7 & 68 & $4.21\pm0.20$ &0.047 &0.98 \\
        &   & 2012 July 20 &  3.8 & 25 & $2.50\pm0.20$ &0.081 &0.69 \\
         &   & 2012 July 21 &  2.3 & 24 & $4.13\pm0.12$ &0.030& 0.51 \\
        &   & 2012 July 22 &  4.0 & 32 & $3.29\pm0.17$ &0.052 &0.73 \\
        &   & 2012 July 23 &  2.5 & 22 & $2.90\pm0.34$ &0.117 &0.99 \\
        \hline
     \citet{fazio:2018} & SMA  & 2015 May 14 &  5.7 & 439 & $5.25\pm0.36$ &0.069 & 1.56 \\
        \hline
     \citet{Bower2018} & ALMA  & 2016 March 3 &  3.0 & 40 & $3.92\pm0.16$ &0.040 & 0.72 \\
      & 
      & 2016 May 3 &  3.0 & 45 & $3.21\pm0.17$ &0.054 &0.62 \\
      &   & 2016 August 13 &  6.4 & 26 & $2.66\pm0.28$ &0.103 & 0.82 \\
        \hline
          \citet{Witzel2021}    & ALMA  & 2016 July 12 &  4.7 & 78 & $4.00\pm0.09$ &0.023 & 0.34 \\
      & 
      & 2016 July 18 &  6.5 & 88 & $3.20\pm0.34$ &0.106 &1.13 \\
      \cline{2-8}
      & SMA  & 2016 July 13 &  6.5 & 798 & $3.71\pm0.22$ &0.059 & 0.82 \\
      & & 2017 July 15 &  7.7 & 1671 & $2.75\pm0.27$ &0.098 &1.45 \\
      &   & 2017 July 25 &  6.0 & 1280 & $2.37\pm0.15$ &0.064 & 0.91 \\
        \hline
\citet{Iwata2020} & ALMA  & 2017 October 5 &  1.2 & 44 & $2.85\pm0.07$ &0.023 & 0.21 \\
 &  
 & 2017 October 7 &  1.2 & 45 & $3.20\pm0.08$ &0.025 &0.23\\
  &  & 2017 October 8 &  1.2 & 44 & $3.11\pm0.08$ &0.027 & 0.30 \\
    &  & 2017 October 10 &  1.2 & 45 & $3.27\pm0.05$ &0.017 & 0.18 \\
     &  & 2017 October 11a &  1.2 & 45 & $3.25\pm0.06$ &0.020 &  0.23 \\
     &  & 2017 October 11b &  1.2 & 44 & $2.96\pm0.05$ &0.016 & 0.16 \\
     &  & 2017 October 14 &  1.2 & 44 & $3.00\pm0.03$ &0.009 & 0.11 \\
     &  & 2017 October 17 &  1.2 & 45 & $2.63\pm0.12$ &0.047 & 0.36 \\
     &  & 2017 October 18 &  1.2 & 44 & $2.08\pm0.08$ &0.040 & 0.23\\
     &  & 2017 October 19 &  1.2 & 45 & $3.04\pm0.05$ &0.018 & 0.18 \\
  \hline
\citet{Murchikova2021} & ALMA  & 2019 June 12 &  1.2 & 1252 & $4.62\pm0.06$ &0.013 & 0.27 \\
 &  
 & 2019 June 13 &  1.2 & 1286 & $3.14\pm0.05$ &0.015 &0.22\\
  &  & 2019 June 14 &  1.2 & 1267 & $3.33\pm0.13$ &0.039 & 0.49 \\
    &  & 2019 June 20 &  1.2 & 1305 & $3.59\pm0.07$ &0.021 & 0.25 \\
&  & 2019 June 21 &  4.6 & 3913 & $3.85\pm0.19$ &0.050 & 0.68
\enddata
 \tablecomments{$^a$Combined Array for Research in Millimeter-wave Astronomy, Cedar Flat, California, USA}
\end{deluxetable*}

\section{Variability characterization}
\label{sec:variability}

The compact radio source \sgra is associated with a supermassive black hole of mass $\sim 4\times10^6\,M_\odot$ \citep{Gravity2019,Do2019}. The mm synchrotron emission unresolved by the ALMA and SMA arrays originates predominantly in the hot innermost part of the accretion flow, on a scale of a few Schwarzschild radii (\citealt{doeleman2008,Fish2011}; \citetalias{SgraP1})\footnote{It is prudent to notice that in our observations the longest projected ALMA baselines reach about 0.2 M$\lambda$ ($\sim 1^{\prime\prime}$), ALMA-APEX 2 M$\lambda$ ($\sim$100 mas), while the shortest EHT non-intrasite VLBI baselines reach 500 M$\lambda$ ($\sim$400 $\mu$as). Thus, there is a range of mas angular scales to which we are blind, and additional extended source structure could be hidden. However, there is no evidence for a significant missing flux, either in the EHT data or in the lower frequency VLBI observations. This is discussed more extensively in \citetalias{SgraP2,SgraP5}.}. Given that the light crossing timescale, $t_{M} = GM/c^3$, is only about 20\,s for \sgra, brightness variability on timescales as short as $\sim1$\,min can be expected. A characterisation of the variability can be achieved through the estimation of the associated magnitudes and timescales. By comparing
the variability analysis results for different days, we aim to establish
whether the estimated variability properties persist over the whole period covered
by the observations or if they change with time. Such changes, if
detected, could indicate a variation in the state of the source and be compared with other observables, such as the simultaneous VLBI observations \citepalias{SgraP1}, for a deeper understanding of the emission process.

\subsection{Comparison to historic data}

In Table \ref{tab:detections_other_papers}, we present the previously published \sgra light curve data sets at frequencies close to 230\,GHz (that is, closest in frequency to our HI band). We only consider observations with radiointerferometric arrays, where reliable extraction of the compact source light curve component is feasible. Compared to data sets published in this paper, summarized in Table \ref{tab:detections}, the archival data sets typically have lower cadence and a far lower number of collected data points. Thus, more reliable studies of the source variability, particularly on short timescales, are enabled by our new data sets.

All data sets given in Table \ref{tab:detections} and Table \ref{tab:detections_other_papers}, spanning a total period of about 14 years, can be divided into 18 observing epochs no longer than 16 days, where the EHT observations constitute a single epoch of 2017 April 5--11. Normalized histograms of the 230\,GHz flux density observed in these epochs are shown in the top left panel of Figure \ref{fig:distributions}. The flux density remains remarkably consistent across all these epochs, with all measurements in agreement with $4.0$\,Jy within about 50\%.

We also show a (differently normalized) generalized $\lambda$ distribution \citep[G$\lambda$D; ][]{freimer1988study} fit to all of the 2005--2019 data sets, computed using the \texttt{gldex} package \citep{su2007fitting}. It approximates the full distribution of the \sgra flux density at 230\,GHz across all of the observing epochs. The G$\lambda$D fit corresponds to flux density values within $3.24^{+0.68}_{-0.60}$\,Jy at 68\,\% confidence, indicating a weak positive tail driven primarily by the record high flux densities observed by \citet{fazio:2018}. All measurements given in Tables \ref{tab:detections} and \ref{tab:detections_other_papers} are also shown in the bottom panel of Figure \ref{fig:distributions} as a function of the observing date.

These ranges are also consistent with \sgra monitoring with ALMA and SMA in 2013 June -- 2014 Nov presented in \citet{Bower2015}\footnote{The data set of \citet{Bower2015} consists of at most a single measurement per day and is not reported in Table \ref{tab:detections_other_papers}. ALMA detections reported by \citet{Bower2015} are shown in the bottom panel of Figure \ref{fig:distributions}.}.
The relative calmness of \sgra is in strong contrast to the X-ray and IR behavior, where flux densities may vary by orders of magnitude during the flaring events \citep{Porquet2003, Do2019_IRflare}. We notice that the 2017 April epoch is characterized by the lowest mean flux density among all 2005--2019 observations. 
Within several hours of a single observing epoch, the mm flux density of \sgra may fluctuate by $\sim$1\,Jy.

\subsection{Modulation index}
\label{sec:modulation_index}

We quantify the variability with the modulation index, $\sigma/\mu$, corresponding to the ratio between the signal standard deviation, $\sigma$, and its mean value, $\mu$. It is related to the rms-flux relation, abundantly used in IR and X-ray studies of variability. We notice, however, that unlike in the IR \citep[e.g.,][]{Gravity2020}, we did not find strong indications of a linear rms-flux relation in the mm light curves. For a red noise stochastic process, which we expect to describe the variability of \sgra mm light curves well \citep{Dexter2014}, most variability manifests on the longest timescales. Hence, the modulation index, $\sigma_T/\mu_T$, calculated from a chunk of data of a finite duration $T$, is biased with respect to the asymptotic modulation index, $\sigma_\infty/\mu_\infty$. On the other hand, for a particular light curve realization, the influence of sparse or nonuniform sampling on $\sigma/\mu$ is small. We verified the robustness of the $\sigma/\mu$ estimation against the measurement noise and irregular sampling by comparing our findings with the results of the intrinsic modulation index algorithm of \citet{Richards2011}, finding an excellent agreement. The influence of the light curve duration $T$ can be seen in the top right panel of Figure \ref{fig:distributions}, where light curves of longer duration generally exhibit a larger modulation index. The figure presents all of the observations listed in Table \ref{tab:detections} and Table \ref{tab:detections_other_papers}. For data sets spanning several days, we also show the modulation index calculated for the entire campaign, corresponding to the histograms in the top left panel of Figure \ref{fig:distributions}. The variability measurements are compared with the expectations from a Gaussian process model (damped random walk; black line indicating the expected values, with 1$\sigma$, 2$\sigma$, and 3$\sigma$ error ranges plotted as blue bands) best-fitting the full combined 2005--2019 data set, see Section \ref{sec:modeling} for details and discussion.

The 230\,GHz light curves collected in 2005--2019, in particular the high quality EHT light curves from 2017 April, indicate rather low modulation index, $\sigma/\mu$, typically below 0.10. Hence, we conclude that on 2017 April 6 and 7 the source displayed an amount of variability consistent with historical measurements. On 2006 July 17 \citep{Marrone2008}, 2015 May 14 \citep{fazio:2018}, and 2017 April 11 \citepalias[this work and][]{SgraP2} increased variability metrics can be connected to flares detected in the X-ray, however, the variability enhancement is particularly clear only in the case of the 2017 April 11 observations. We expect that modulation index values above $\sim$\,0.15 seen in the top right panel of Figure \ref{fig:distributions} may possibly be outliers suffering from calibration errors -- it is generally far easier to increase the apparent variability with the calibration errors than to reduce it (e.g., erroneous amplitude gains, coherence losses, pointing issues).


The modulation index measured in General Relativistic Magnetohydrodynamic (GRMHD) simulations was found to be generally larger than what the observations indicate \citepalias[\citealt{Koushik2021},][]{SgraP5}. For comparisons between observations and simulations, a $T=3$\,h $\approx 540\,GM/c^3$ window for computing the modulation index was used in \citetalias{SgraP5}. This duration is justified by the synthetic observations decorrelation argument -- separate 3\,h segments are expected to behave like statistically independent draws from the modulation index statistic. In Table \ref{tab:sigma_mu_3h}, we give non-overlapping values of $\left(\sigma/\mu\right)_{\rm 3h}$ from all days and sites / pipelines (3 non-overlapping samples for ALMA 2017 April 7, a single 3\,h modulation index measurement for all of the other light curves). The measurements presented in Table \ref{tab:sigma_mu_3h} show a factor of 2 enhancement of the 3\,h modulation index on the X-ray flare day of 2017 April 11. On the remaining days, the modulation index varies between 0.024 and 0.051, while the damped random walk model fitted to all of the 2005--2019 data sets predicts $ \left( \sigma/\mu \right)_{\rm 3h} = 0.03^{+0.02}_{-0.01}$, as shown in the top right panel of Figure \ref{fig:distributions}.
\begin{table}[htp]
\caption{Independent measurements of $\left(\sigma/\mu\right)_{\rm 3h}$}
\begin{center}
\tabcolsep=0.25cm
\begin{tabularx}{0.98\linewidth}{lccccc}
\hline
\multicolumn{6}{c}{ALMA A1}\\
\hline
Band & April 6 & \multicolumn{3}{c}{April 7}  & April 11 \\
\hline
B1    &   0.026 &  \multicolumn{3}{c}{0.026,\ 0.048,\ 0.044} &  0.098\\
B2    &   0.025 &  \multicolumn{3}{c}{0.025,\ 0.050,\ 0.044}  &  0.099\\
LO    & 0.028 &  \multicolumn{3}{c}{0.030,\ 0.051,\ 0.040} &  0.097\\
HI     &  0.029 &  \multicolumn{3}{c}{0.024,\ 0.051,\ 0.044} & 0.099\\
\hline
\multicolumn{6}{c}{ALMA A2}\\
\hline
Band & April 6 & \multicolumn{3}{c}{April 7}  & April 11 \\
\hline
B1    &   0.043 &  \multicolumn{3}{c}{0.035,\ 0.044} &  0.097\\
B2    &   0.044 &  \multicolumn{3}{c}{0.035,\ 0.046}  &  0.098\\
LO    & 0.044 &  \multicolumn{3}{c}{0.038,\ 0.048} &  0.084\\
HI     &  0.045 &  \multicolumn{3}{c}{0.039,\ 0.050} & 0.079\\
\hline
\multicolumn{6}{c}{SMA}\\
\hline
Band &April 5 & April 6 & April 7 & April 10 & April 11 \\
\hline
LO    &  0.049  & 0.030 &  0.042 &  0.039  & 0.117\\
HI     & 0.049   & 0.029 &  0.040 &  0.040  & 0.115\\
\hline
\label{tab:sigma_mu_3h}
\end{tabularx}
\end{center}
\end{table}

\subsection{Structure Function analysis}
\label{sec:SF}

To investigate the possible existence of characteristic variability timescales in the \sgra light curves, a second order structure function
(SF) analysis \citep{Simonetti1985} has been applied to the data. The
SF of a time series $\{x_i\} = x_1,x_2, ..., x_n$, observed at times $\{t_i\} = t_1,t_2,..., t_n$, at time-lag $\Delta t$, is defined as,
\begin{equation}
{\rm SF}(\Delta t)=\frac{1}{M_{\Delta t}}\sum_{i,j}(x_i-x_j)^2 \ ,
\label{eq:structure_function}
\end{equation}
where the sum is extended to all $M_{\Delta t}$ pairs $(t_i, t_j)$ for which
$\Delta t -\Delta t_0/2<(t_i-t_j)<\Delta t +\Delta t_0/2$, and $\Delta t_0$ is the shortest time-lag for which the structure function is calculated. The SF informs us about the signal variance across a range of timescales. A noise contribution has been neglected in Equation \ref{eq:structure_function}, given the very high reported data \textit{S/N}. For this analysis, we use the data cadence reported in Table \ref{tab:detections} as $\Delta t_0$. Assuming that the observed variability can be described as a sum of the random error (measurement/calibration error) and the true signal, possibly resulting
from a complex superposition of processes with different spectral
properties and characteristic timescales, the SF is
expected to show the following:
\begin{itemize}
    \item a flat slope at the shortest timescales, when the random
      error amplitude dominates over the source signal,
    \item a steepening increase on a range of timescales for which the
      random error amplitude is non-negligible compared to the signal,
    \item a steep increase with a constant slope, on timescales for
      which the contribution of the random error to the flux density
      measurements is negligible compared to the variations induced by
      the source. If the signal can be modelled in the
      spectral domain as a power law with the PSD exponent ($\alpha_{\rm PSD}$) steeper than $\approx -1.5$, but not steeper than $-3$ \citep{Emmanoulopoulos2010}, the SF slope ($\alpha_{\rm SF}$) should have a value of
      $ \alpha_{\rm SF} \approx -(1+ \alpha_{\rm PSD})$,
    \item a change of slope at the characteristic timescale of the
      source signal, which corresponds to $0.5/f_\mathrm{b}$, where
      $f_\mathrm{b}$ is the frequency at which the power-law PSD shows
      a break. In case the signal is a superposition of
      multiple components, each characterised by its own timescale and
      power-law slope, these should be reflected as slope changes in
      the SF,
    \item a plateau at a time-lag corresponding to the maximum
      characteristic timescale of the source,
    \item a flat slope at larger timescales, where the SF should oscillate
      around a value of twice the sum of the variances of the signal and the measurement/calibration noise. 
      
\end{itemize}

An SF analysis is prone to identifying spurious characteristic timescales resulting from random fluctuations in a finite realization of a red-noise process, possibly interacting with the sampling window function. To determine whether a detected characteristic timescale is real, it is necessary to sample at least several cycles of the variability (that is, to observe for a duration of at least several times longer than the timescale in question).
The significance of timescales larger
than 0.2 times the total duration of the observations is low; it
slowly increases as this ratio becomes smaller. 
At the shorter timescales, the SF results reflect quite accurately the properties of
the signal realised in the light curves. The SF slope can therefore be
a faithful estimator for the PSD slope ($\alpha_{\rm PSD}$).

\subsubsection{Estimating intrinsic noise}

As a first step of the SF analysis, we isolated the random noise component by applying a denoising algorithm, which works as a low-pass filter with a cut-off timescale
of 0.01\,h $< 2 GM/c^3$.
To verify the correct separation between the
source signal and the random noise contribution to the variability,
we applied the SF to both the denoised signal and the noise component. In the first case, we checked that the slope at the shortest timescales follows the same trend as at the intermediate ones, where the random
noise is negligible. For the noise component, we verified that the SF
slope is approximately zero, in agreement with properties of white noise. The noise component becomes subdominant for timescales longer than $\sim$1\,min, see Figure \ref{fig:sf}. As a by-product of this step, we obtain a realistic estimate of the flux
density uncertainties. These uncertainties turn out to be generally larger than the
statistical errors reported in the data sets, with values on the order of $\sim 0.5\%$ of the
measured flux densities, or about $0.01$\,Jy.
A cross-correlation of the ALMA noise component extracted from the two
independent calibration procedures shows that for all epochs and
frequencies there is no correlation between them. This result allows us to
conclude that the random noise is mainly due to the calibration-specific
uncertainties. 

\subsubsection{SF analysis results}

\begin{figure*}
    \centering
    \includegraphics[width=0.49\textwidth,trim=0 0 0cm 0,clip]{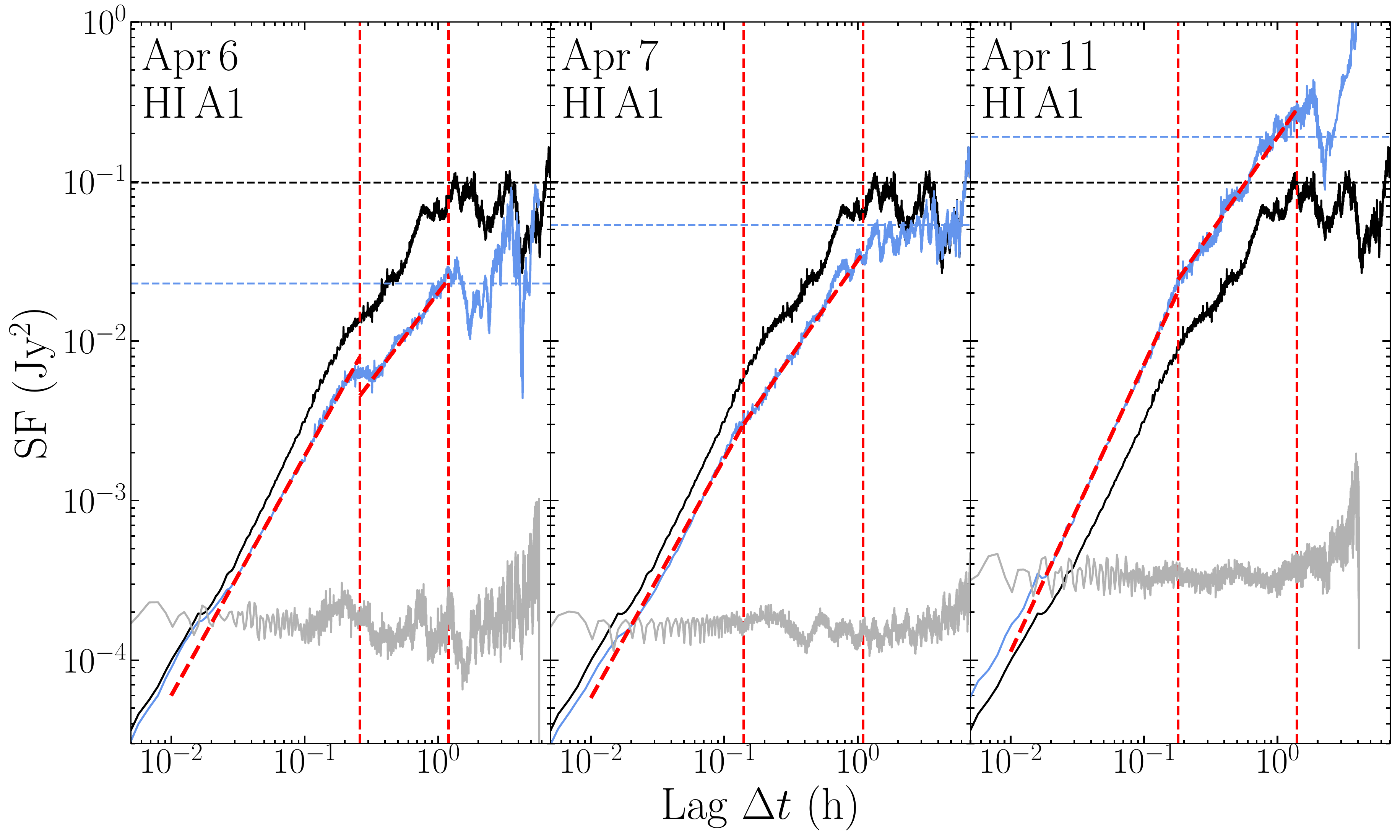}
    \includegraphics[width=0.49\textwidth,trim=0 0 0cm 0,clip]{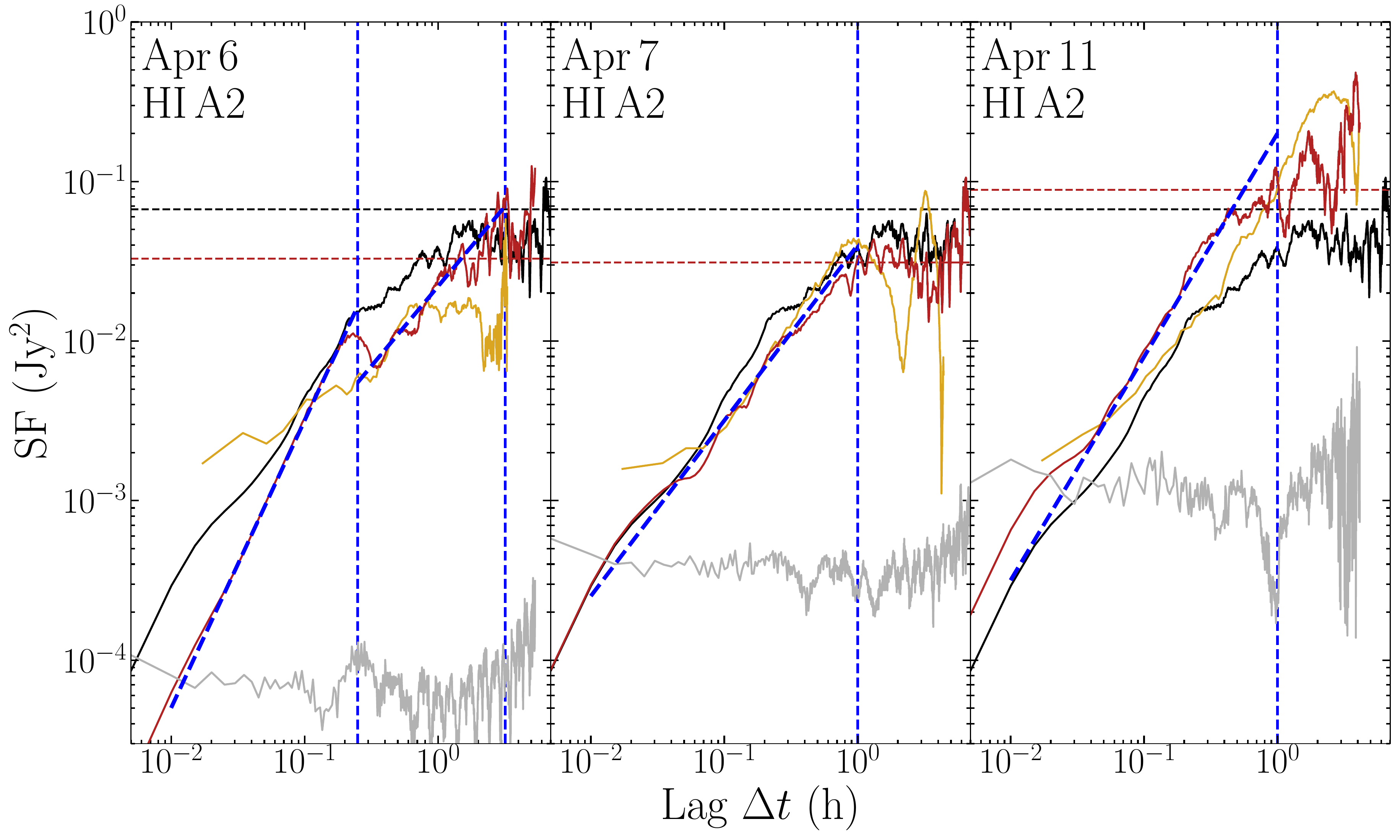}
    \caption{\textit{Left:} Structure function plots for the ALMA A1 pipeline HI band
      data. Results are shown for 2017 April 6, 7, and 11 (from left
      to right). Results from the joined data sets are shown in black. The dashed lines indicate the estimated timescales and slopes reported in Table \ref{tab:SFresultsALMA}, and the value of twice the variance, where the structure
      function is expected to asymptote for long lags in the case of stationary
      signals. The gray lines represent the structure function calculated for the extracted noise component; its flat slope confirms the white noise characteristic of this component. \textit{Right:} Same as above, but for the ALMA A2 pipeline HI band data (red lines). Additionally, the SMA HI band structure functions are shown (orange curves). An excess variance on 2017 April 11 can be seen across all pipelines, particularly at the longer timescales.
    }
    \label{fig:sf}
\end{figure*}

The results of the SF analysis applied to the denoised light curves and cross-checked on the original ones,
are reported in Tables \ref{tab:SFresultsALMA}--\ref{tab:SFresultsSMA}. The variability characteristics inferred through the SF appear to be
nearly identical across all of the frequency bands, while they show a noticeable variation with both the observing day and the instrument/calibration pipeline (e.g., the SFs of the ALMA B1 band A1 and A2 pipeline light curves, plotted in Figure \ref{fig:sf}). Despite these differences, the SF results seem to
converge toward a description of the variability as a superposition of two power-law components. The faster component corresponds to a timescale, $t_1$,  between 0.14 and 0.30\,h and is characterised by a steeper slope for delays $\Delta t < t_1$ ($ \alpha_{\rm SF} \sim 1.6$). The slower component is characterized by a milder slope ($\alpha_{\rm SF} \sim 1.1$) for lags between $t_1$ and timescale $t_2 \approx$1.0--1.5\,h. In most of the epochs, it is possible to observe a long-term trend which exceeds the duration of the observations; this explains the episodic detection of a further SF timescale for which we can only derive a lower limit between 3 and 6\,h.
\begin{table}[htp]
\caption{Structure function analysis results for the ALMA light curves. The first reported slope corresponds to lags shorter than the first reported timescale and so on.}
\begin{center}
\tabcolsep=0.1cm
\begin{tabularx}{0.94\linewidth}{lcccc}
\hline
Data set & Timescales & Slopes $\alpha_{\rm SF}$  & Noise\\
  &  (h) &  $\alpha_{\rm PSD} \approx -(1+\alpha_{\rm SF})$ & (Jy)    \\    
\hline
\multicolumn{4}{c}{2017 April 6}\\
\hline
A1 B1    &   0.26$\pm$0.05, $>1.2$  & 1.5, 1.1  &  0.008 \\
A1 B2    &   0.26$\pm$0.05, $>1.2$  & 1.5, 1.1  & 0.009\\
A1 LO    &   0.26$\pm$0.05, $>1.2$  & 1.4, 1.1  &  0.010 \\
A1 HI    &   0.26$\pm$0.05, $>1.2$  & 1.5, 1.1 &  0.009 \\
A2 B1    &   0.25$\pm$0.05, $>3.2$  &  1.8, 1.0 &  0.006\\
A2 B2    &   0.25$\pm$0.05, $>3.2$  &  1.8, 1.0 &  0.006\\
A2 LO    &   0.25$\pm$0.05, $>3.2$  & 1.8, 1.0  &  0.006\\
A2 HI    &   0.25$\pm$0.05, $>3.2$  & 1.8, 1.0 & 0.006\\

\hline
\multicolumn{4}{c}{2017 April 7}\\
\hline
A1 B1     & 0.14$\pm$0.03, 1.1$\pm$0.3, $>6$ & 1.5, 1.2, --- & 0.009\\
A1 B2     & 0.14$\pm$0.03, 1.1$\pm$0.3, $>6$ & 1.5, 1.2, ---&      0.009\\
A1 LO     & 0.14$\pm$0.03, 1.1$\pm$0.3, $>6$ & 1.5, 1.2, ---&     0.010\\
A1 HI     & 0.14$\pm$0.03, 1.1$\pm$0.3, $>6$ & 1.5, 1.2, ---&    0.010\\
A2 B1     &  1.1$\pm$0.3, $>6$  &  1.1, --- &    0.013\\
A2 B2      &  1.1$\pm$0.3, $>6$  &  1.1, ---&     0.013\\
A2 LO     &   1.1$\pm$0.3, $>6$  &  1.1, ---&      0.014\\
A2 HI      &  1.1$\pm$0.3, $>6$  &  1.1, ---&     0.014\\

\hline
\multicolumn{4}{c}{2017 April 11}\\
\hline
A1 B1     &  0.18$\pm$0.03, 1.4$\pm$0.4 & 1.8, 1.2 &    0.013\\
A1 B2     &  0.18$\pm$0.03, 1.4$\pm$0.4 & 1.8, 1.2&      0.013\\
A1 LO     &  0.18$\pm$0.03, 1.4$\pm$0.4 & 1.8, 1.2&      0.014\\
A1 HI     & 0.18$\pm$0.03, 1.4$\pm$0.4 & 1.8, 1.2&     0.013\\
A2 B1     &  1.4$\pm$0.4 & 1.4&      0.019\\
A2 B2     &  1.4$\pm$0.4 & 1.4&     0.019\\
A2 LO     &  $>1.$ & 1.4&      0.024\\
A2 HI    &  $>1.$  & 1.4 &     0.022\\

\hline
\multicolumn{4}{c}{All days}\\
\hline
A1 B1        &      0.23, 1.2$\pm$0.4, $>6$ & 1.7, 1.2& ---\\
A1 B2        &        0.23, 1.2$\pm$0.4, $>6$& 1.7, 1.2& ---\\
A1 LO       &        0.23, 1.2$\pm$0.4, $>6$& 1.6, 1.2& ---\\
A1 HI       &       0.23, 1.2$\pm$0.4, $>6$& 1.6, 1.2 &  ---\\
A2 B1        &        0.26, 1.5$\pm$0.3 &1.3, 1.0& ---\\
A2 B2        &        0.26, 1.5$\pm$0.3& 1.3, 1.0& ---\\
A2 LO        &        0.3, 1.4$\pm$0.4 & 1.3, 1.1& ---\\
A2 HI       &       0.3, 1.4$\pm$0.4 & 1.3, 1.1& --- \\

\hline
\label{tab:SFresultsALMA}
\end{tabularx}
\end{center}
\end{table}
\begin{table}[htp]
\caption{Structure function analysis results for the SMA light curves.}
\begin{center}
\tabcolsep=0.14cm
\begin{tabularx}{0.8\linewidth}{lcccc}
\hline
Data set & Timescales & Slopes $\alpha_{\rm SF}$  & Noise\\
  &  (h) &  $\alpha_{\rm PSD} \approx -(1+\alpha_{\rm SF})$ & (Jy)    \\    

\hline
\multicolumn{4}{c}{2017 April 5}\\
\hline

SM LO   &   1.7$\pm$0.3  &    0.7  & 0.060 \\
SM HI   &   1.7$\pm$0.3  &    0.7 &  0.060\\
\hline
\multicolumn{4}{c}{2017 April 6}\\
\hline
SM LO      &   0.9$\pm$0.2    &  1.2 &      0.030 \\
SM HI       &  0.8$\pm$0.2    &  1.0&     0.030  \\

\hline
\multicolumn{4}{c}{2017 April 7}\\
\hline
SM LO     &    1.0$\pm$0.1   &   1.4  &  0.020\\
SM HI     &   1.0$\pm$0.1   &   1.4 &    0.020 \\

\hline
\multicolumn{4}{c}{2017 April 10}\\
\hline
SM LO    &    1.0$\pm$0.1   &   1.5 &    0.017\\
SM HI    &    1.0$\pm$0.1   &   1.5  &   0.016 \\

\hline
\multicolumn{4}{c}{2017 April 11}\\
\hline
SM LO   &      2.3$\pm$0.1  &   1.5 &  0.020\\
SM HI   &      2.3$\pm$0.1  &   1.5 &  0.030  \\

\hline
\multicolumn{4}{c}{All days}\\
\hline
SM LO     &     3.3$\pm$0.1   &   1.1 & ---\\
SM HI       &   3.3$\pm$0.1    &  1.1 & ---\\

\hline
\label{tab:SFresultsSMA}
\end{tabularx}
\end{center}
\end{table}

Both timescales above should be taken with some caution. The 0.14--0.30 h timescale is highly significant given the number of variability cycles available across the entire period of the observation. However, the fact that this timescale is similar to the scan segmentation timescale, raises the suspicion of a sampling effect. This suspicion is corroborated by some discrepancies in the SF shape at short timescales for the A1 and A2 pipeline data, although the combined A2 light curves do indicate a similar break, see Figure \ref{fig:sf}. The fact that the SMA data never show evidence of such a fast variability component is less significant because of the higher noise and worse sampling of the light curves, which could make its detection very difficult. Additionally, we verified for the synthetic light curves modeled in the Gaussian process framework (see Section \ref{sec:model_select}) that the sampling of the ALMA observations is sufficient to measure the SF slopes robustly and without any persistent spurious characteristic timescales shorter than $\sim$1\,h. Finally, indication of the SF slope flattening on a timescale of $\sim 0.3$\,h was also reported by \citet{Iwata2020}. We measured the slope in their data set to be $\alpha_{\rm SF} \approx 1.8$. Recently, the analysis of high cadence ALMA light curves was reported by \citet{Murchikova2021}, who found $\alpha_{\rm SF} \approx 1.6$ (when adopted to our conventions) for timescales shorter than 0.4\,h. Overall, we see suggestive evidence that the SF slope, $\alpha_{\rm SF}$, is steeper than 1.0 for short timescale variability, and closer to 1.6. This is inconsistent with the damped random walk model, see more discussion in Section \ref{sec:model_select}. Within the power-law PSD model assumption, these findings correspond to a PSD slope of $\alpha_{\rm PSD} \approx -2.6$, flattening to about $-2$ for variability on timescales longer than 0.15--0.30\,h, comparable to the dynamical timescale of the innermost part of the accretion flow.

\begin{figure*}[t]
    \centering
    \includegraphics[width=\textwidth]{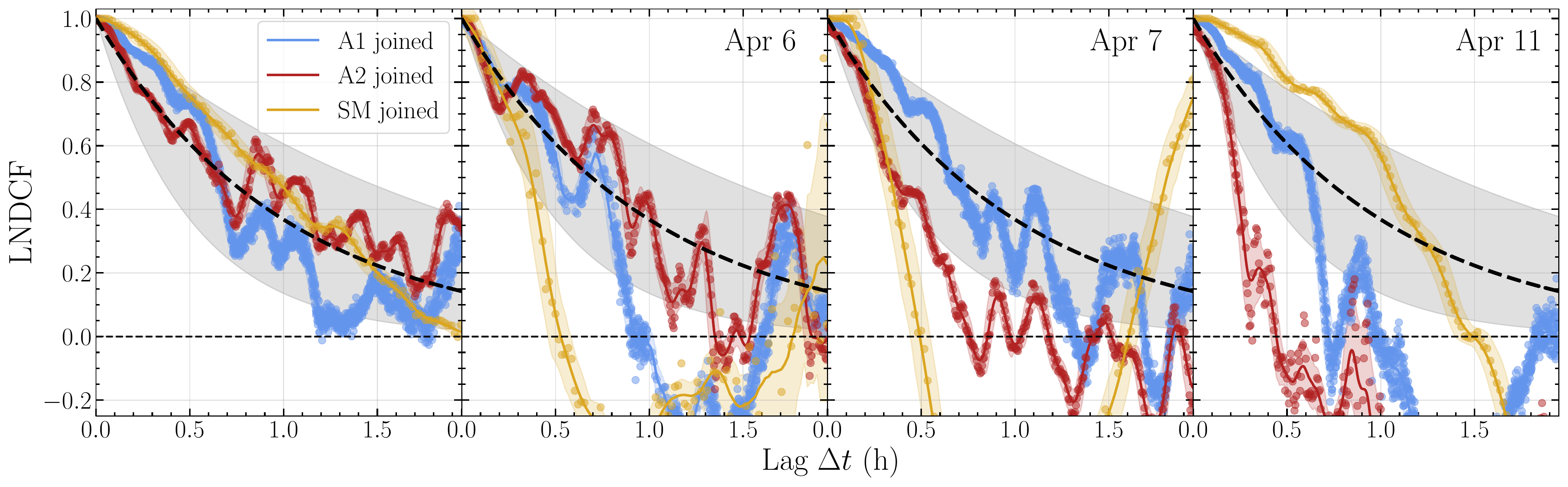}
    \caption{Estimated autocorrelation of the \sgra light curves in the HI band. The black dashed line corresponds to an exponential decay with 1\,h timescale and the shaded region corresponds to autocorrelation timescales between 0.5\,h and 2\,h. Due to the irregular sampling, we show the actual values of the measured autocorrelation along with the running mean and the running standard deviation uncertainty band for each day.}
    \label{fig:LNDCF_panel}
\end{figure*}

\begin{figure}[h!]
    \centering
    \includegraphics[width=\columnwidth]{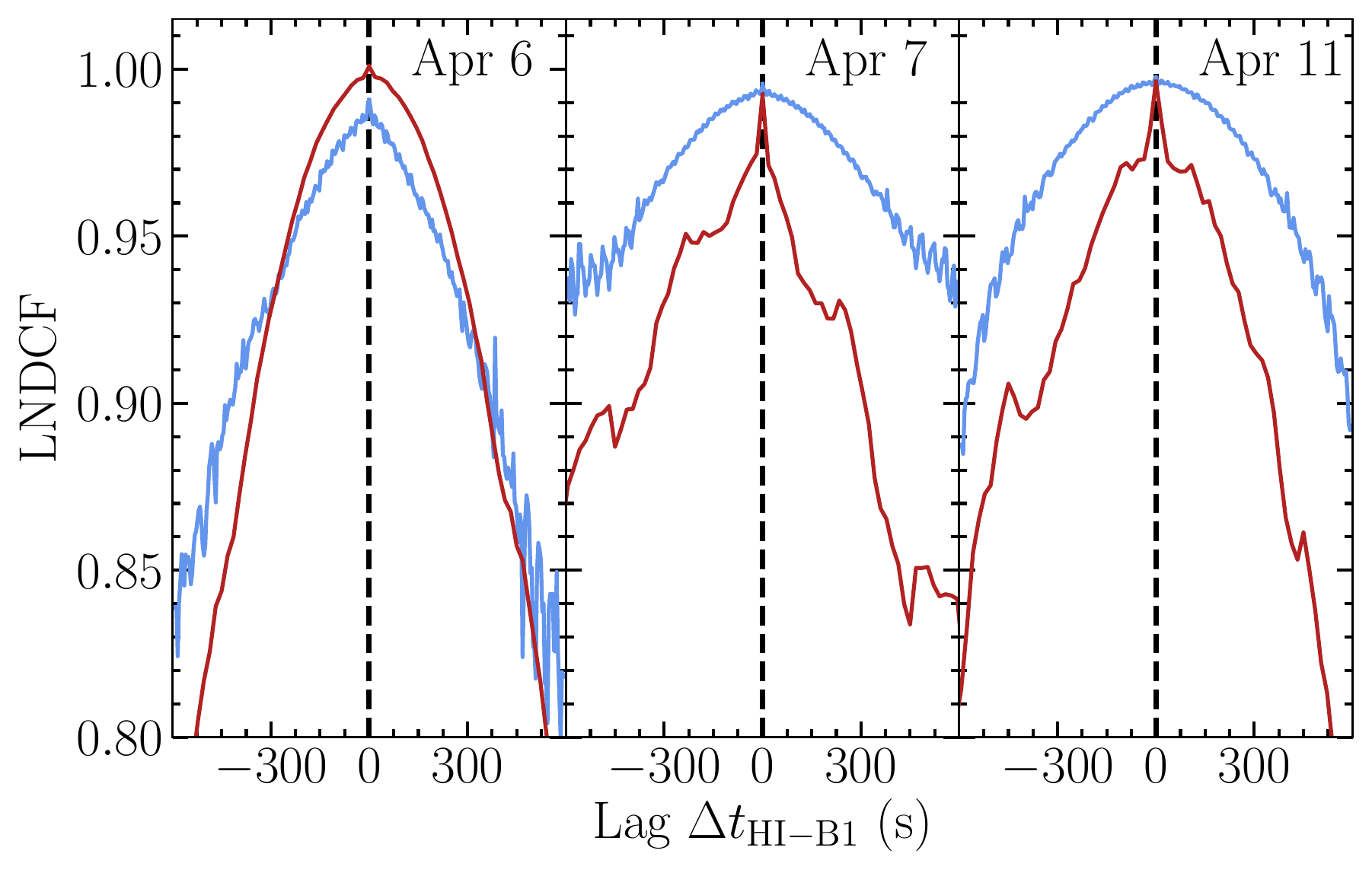}
    \caption{The cross-correlation between the HI and B1 bands for the time-lags $\pm 600$\,s. The results for the A1 (A2) pipeline are shown in blue (red). There is no indication of delays between the frequency bands.}
    \label{fig:LNDCF_panel_bands}
\end{figure}

The 1.0--1.5\,h characteristic timescale falls into a time range for which the number of measured variability cycles is still not large enough to ensure the significance of the detection. Its identification in the all-epoch light curves corroborates the detection, but the sampling effects are still too important to reach a very confident conclusion.

With the presented SF analysis, we confirm the intrinsic source variability on timescales as short as 1\,min $\approx 3 GM/c^3$, which is generally less than the expected emission region diameter \citepalias{SgraP5}. This implies that at least the variability on shortest timescales must have a structural characteristic in the compact source resolved by the EHT, and it hence requires mitigation in the analysis of the VLBI data beyond the simple light curve normalization (\citetalias{SgraP4}; \citealt{Broderick2022, Georgiev2021}).

Another important observation concerns the difference between 2017 April 11 and other observing days. The SF values on 2017 April 6 and 7, as well as the SMA-only 2017 April 5 and 10, are reasonably consistent. However, we see that SF values for 2017 April 11 are significantly larger than those found for the other days. This is consistent with Table \ref{tab:detections}, reporting standard deviations larger by a factor of $\sim$2--3 on 2017 April 11, compared to those measured on the other days. The SF analysis allows us to see that this effect of enhanced variability is present across all timescales, although it becomes more prominent for the longer ones; for a minute timescale the ratio between the 2017 April 11 SF and the 2017 April 6/7 SF is $\sim$ 2, and it becomes $\sim$10 for timescales longer than 1\,h. We connect this significantly enhanced variability to the flaring event preceding the ALMA observations on 2017 April 11. Interestingly, this enhanced variability effect is seen also in the SMA light curves, despite the fact that the SMA started observing 2\,h after the X-ray flare peak.

\subsection{Autocorrelations}
\label{subsec:correlations}

In the case of stationary signals, there is a unique relationship between the autocorrelation and the SF (see Section \ref{sec:modeling}). Nevertheless, apart from the uncertainty in the stationarity assumption, studying autocorrelations separately offers a different perspective into the data. We study the signal autocorrelation using the LNDCF method \citep{Lehar1992}, Equation \ref{eq:LNDCF}. A summary of these results is shown in Figure \ref{fig:LNDCF_panel}. In this plot, we indicate all of the contributing autocorrelation measurements (circles), with the running mean (colored line) and the running one standard deviation bands. In the first panel of Figure \ref{fig:LNDCF_panel}, we show the autocorrelation calculated using all of the available observations in each
pipeline, in the HI band (the other bands are very consistent). The
data indicate autocorrelations decreasing roughly on a timescale of
$\sim$ 1\,h; the black dashed line corresponds to $\exp{(-\Delta t /
  1\rm{h})}$. This is significantly less than one would expect based on the results of \citet{Dexter2014}. In the subsequent panels of Figure \ref{fig:LNDCF_panel},
we show autocorrelations for 2017 April 6, 7, and 11. The non-monotonic structure of the autocorrelation functions is not detected confidently, given the associated uncertainties. The persistence of such features may be established with more observations, e.g., pipeline A1 results show a bump at $\sim$30\,min for both 2017 April 7 and 11. This resembles the innermost stable circular orbit period for a Schwarzschild black hole with $4 \times 10^6 M_\odot$ mass, but there is very little confidence in such an association at this point. Significant biases that may affect the autocorrelation measurement prevent us from drawing strong conclusions based on this analysis, see the discussion in Section \ref{sec:model_select}.

Several authors have recently considered signatures of multiple-path propagation of photons traveling through a strongly curved spacetime in a black hole vicinity and reaching the observer with a delay \citep[e.g.,][or \citealt{Wielgus2020} for the case of exotic spacetimes of black hole mimickers]{Moriyama2019,Hadar2020,Chesler2020,Wong2021}. While observing a feature related to a photon shell around a black hole is a tantalizing possibility, such an observation does not seem feasible yet, given the model simplifications and the limited duration of the observations. There are no significant signatures of autocorrelations detected at the relevant time-lags of $\sim 20 GM/c^3 \approx 400$\,s in our presented data sets.

\subsection{Lags between the frequency bands}

The presence of time-lags between frequency bands in observations of \sgra has been theoretically predicted. In the optically thick regime of radio observations at frequencies below 100\,GHz, theoretical models attribute those lags to the adiabatic spherical expansion of plasma blobs \citep{Laan1966,yusef-zadeh:08a,eckart:12a}, or to a bulk outflow \citep{Falcke2009}. Such lags could be detected across the spectrum, with the higher frequency signal typically leading the lower frequency signal \citep{Yusef2009,Brinkerink2015,Brinkerink2021}. Hints of a similar delay structure have been seen in some numerical GRMHD models of \sgra compact emission \citep{Chan2015}. In this theoretical framework, we could expect lags of 1--2 min between the HI and B1 band, easily detectable with the cadence of the ALMA data. However, no indication of a correlation lag between the bands in any of our data sets is found, with the uncertainty no larger than 20\,s. The cross-correlation function very clearly peaks at zero for all days and for both ALMA reduction pipelines, as shown in Figure \ref{fig:LNDCF_panel_bands}. We interpret this lack of detectable delays as a signature that the emission region in the 213--229\,GHz range is already optically thin all the way to the horizon, possibly with patches of higher optical depth material formed in the turbulent accretion flow, necessary to explain the intermediate spectral index $\alpha \approx 0$ (identified in Section \ref{sec:spectral_index}). This is particularly likely given that in 2017 April \sgra was in a rather low mm flux density state \citep[Section \ref{sec:variability};][]{Moscibrodzka2012}. Historically, no delays were reported by \citet{Iwata2020} across similar frequency bands, there was also no significant delay between 230\,GHz and 345\,GHz reported by \citet{Marrone2008}, and no delay between 134 and 146\,GHz or between 230 and 660\,GHz reported by \citet{Yusef2009}. Finally, the conclusion of a low optical depth is consistent with the interpretation of the first EHT images of \sgra, reported in \citetalias{SgraP3}, as the observable shadow of a supermassive black hole.

%% file: S5_Modeling.tex
\section{Modeling light curve variability}
\label{sec:modeling}

We attempt to represent the variable behavior of the \sgra light curves using statistical models within a Gaussian process (GP) framework \citep{GP2006}. The GP assumption is restrictive by itself. To a degree its impact was investigated by \citet{Dexter2014}, who compared fits to light curves in linear and in logarithmic space, finding reasonably consistent results. We consider the low relative variability of the 230\,GHz light curves (see Section \ref{sec:variability}), with only weak tails of the flux density distributions, to be a convincing motivation for limiting the modeling efforts to GP. To fit the models and explore the associated posterior probability space, we use the dynamic nested sampling algorithm implemented in \texttt{dynesty} \citep{Speagle2020}.

\subsection{Damped random walk (DRW)}

DRW (or the Ornstein–Uhlenbeck process) is a unique Markovian stationary GP \citep{GP2006}. Application of the DRW as a mathematical model to describe the optical variability of quasars was proposed by \citet{Kelly2009}. The DRW with variance $\sigma^2$ is characterized by a covariance function,
\begin{equation}
    k_{\rm DRW}(\Delta t) = \sigma^2 \exp{\left(- \left| \frac{\Delta t}{\tau} \right|  \right)} \ ,
\label{eq:covariance}
\end{equation}
where a characteristic timescale, $\tau$, is a model parameter. For stationary processes there is a general relation between the covariance and the structure function defined in Equation~\ref{eq:structure_function},
\begin{equation}
    {\rm SF}(\Delta t) = 2 \sigma^2 - 2k(\Delta t) \ .
    \label{eq:SF_from_ACF}
\end{equation}
The corresponding power spectral density is related to the covariance function through the Wiener-Khinchin theorem, and in the case of a DRW process it becomes
\begin{equation}
    {\rm PSD}(f) = \frac{4 \sigma^2 \tau}{1 + (2 \pi f \tau)^2} \ ,
\end{equation}
which corresponds to a red noise spectrum with an index of $\alpha_{\rm PSD} = -2$ in a high frequency limit ($f \tau \gg 1$), and a flat white noise spectrum at low frequencies ($f \tau \ll 1$). In the context of the Galactic Center, \citet{Dexter2014} modeled the \sgra variability at mm wavelengths with the DRW, following the procedure outlined by \citet{Kelly2009}. By fitting several years of \sgra observations (Section \ref{sec:variability}), they identified a poorly constrained DRW timescale of $\tau \sim$\,8\,h.

Our model differs from that of \citet{Kelly2009} and \citet{Dexter2014} only by the inclusion of an additional parameter, $\sigma_0$, representing the noise floor.
Then, the full model consists of 4 variables,
\begin{equation}
    \theta = (\tau,\mu,\sigma,\sigma_0), 
\end{equation}
the correlation timescale $\tau$, the mean value of the process $\mu$, the standard deviation $\sigma$, and
the noise floor $\sigma_0$. The timescale $\tau$ is not necessarily related to the timescales estimated in Section \ref{sec:SF} -- the SF timescales may be indicative of the presence of multiple stochastic components in the real signal. Because the DRW is a Markovian process, the likelihood function for observations,
$\{x_i\} = x_1,x_2, ..., x_n$, observed at times $\{t_i\} = t_1,t_2,
..., t_n$, can be calculated directly as
\begin{equation}
    p \left( \{ x_i \} | \theta \right) = \prod^n_{i=1} \frac{\exp \left[-0.5\left(\hat{x}_i - x^\ast_i \right)^2/\widetilde{\Omega}_i \right] }{ \left( 2 \pi  \widetilde{\Omega}_i \right)^{1/2}} \ ,
    \label{eq:DRWlikelihood}
\end{equation}
where
\begin{align}
    \widetilde{\Omega}_i &= \Omega_i + \sigma_0^2 + \sigma_i^2 \ , \\
    x^\ast_i &= x_i - \mu \ .
\end{align}
The indexed $\sigma_i$ represents measurement uncertainties and is distinct from the estimated process standard deviation $\sigma$, and the noise floor $\sigma_0$. The quantities $\Omega_i$ and $\hat{x}_i$, are calculated through an iterative procedure,
\begin{align}
    \hat{x}_i &= a_i \hat{x}_{i-1} +  \frac{a_i \Omega_{i-1}}{\widetilde{\Omega}_{i-1}} \left(\hat{x}_{i-1} - x^\ast_{i-1} \right) \ , \\
   \Omega_i &= \Omega_1 \left(1 - a^2_i \right) + a^2_i \Omega_{i-1} \left(1 - \frac{\Omega_{i-1}}{\widetilde{\Omega}_{i-1}} \right) \ , \\
    a_i &= \exp{\left[-(t_i - t_{i-1})/\tau\right] \ ,}
\end{align}
with the initial conditions
\begin{equation}
   \Omega_1 = \sigma^2 \  ; \ \hat{x}_1 = 0 \ .
\end{equation}
Note that we use a slightly different parametrization of the DRW model than \citet{Kelly2009}, with $\sigma^2$ representing the variance of the DRW, related to their parametrization by $\sigma_{\rm Kelly} = \sigma \sqrt{2/\tau}$. Since the procedure outlined above allows us to explicitly compute the best-fitting DRW realization for a given vector of parameters, $\theta$, we can assess the fit quality by computing the reduced-$\chi^2$ statistic for the residuals,
\begin{equation}
    \chi^2_n =\frac{1}{n}\sum^{n}_{i=1} \frac{\left( x^\ast_i - \hat{x}_i \right)^2}{\widetilde{\Omega}_i} \ .
\end{equation}

\subsection{Mat\'{e}rn covariance model}

The DRW model fixes the high frequency limit PSD slope to $\alpha_{\rm PSD}=-2$. This is a rather strong assumption, and there are indications of a steeper PSD slope both in the context of optical variability of quasars \citep{Mushotzky2011, Zu2013} and the variability of X-ray binaries \citep[e.g.,][]{Tetarenko2021}. The high frequency PSD slope may be a relevant parameter to extract, less affected by the sampling and observation duration limitations than the timescale $\tau$, and having the potential to constrain theoretical models of \sgra. Moreover, observations presented in this paper sample the high frequency regime, relevant for constraining $\alpha_{\rm PSD}$ uniquely well. Hence, we employ a more general statistical model of a GP with a Mat\'{e}rn covariance function \citep[see, e.g.,][]{GP2006},
\begin{equation}
    k_{\rm Mat}(\Delta t) = \sigma^2 \frac{2^{1-\nu}}{\Gamma(\nu)}\left(\sqrt{2\nu}\frac{\Delta t}{\tau}\right)^\nu K_\nu\left(\sqrt{2\nu}\frac{\Delta t}{\tau}\right) \ ,
    \label{eq:covariance_matern}
\end{equation}
where $K_\nu$ is the modified Bessel function of the second kind. The
parameter $\nu$ defines the order of the Mat\'{e}rn process and
subsequently controls the smoothness of the resulting curve. The
PSD of the Mat\'{e}rn process is,
\begin{equation}
    {\rm PSD}_{\rm Mat}(f) \propto \left[ 1 + \frac{(2 \pi f \tau)^2}{2\nu} \right]^{-(\nu + 1/2)} \ ,
    \label{eq:psd_matern}
\end{equation}
so in the high frequency limit $f \tau \gg 1$, we find the PSD slope of $ \alpha_{\rm PSD} = -2\nu -1$. The DRW is recovered as a special case of the Mat\'{e}rn process with $\nu = 0.5$.
As an arbitrary $\nu$, the Mat\'{e}rn covariance represents a non-Markovian process, and the likelihood can not be evaluated explicitly as in case of the DRW. Instead, we evaluate it numerically using the \texttt{Stheno} library.\footnote{\url{https://github.com/JuliaGaussianProcesses/Stheno.jl}}

\subsection{Modeling set-up}

Given the low computational cost of the DRW model fitting, we were
able to perform a survey of different modeling parameters, such as
the type and range of priors, subsets of data to be used, and treatment of the systematic uncertainties and the noise floor. Our general conclusion is that the timescale $\tau$ can not be well constrained and its posterior distributions are dominated by the assumed priors. As an example, \citet{Dexter2014} used log-uniform priors, reducing the distribution tails for large $\tau$. We find that for our data sets $\tau$ remains poorly constrained, and with uniform priors very large timescales are permitted. As noted by \citet{Kozlowski2017}, the duration of the light curve needs to be significantly longer than the timescale $\tau$ to constrain it reliably. The duration and sampling of the 2017 April data may not be sufficient. On the other hand, when fitting data spanning several years (such as in the case of \citealt{Dexter2014}), one needs to consider if the underlying process can be assumed to be stationary on such long timescales (e.g., as a consequence of the mass accretion rate modulation).
\begin{table*}
     \caption{Gaussian process modeling results (ML estimators with 68\% confidence intervals).}
\begin{center}
\tabcolsep=0.12cm
\begin{tabularx}{0.98\linewidth}{lccccccccccccc } 

 \hline
 Data set & \multicolumn{6}{c}{DRW} &  &\multicolumn{6}{c}{Mat\'{e}rn}  \\
 
 \cline{2-7}\cline{9-14}
  & $\mu$ (Jy) & $\sigma$ (Jy) & $\tau$ (h)  & $\sigma_0$ (Jy) & $\chi^2_n$ & $\log Z_{\rm DRW}$ & & $\mu$ (Jy) & $\sigma$ (Jy) & $\tau$ (h) & $\alpha_{\rm PSD}$  & $\sigma_0$ (Jy) & $\log Z_{\rm Mat}$  \\
 
 \hline
 SM all LO  & $2.45^{+0.14}_{-0.13}$ & $0.20^{+0.13}_{-0.02}$  & $3.57^{+5.63}_{-0.62}$ & $< 0.005$ & 0.77 & 1389 &  & $2.47^{0.31}_{-0.42}$ & $0.20^{+0.55}_{-0.02}$ & $0.87^{+3.60}_{-0.11}$ & $-3.25^{+0.61}_{-0.47}$ & $< 0.005$ & 1412  \\
 
 SM all HI  & $2.46^{+0.16}_{-0.13}$& $0.21^{+0.13}_{-0.02}$ & $3.86^{+5.62}_{-0.73}$& $< 0.005$ & 0.72 & 1356 &  & $2.48^{+0.31}_{-0.42}$ & $0.21^{+0.57}_{-0.02}$ & $0.93^{+2.85}_{-0.37}$ & $-3.19^{+0.65}_{-0.43}$ &$< 0.005$ & 1377 \\

  A1 all LO &  $2.37^{+0.23}_{-0.18}$ & $0.32^{+0.11}_{-0.04}$  & $10.56^{+6.84}_{-2.66}$ & 0.010 & 1.01 & 5128$^{b}$ &  &  $2.39^{+1.12}_{-0.41}$ & $0.29^{+0.62}_{-0.06}$ & $1.96^{+5.41}_{-0.63}$ & $-2.60^{+0.33}_{-0.41}$ &0.011 & 4967 \\
  A1 all HI & $2.42^{+0.25}_{-0.22}$ & $0.32^{+0.12}_{-0.04}$  & $10.36^{+6.27}_{-2.95}$ & 0.010 & 1.01 & 5025$^{b}$ &  &  $2.46^{+0.48}_{-0.69}$ & $0.31^{+0.63}_{-0.05}$ & $1.92^{+5.26}_{-0.61}$ & $-2.65^{+0.44}_{-0.33}$ &0.010 & 5054 \\
  \hline

   A2 all LO  & $2.20^{+0.23}_{-0.18}$ & $0.24^{+0.18}_{-0.02}$  & $3.46^{+6.51}_{-0.66}$ & 0.014 & 1.09 & 4085 &  &  $2.22^{+0.97}_{-0.31}$ & $0.23^{+0.55}_{-0.03}$ & $1.56^{+5.67}_{-0.46}$ & $-2.31^{+0.34}_{-0.30}$ & 0.012 & 4029 \\
  A2 all HI & $2.30^{+0.32}_{-0.19}$ & $0.25^{+0.18}_{-0.02}$  & $3.65^{+6.60}_{-0.71}$& 0.014 &1.10 & 3898 &  &  $2.31^{+0.36}_{-0.66}$ & $0.24^{+0.56}_{-0.03}$ & $1.53^{+5.38}_{-0.53}$ & $-2.36^{+0.35}_{-0.28}$ &0.013 & 3864  \\
  \hline

  FULL LO& $2.38^{+0.22}_{-0.12}$ & $0.27^{+0.10}_{-0.03}$ & $7.37^{+5.83}_{-1.50}$ & 0.010 & 0.95 & 5716$^{b}$ & & $2.39^{+0.36}_{-0.62}$ & $0.26^{+0.54}_{-0.04}$ & $1.73^{+5.05}_{-0.39}$ & $-2.58^{+0.32}_{-0.30}$ & 0.011 & 5748 \\
  FULL HI$^{a}$  & $2.43^{+0.15}_{-0.18}$ & $0.29^{+0.10}_{-0.03}$ & $8.07^{+5.79}_{-1.70}$ & 0.010 & 0.84 & 5786$^{b}$ & & $2.44^{+0.86}_{-0.37}$ & $0.28^{+0.58}_{-0.03}$ & $1.82^{+5.02}_{-0.43}$ & $-2.60^{+0.32}_{-0.31}$ & 0.010 & 5817 \\
 \hline
 
  
  2005-2019  & $3.22^{+0.09}_{-0.10}$ & $0.62^{+0.04}_{-0.05}$ & $20.72^{+3.16}_{-3.53}$ & 0.009 & 1.05 & 31926 & &  &  &  &  & &  \\
 
  \hline
 
 \label{tab:GPresults}
\end{tabularx}
\end{center}
\vspace{-0.4cm}
$^{a}$selected as a fiducial DRW fit. \  $^{b}$in these fits the DRW evidence and Mat\'{e}rn model results correspond to data sets with the A1 data sub-sampled with a factor of 4 to facilitate the computationally expensive model fitting. A high degree of consistency between the DRW fits for normal and sub-sampled data sets has been verified.

\end{table*}
As a result of the DRW survey, we selected the following set of priors, $\pi(\theta)$,
\begin{align}
    &\tau \ (\text{h})  \sim \mathcal{N}_{\rm T}(0,8) ,\nonumber \\
    &\mu \ (\text{Jy}) \sim \mathcal{N}_{\rm T}(2,1) ,\nonumber \\
    &\sigma \ (\text{Jy})  \sim \mathcal{N}_{\rm T}(0,1) ,\nonumber \\
    &\sigma_0 \ (\text{Jy})  \sim \mathcal{U}(0.0, 0.1) \, , \label{eq:DRWpriors} \\
    \nonumber
\end{align}
where $\mathcal{N}_{\rm T}(a, b)$ is a normal distribution of mean $a$ and standard deviation $b$ truncated to positive values, and $\mathcal{U}(a,b)$ is a uniform distribution with a range between $a$ and $b$. For the Mat\'{e}rn model fitting, we adopt identical priors as given by Equation \ref{eq:DRWpriors}, with an additional prior on the PSD slope, $\alpha_{\rm PSD}$,
\begin{equation}
    \alpha_{\rm PSD} = -2\nu -1 \sim \mathcal{U}(-9,-1) \ .
\end{equation} 

\subsection{Modeling results}

An overview of the fitting results for different data subsets is shown in
Table \ref{tab:GPresults}, where the max-likelihood (ML) estimator parameters are given, along with the 68\% confidence intervals. These results establish a good consistency between bands, less so between the pipelines. The estimated noise floor, $\sigma_0$, is comparable to the noise amplitudes estimated in Section \ref{sec:SF}. 

For the fiducial fit, a combined data set was prepared, merging light curves from the A1 and SM pipelines for increased time coverage (FULL data set in Table \ref{tab:GPresults}, see also Table \ref{tab:detections} and Appendix \ref{appendix:corns} for the fits cornerplots) for a light curve with a total time span of 148\,h. In the overlapping time periods, we only use the A1 data. Additionally, since we found constant scaling biases between the pipelines, we correct the SM data by applying small constant (per day / band) scaling factors, reported in Table \ref{tab:medians}, in order to ensure continuity. The DRW and Mat\'{e}rn fits generally yield consistent ML estimators of mean $\mu$, standard deviation $\sigma$, and noise floor $\sigma_0$. The Mat\'{e}rn fit has a clear preference for a shorter timescale $\tau$. This may possibly be a demonstration of a DRW bias reported by \citet{Kozlowski2016}; the DRW may fit data drawn from different processes well (notice good $\chi^2_n$ values reported in Table \ref{tab:GPresults}), while biasing timescales toward larger values if the true underlying process has a steeper PSD slope. On the other hand, the DRW timescale fitted to the FULL data sets is consistent with the $\sim 8$\,h found by \citealt{Dexter2014} (which, however, could be biased just the same if the true underlying process was not a DRW). For comparison, in IR the DRW timescale was found to be $\sim\,3-4$\,h \citep{meyer:09a,witzel:18a}, between the Mat\'{e}rn and DRW values fitted at mm wavelength. We find that estimated timescales, unlike other model parameters, are generally susceptible to details of the priors. We also report a DRW fit to the FULL data sets (Table \ref{tab:detections}) combined with the 2005--2019 non-EHT data sets given in Table \ref{tab:detections_other_papers}. Due to the numerical conditioning issues of the problem, the Mat\'{e}rn fit was not obtained for this data set. For this complete data set, the fit needs to accommodate a larger historical mean flux density of \sgra and a wider range of historically measured values, hence the mean flux and standard deviation are driven up. It is interesting to notice that the estimated parameters of the 2005--2019 DRW fit are reasonably consistent with the properties of the G$\lambda$D fit shown in the top left panel of Figure \ref{fig:distributions}, the former corresponding to $3.22\pm 0.62$\,Jy, and the latter corresponding to $3.24^{+0.68}_{-0.60}$\,Jy. This confirms that the GP models are capable of describing the source dynamics reasonably well. 

\subsection{Model selection and the PSD slope}
\label{sec:model_select}

\begin{figure}[t]
    \centering
    \includegraphics[width=\columnwidth]{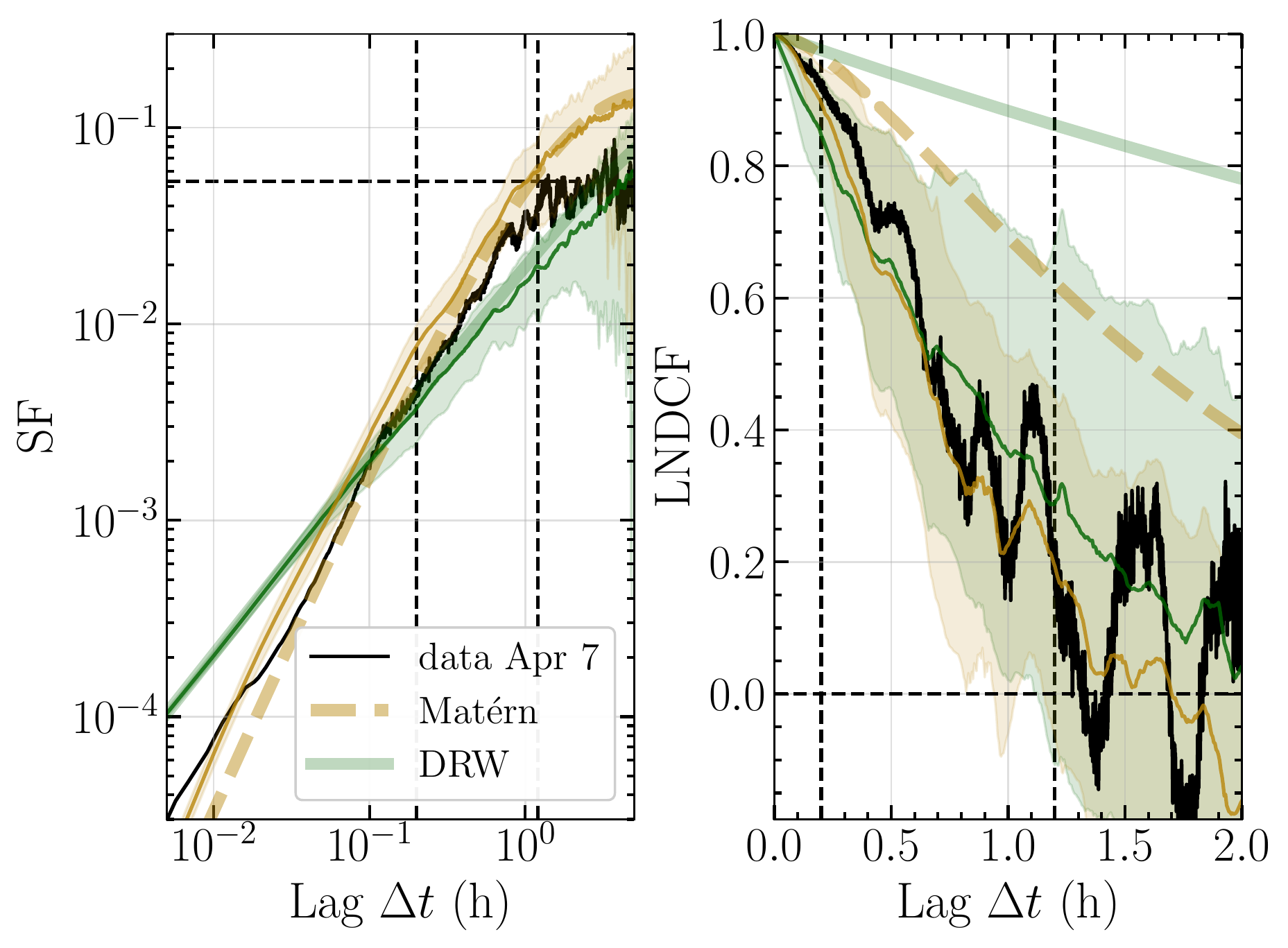}
    \caption{Analytic SF and autocorrelation of the fiducial fits with the DRW and Mat\'{e}rn process compared with the observations (black; A1 pipeline, HI band, 2017 April 7 data). Thin lines and color bands show the median and 1\,$\sigma$ ranges for the estimates of the SF and autocorrelation in a random realization of a best-fitting process, given the actual sampling and duration of the ALMA observations.}
    \label{fig:SF_ACF}
\end{figure}

\begin{figure}[t]
    \centering
    \includegraphics[width=\columnwidth]{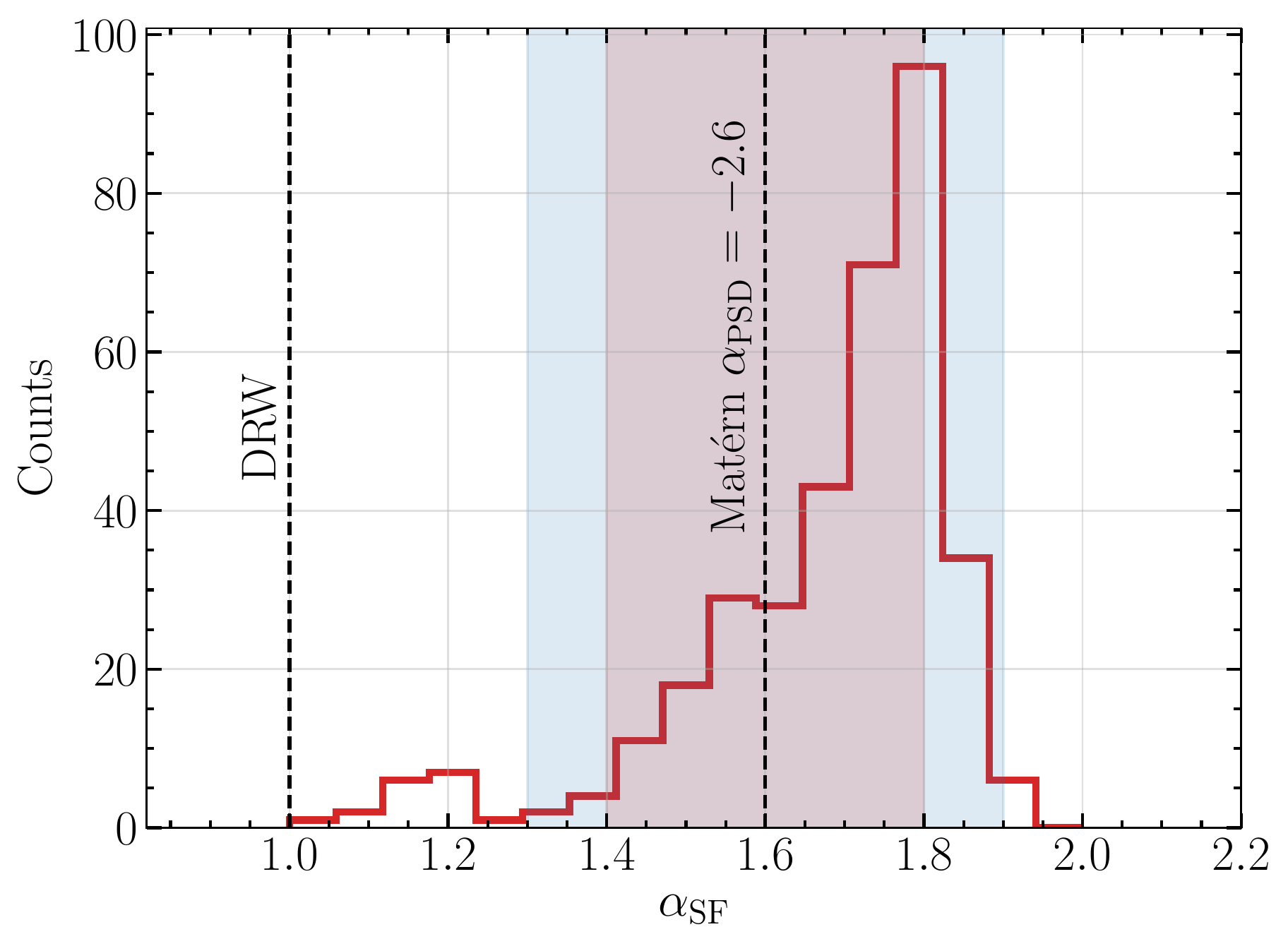}
    \caption{Histogram of the SF slopes ($\alpha_{\rm SF}$) calculated for over 350 GRMHD simulations of \sgra. The red shaded region corresponds to the range of slopes measured in the 2017 April observations, the blue shaded region corresponds to the uncertainty of the slope estimated with the Mat\'{e}rn covariance model. Both the GRMHD simulations and observations disfavor the DRW model and show consistency with the best-fitting Mat\'{e}rn process model.}
    \label{fig:SFslopeGRMHD}
\end{figure}

Since we explore the posterior space with a nested sampling algorithm, we obtain the Bayesian evidence along with the posterior distributions \citep{Speagle2020}, representing the total likelihood of a given model, $\Theta$,
\begin{equation}
    Z_{ \Theta} = \int p \left( \{ x_i \} | \theta \right) \pi(\theta) {\rm d} \theta \ .
\end{equation}
By directly comparing Bayesian evidences obtained for the same data sets with the DRW and Mat\'{e}rn models, we may select a more likely model. In Table \ref{tab:GPresults}, we compare the logarithm of Bayesian evidence for the DRW ($\log Z_{\rm DRW}$) and the Mat\'{e}rn ($\log Z_{\rm Mat}$) models. While the comparison results generally vary depending on the data subset used, the fiducial fit shows the advantage of the Mat\'{e}rn over the DRW model. 

We further verify this conclusion by considering a consistency test for the best-fitting DRW and Mat\'{e}rn processes. In this test, we generate a collection of synthetic light curves corresponding to random realizations of the models described by the FULL HI fiducial fits reported in Table \ref{tab:GPresults}. These synthetic data sets were generated with the exact sampling of the ALMA A1 light curves from 2017 April 7. We then calculate the analytic SFs for both processes (Equation \ref{eq:SF_from_ACF}), empirically measured SFs for the synthetic light curves (Equation \ref{eq:structure_function}), analytic autocorrelation functions (Eqs. \ref{eq:covariance} and \ref{eq:covariance_matern}), and autocorrelations measured on the synthetic light curves (Equation \ref{eq:LNDCF}). The results are shown in Figure \ref{fig:SF_ACF}. We see that the bias between the analytic results and what we measure, given the limited time coverage, is more significant for autocorrelations than for SFs. When these synthetic data tests are compared with the actual SFs and autocorrelations measured from the 2017 April 7 ALMA A1 observations, we see a low constraining power of the autocorrelation measurements. On the other hand, the SFs results show much higher consistency with observations for the Mat\'{e}rn model than for the DRW, as the former reproduces the steep observed SF slope reported in Section \ref{sec:SF}.

We can also make use of the EHT GRMHD library, consisting of over 350 simulations of \sgra exploring a variety of black hole spins, observer inclinations, plasma heating parameters, and accretion flow magnetization states \citepalias{SgraP5}. In Figure \ref{fig:SFslopeGRMHD}, we show a histogram of the high frequency SF slopes ($\alpha_{\rm SF}$) for the simulation library. The slopes were measured with linear regression on the logarithm of the SF for the timescales shorter than 25\,$GM/c^3 \approx$\,500\,s. The DRW slope is always $\alpha_{\rm SF} = 1$, while for the best-fitting Mat\'{e}rn model with $\alpha_{\rm PSD} = -2.6$, we found the approximated formula given in Section \ref{sec:SF} to be very consistent with a numerical evaluation, hence $\alpha_{\rm SF} \approx 1.6$. The range of high frequency slope values measured for ALMA (reported in Table \ref{tab:SFresultsALMA}) is shaded in red in Figure \ref{fig:SFslopeGRMHD}. The uncertainty of the Mat\'{e}rn process slope estimation, as reported in Table \ref{tab:GPresults}, is shaded in blue. Remarkably, GRMHD simulations, the SF calculated on observations, and the Mat\'{e}rn fit are reasonably consistent. All GRMHD models indicate $\alpha_{\rm SF} > 1$, steeper than the DRW value. 

We also notice that recent results suggest that the mm PSD on shorter timescales may be steeper than $\alpha_{\rm PSD}= -2$ \citep{Iwata2020,Murchikova2021}, and some indications of a steeper 230\,GHz slope were also reported by \citet{Dexter2014}, who gave a value of $\alpha_{\rm PSD} = -2.3^{+0.6}_{-0.8}$. Along with other hints, all these allow us to conclude that the Mat\'{e}rn covariance model captures the short timescale variable behavior of \sgra light curves better than the DRW model. Note, however, that if the PSD break timescales discussed in Section \ref{sec:SF} are real, they would not be properly represented within the Mat\'{e}rn covariance model of a single stochastic process with a smooth PSD -- the presence of a sharp break in the PSD would indicate superposition of at least two stochastic processes. 

%% file: S6_PSD.tex
\section{Periodicity search and PSD}
\label{sec:periodicity}

\begin{figure}[h!]
    \centering
    \includegraphics[width=\columnwidth]{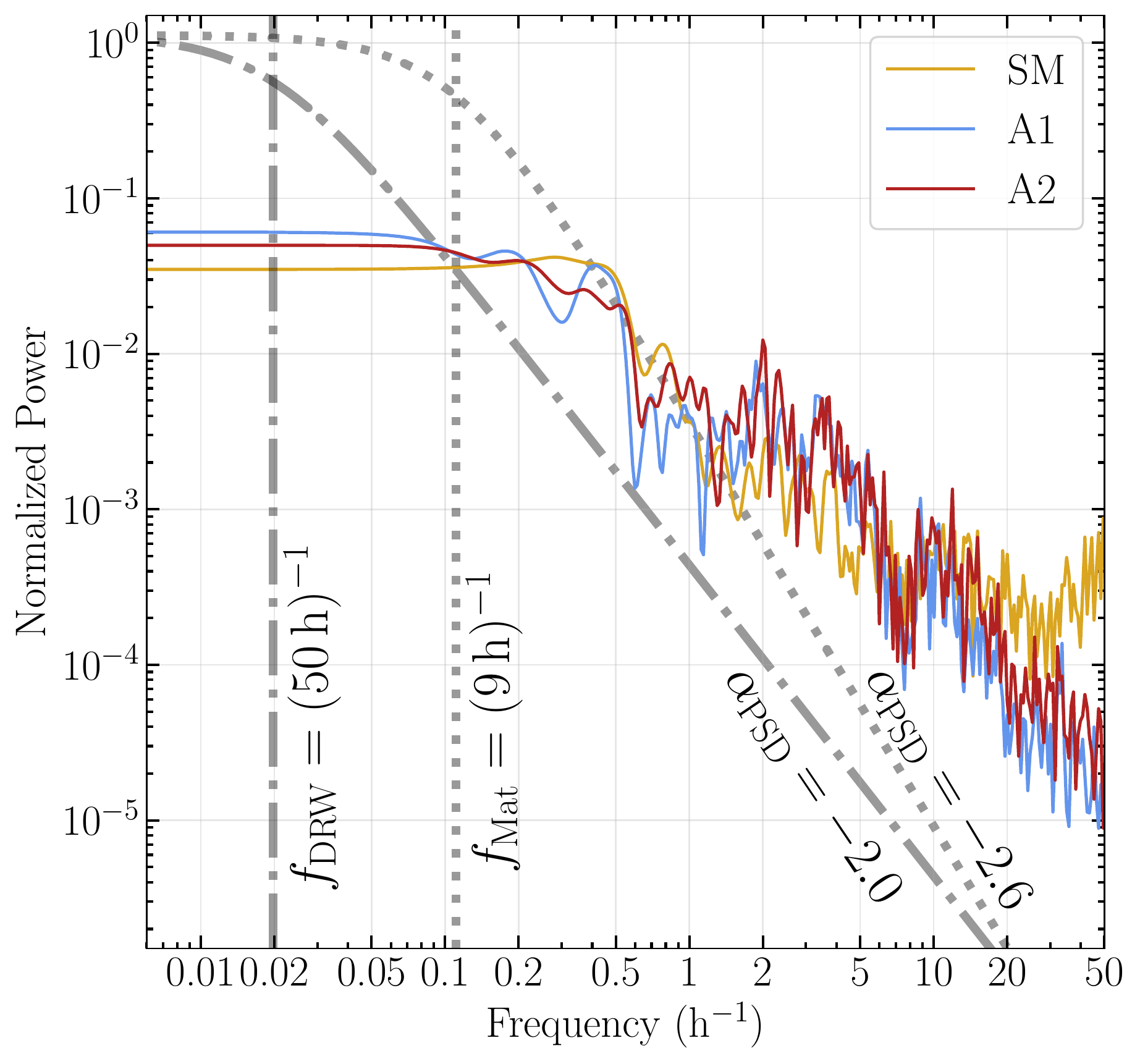}
    \caption{Incoherently averaged, normalized L-S periodograms for all data sets available within each pipeline (averaging all days and bands). The PSD of the best-fitting DRW ($\alpha_{\rm PSD} = -2.0$) and Mat\'{e}rn ($\alpha_{\rm PSD} = -2.6$) models are shown with dash-dot and dotted lines, respectively. The characteristic frequencies of the transition to white noise are shown for both fitted models. }
    \label{fig:LSsummary}
\end{figure}

\begin{figure*}[t]
    \centering
    \includegraphics[width=0.998\textwidth,trim=0.0cm 0.96cm -0.16cm 0cm,clip]{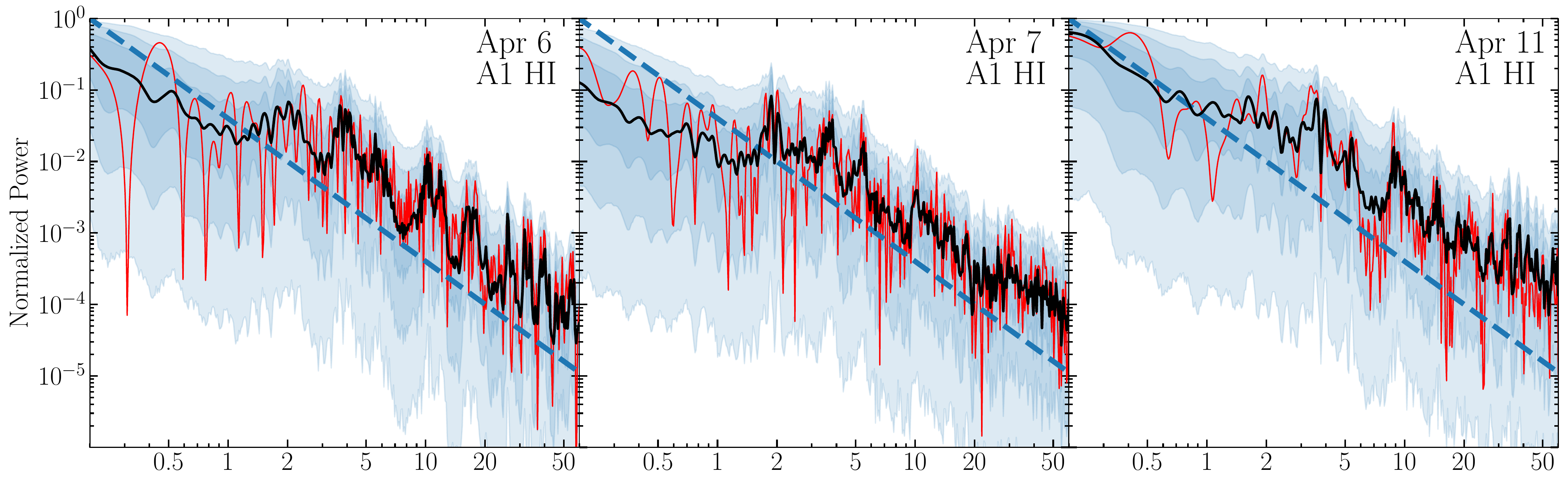}
    \includegraphics[width=0.985\textwidth,trim=-0.12cm 1.85cm 0.12cm 0.25cm,clip]{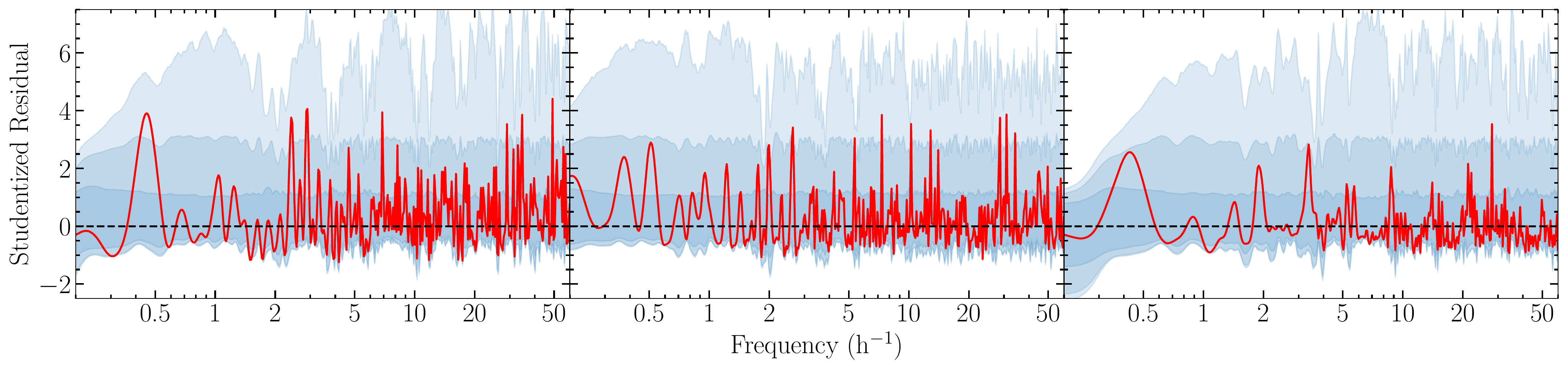}
    \includegraphics[width=0.985\textwidth,trim=-0.12cm 0 0.12cm 0.25cm,clip]{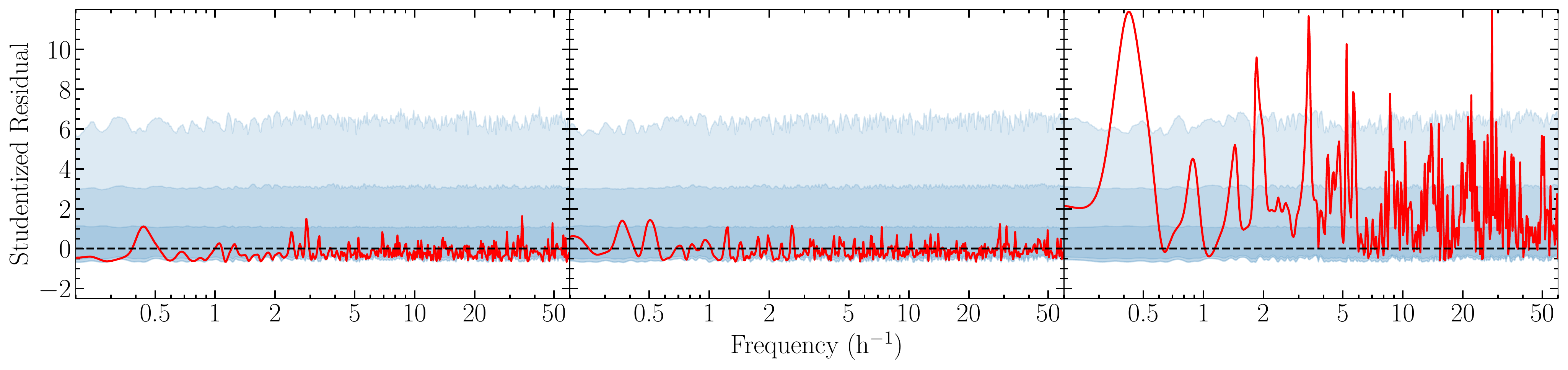}
    \caption{\textit{Top:} The normalized L-S periodograms for the A1 pipeline HI band data are shown with red lines. The median value of the L-S periodogram, corresponding to a DRW fit to all 2005--2019 data, is shown with black lines. Shaded areas indicate 68.0\%, 95.0\% and 99.7\% intervals for the L-S periodogram of the DRW model, evaluated with a Monte Carlo procedure. The dashed line represents the ideal PSD of the considered DRW model.
\textit{Middle:} Studentized residuals between the data and the median DRW power for a normalized L-S periodogram. The vertical axis is in units of the standard deviation.
    \textit{Bottom:} Similar to the middle row, but with an unnormalized (physical units) L-S periodogram, showing the overall excess of power on 2017 April 11.}
    \label{fig:LSperiodicity}
\end{figure*}

Identifying a periodic component in radio astronomical data is particularly challenging with the presence of red noise, and given the non-uniform sampling. It is common to misinterpret the uncertainty budget, e.g., by imprinting the white noise background model in the analysis. Unfortunately, the properly calculated uncertainties related to a particular realization of a stochastic process, along with the non-uniform sampling biases, may prevent one from confidently detecting real periodicity, unless a large number of periods is sampled. In this section, we discuss the PSD estimated from the observational \sgra light curve data with a Lomb-Scargle algorithm \citep[L-S,][]{Scargle1982}. In particular, our aim is to determine whether there are any frequencies excited significantly more than the expectations from the fitted aperiodic GP models, and thus indicating a difficulty in interpreting them in the purely stochastic framework discussed in Section \ref{sec:modeling}. Similar investigations in the IR, presented in \citet{do:09a}, concluded that the light curves are consistent with a stochastic red noise process.

If we consider an L-S periodogram of the EHT observations, the red noise characteristic is apparent, see Figure \ref{fig:LSsummary}. For ALMA, we can trace the negative slope all the way to a single minute timescale before the observational noise takes over, which was discussed in Section \ref{sec:SF}. The SMA periodograms flatten for timescales shorter than about 3--5\,min because of the residual noise. For the fiducial fits (FULL HI in Table \ref{tab:GPresults}), the transition frequency separating the white and red noise parts of the DRW PSD is $f_{\rm DRW} = (2 \pi \tau_{\rm DRW})^{-1} = (50\,{\rm h})^{-1}$, while for the Mat\'{e}rn process fit it is $f_{\rm Mat} = \sqrt{2 \nu}(2 \pi \tau_{\rm Mat})^{-1} = (9\,{\rm h})^{-1}$. When the L-S periodogram is compared with the analytic PSDs (Figure \ref{fig:LSsummary}), neither of the best-fitting models appear to be in good agreement with the data. The reason is the corruption related to the sampling window. To study whether the stochastic model can reproduce the data periodogram, we need to incorporate the real data sampling into the discussion. Hence, we take a Monte Carlo approach, similar to the procedure employed by \citet{Haggard2019}. We generate $5\times 10^4$ realizations of the light curves from the best-fitting models, sampled with the original sampling windows of the ALMA A1 data on 2017 April 6, 7, and 11. We use \texttt{astroML} \citep{astroML} for the DRW and \texttt{Stheno} for the Mat\'{e}rn process light curve generation. We then calculate the L-S periodogram for each synthetic light curve with \texttt{Astropy} \citep{astropy2018}, and compare the results with the L-S periodograms calculated for the actual \sgra light curve data sets. We performed this test for the FULL HI DRW and Mat\'{e}rn fits, as well as for the 2005--2019 combined DRW fit. The example periodograms for the latter model are presented in the top row of Figure \ref{fig:LSperiodicity}, along with the residuals between the median L-S value for the DRW model and the value estimated for the observations (middle and bottom rows). The ideal PSD of the DRW model is shown with blue dashed lines. Given the large model correlation timescale with respect to light curve duration, the ideal DRW PSD effectively corresponds to an almost constant slope of $\alpha_{\rm PSD}=-2$. Hence, all of the intricate structure of the median DRW periodogram inferred from synthetic light curves (black curve), clearly reflected also in the L-S periodograms of the real observations (red curve), can be attributed to the limited and irregular sampling alone. This is visible more clearly in the middle row of Figure \ref{fig:LSperiodicity}. No L-S normalized periodogram peak on either of the observing nights indicates deviation by more than 3\,$\sigma$ from the aperiodic model predictions. However, when we consider residuals of an unnormalized periodogram (Figure \ref{fig:LSperiodicity}, bottom row), in which the PSD represents the amount of variability in physical units (and hence the periodogram test is sensitive to the overall scaling of variance), we see big differences between the days. While 2017 April 6 and 7 are rather calm in comparison to the global fit predictions, the flaring day of 2017 April 11 now shows far more variability than the fit would predict. This variability increase is, however, affecting the whole PSD, not just any selected characteristic frequency. While the fit to all of the 2005--2019 data is shown in Figure \ref{fig:LSperiodicity}, these findings are consistent for the other considered models (Mat\'{e}rn and DRW fitted to the EHT 2017 data). We conclude that the amount of variability on the flaring day is not properly described with any of the best-fit models fitted to the broader data sets.

The variability increase on 2017 April 11 was seen already in Table \ref{tab:detections} (standard deviations on 2017 April 11 increased 2--3 times) and in Section \ref{sec:SF} (long timescale variance was enhanced by a factor of $\sim$10). If we quantify the periodogram consistency with the Monte Carlo model periodogram test, described by \citet{Uttley2002}, the DRW model fitted to all of the 2005--2019 data is 99.90\% inconsistent with the 2017 April 11 data (the FULL DRW fit to 2017 April data is inconsistent at 100.00\% and the Mat\'{e}rn fit to 2017 April data at 98.71\%). All best-fitting models are consistent with all the remaining observing days, bands, and reduction pipelines. This particularly strong variability of \sgra on 2017 April 11 motivated restricting the first analysis of the EHT VLBI observations to 2017 April 6 and 7, where static imaging \citepalias{SgraP3} and modeling \citepalias{SgraP4} techniques are more straightforwardly applicable.

%% file: S7_Summary.tex
\section{Summary and Discussion}
\label{sec:discussion}

We have developed algorithms to generate light curve data from observations with phased interferometric arrays, enabling simultaneous participation in VLBI observations. We apply them here and present the high cadence and high \textit{S/N} 1.3\,mm light curves of \sgra obtained during the EHT observing campaign in 2017 April with ALMA and SMA. There are several noteworthy conclusions: 
\begin{enumerate}

\item With the very high \textit{S/N} of ALMA, thermal noise is not limiting in the analysis. However, significant systematic uncertainties related to the data calibration persist. We elucidate that issue by comparing three independent data reduction pipelines (2 for ALMA, 1 for SMA). While we show general consistency between them, some results, such as the GP correlation timescales ($\tau$), or the presence of structure function break timescales, are sensitive to the pipeline choice. We notice overall better performance from the intra-field calibration method A1 (more robust against uncertainties related to low elevation, better consistency with the independently measured SMA flux densities), and conclude that the A1 method should be preferred for future analyses of this nature.

\item During the EHT observations on 2017 April 5--11, \sgra exhibited a low flux density of $2.4\pm0.2$\,Jy, and overall low variability, $ \sigma/\mu < 10$\%. The modulation index, $\sigma/\mu$, is consistent with other observations in 2005--2019. On 2017 April 11, the ALMA observations immediately followed an X-ray flare, with the mm flux density growing by about 50\% and reaching a peak flux density $\sim$2.2\,h after the X-ray flare maximum. We observe strongly enhanced variability across multiple timescales on that day, with a near order of magnitude increase in the variance. The statistical PSD properties of the 2017 April 11 observations are inconsistent with those of the GP models fitted to the \sgra light curve data sets.

\item We measure the average spectral index at 220\,GHz to be $\alpha = 0.0\pm0.1$, where the uncertainties are dominated by the calibration systematics and by the rapid time variability of the spectral index, wandering between $\pm 0.2$ on a timescale of $\sim$1\,h. The spectral index immediately following the X-ray flare of 2017 April 11 is significantly lower, $-0.25\pm{0.10}$.

\item No statistically significant autocorrelations are found. If detected, these persistent correlations could be attributed to the presence of the photon shell in the strongly curved spacetime around the black hole. They continue to be expected if sufficiently long observations are aggregated, e.g., from stacking high cadence ALMA observations over multiple years.

\item There are no time-lags detected between the observed frequency bands (between 213\,GHz and 229\,GHz), indicating that the source is essentially optically thin all the way to the event horizon at the observing frequencies, possibly with irregular patches of higher optical depth evolving on dynamical timescales. This is also consistent with the spectral index variability signatures.

\item With the high cadence of our light curves, we are able to track the short-timescale variability of \sgra, confirming a red noise characteristic across timescales from a single minute to several hours. Furthermore, we see a convincing indication of a PSD slope of $2.6\pm0.3$ for short timescales, steeper than the commonly employed DRW model. There is a mutual consistency between the Mat\'{e}rn process fit to the observations, the structure function analysis results, and the predictions from the GRMHD simulations. In the structure function analysis, we additionally observe a potential power-law break at a 0.15--0.30\,h timescale, that may approximate the steepening PSD slope of the Mat\'{e}rn process, or indicate a superposition of distinct stochastic processes.

\item  Aperiodic GP models fitted to the data provide good quality fits, and generally capture the spectral properties of the light curves well. However, the 2017 April 11 observations indicate too much variability to be represented with the same models as the other days. The correlation timescale is not consistently constrained between different considered models. The DRW fit to the collection of observations from 2005--2019 gives $\tau = 20.7^{+3.2}_{-3.5}$\,h, while correlations on even longer timescales are hinted at in the long term monitoring results. For example, 4 different projects observing between August 2016 and October 2017 all report flux densities below the long term mean. At the same time the DRW fit to EHT 2017 data gives $\tau=8.1^{+5.8}_{-1.7}$\,h, while the Mat\'{e}rn process fits find shorter timescales of $\tau =1.8^{+5.0}_{-0.4}$\,h. We conclude that the correlation timescale remains poorly constrained. Along with the 2017 April 11 inconsistency, this may suggest that the assumption of a single stationary statistical process is incorrect when different epochs (or different source activity states) are combined.

\end{enumerate}

Overall, the light curve analysis presented in this paper indicates that during the 2017 EHT observing campaign \sgra was in a low luminosity state with respect to the 2005--2019 average of $3.2 \pm 0.6$\,Jy, implying low optical depth, and thus strengthening the case for event horizon scale imaging with the VLBI data. The source displayed an average amount of variability on 2017 April 5--10. Hence, we expect that the VLBI analysis of 2017 April 6--7 data, presented in \citetalias{SgraP1,SgraP2,SgraP3,SgraP4,SgraP5,SgraP6}, should reveal a representative event horizon scale morphology of the
source during a non-flaring low variability period. Nevertheless, we see intrinsic source variability on timescales as short as 1\,min which may affect the EHT VLBI observations in a non-trivial way and we argue that these impacts must be mitigated at the data analysis stage (\citetalias{SgraP4}; \citealt{Farah2022,Broderick2022}). On 2017 April 11, \sgra displayed significantly enhanced variability in the aftermath of a strong X-ray flare. This different state may impact \sgra's event horizon scale morphology, and the excess variability on that day may undermine our ability to define a mean static image.

The measured source variability is expected to be related mostly to the intrinsic variability of the compact source, with a small subdominant contribution from the interstellar medium scattering screen \citep[order of 1\%;][]{johnson2018}. Hence, it is possible to use GRMHD simulations of \sgra to make a direct comparison of the observed variability metrics reported in this paper with the predictions from numerical models.
This approach has been pursued in \citetalias{SgraP5}, revealing a rather puzzling disagreement -- numerical GRMHD simulations seem to produce systematically more variability than what we measure in the \sgra light curves, see the discussion in \citetalias{SgraP5}.

During this campaign, ALMA also recorded \sgra's total intensity light curves and full polarization data. The analysis of that data set, interesting particularly in the context of polarization loops hypothetically associated with the orbital motion in the innermost accretion flow region \citep{Marrone2006JPhCS,gravity_loops_2018}, will be presented elsewhere. More light curve data of similar or improved quality will be delivered with the subsequent EHT VLBI observing campaigns, advancing our understanding of the statistical properties of \sgra variability at mm wavelengths and of Galactic Center physics.

%% file: acknowledgments.tex
\section*{Acknowledgements}
We thank  Yuhei Iwata, Lena Murchikova, Rebecca Phillipson, and Chris White for comments and discussions, as well as Alexandra Elbakyan for her contributions to the open science initiative. We also thank the anonymous ApJL referee for helpful and constructive comments. 
The Event Horizon Telescope Collaboration thanks the following
organizations and programs: 
National Science Foundation (awards OISE-1743747, AST-1816420, AST-1716536, AST-1440254, AST-1935980);
the Black Hole Initiative, which is funded by grants from the John Templeton Foundation and the Gordon 
and Betty Moore Foundation (although the opinions expressed in this work are those of the author(s) and do
not necessarily reflect the views of these Foundations);
NASA Hubble Fellowship grant
HST-HF2-51431.001-A awarded by the Space Telescope
Science Institute, which is operated by the Association
of Universities for Research in Astronomy, Inc.,
for NASA, under contract NAS5-26555;
the Academy
of Finland (projects 274477, 284495, 312496, 315721); the Agencia Nacional de Investigación y Desarrollo (ANID), Chile via NCN$19\_058$ (TITANs) and Fondecyt 3190878, the Alexander
von Humboldt Stiftung; an Alfred P. Sloan Research Fellowship;
Allegro, the European ALMA Regional Centre node in the Netherlands, the NL astronomy
research network NOVA and the astronomy institutes of the University of Amsterdam, Leiden University and Radboud University;
the China Scholarship
Council;  Consejo
Nacional de Ciencia y Tecnolog\'{\i}a (CONACYT,
Mexico, projects  U0004-246083, U0004-259839, F0003-272050, M0037-279006, F0003-281692,
104497, 275201, 263356);
the Delaney Family via the Delaney Family John A.
Wheeler Chair at Perimeter Institute; Dirección General
de Asuntos del Personal Académico-—Universidad
Nacional Autónoma de México (DGAPA-—UNAM,
projects IN112417 and IN112820); the European Research Council Synergy
Grant "BlackHoleCam: Imaging the Event Horizon
of Black Holes" (grant 610058); the Generalitat
Valenciana postdoctoral grant APOSTD/2018/177 and
GenT Program (project CIDEGENT/2018/021); MICINN Research Project PID2019-108995GB-C22;
the European Research Council for advanced grant 'JETSET: Launching, propagation and 
emission of relativistic 
jets from binary mergers and across mass scales' (Grant No. 884631); 
the Istituto Nazionale di Fisica
Nucleare (INFN) sezione di Napoli, iniziative specifiche
TEONGRAV; the two Dutch National Supercomputers, Cartesius and Snellius  
(NWO Grant 2021.013);
the International Max Planck Research
School for Astronomy and Astrophysics at the
Universities of Bonn and Cologne; 
DFG research grant ``Jet physics on horizon scales and beyond'' (Grant No. FR 4069/2- 1);
Joint Princeton/Flatiron and Joint Columbia/Flatiron Postdoctoral Fellowships, 
research at the Flatiron Institute is supported by the Simons Foundation; 
the Japanese Government (Monbukagakusho:
MEXT) Scholarship; the Japan Society for
the Promotion of Science (JSPS) Grant-in-Aid for JSPS
Research Fellowship (JP17J08829); the Key Research
Program of Frontier Sciences, Chinese Academy of
Sciences (CAS, grants QYZDJ-SSW-SLH057, QYZDJSSW-
SYS008, ZDBS-LY-SLH011); the Leverhulme Trust Early Career Research
Fellowship; the Max-Planck-Gesellschaft (MPG);
the Max Planck Partner Group of the MPG and the
CAS; the MEXT/JSPS KAKENHI (grants 18KK0090, JP21H01137,
JP18H03721, 18K03709,
18H01245, 25120007); the Malaysian Fundamental Research Grant Scheme (FRGS) FRGS/1/2019/STG02/UM/02/6; the MIT International Science
and Technology Initiatives (MISTI) Funds; 
the Ministry of Science and Technology (MOST) of Taiwan (103-2119-M-001-010-MY2, 105-2112-M-001-025-MY3, 105-2119-M-001-042, 106-2112-M-001-011, 106-2119-M-001-013, 106-2119-M-001-027, 106-2923-M-001-005, 107-2119-M-001-017, 107-2119-M-001-020, 107-2119-M-001-041, 107-2119-M-110-005, 107-2923-M-001-009, 108-2112-M-001-048, 108-2112-M-001-051, 108-2923-M-001-002, 109-2112-M-001-025, 109-2124-M-001-005, 109-2923-M-001-001, 110-2112-M-003-007-MY2, 110-2112-M-001-033, 110-2124-M-001-007, and 110-2923-M-001-001);
the Ministry of Education (MoE) of Taiwan Yushan Young Scholar Program;
the Physics Division, National Center for Theoretical Sciences of Taiwan;
the National Aeronautics and
Space Administration (NASA, Fermi Guest Investigator
grant 80NSSC20K1567, NASA Astrophysics Theory Program grant 80NSSC20K0527, NASA NuSTAR award 
80NSSC20K0645); 
the National
Institute of Natural Sciences (NINS) of Japan; the National
Key Research and Development Program of China
(grant 2016YFA0400704, 2016YFA0400702); the National
Science Foundation (NSF, grants AST-0096454,
AST-0352953, AST-0521233, AST-0705062, AST-0905844, AST-0922984, AST-1126433, AST-1140030,
DGE-1144085, AST-1207704, AST-1207730, AST-1207752, MRI-1228509, OPP-1248097, AST-1310896,  
AST-1555365, AST-1615796, AST-1715061, AST-1716327,  AST-1903847,AST-2034306); the Natural Science
Foundation of China (grants 
11650110427, 10625314, 11721303, 11725312, 11933007, 11991052, 11991053);
NWO grant number OCENW.KLEIN.113; a 
fellowship of China Postdoctoral Science Foundation (2020M671266); the Natural
Sciences and Engineering Research Council of
Canada (NSERC, including a Discovery Grant and
the NSERC Alexander Graham Bell Canada Graduate
Scholarships-Doctoral Program); the National Youth
Thousand Talents Program of China; the National Research
Foundation of Korea (the Global PhD Fellowship
Grant: grants NRF-2015H1A2A1033752, 2015-
R1D1A1A01056807, the Korea Research Fellowship Program:
NRF-2015H1D3A1066561, Basic Research Support Grant 2019R1F1A1059721); the Netherlands Organization
for Scientific Research (NWO) VICI award
(grant 639.043.513) and Spinoza Prize SPI 78-409; the
New Scientific Frontiers with Precision Radio Interferometry
Fellowship awarded by the South African Radio
Astronomy Observatory (SARAO), which is a facility
of the National Research Foundation (NRF), an
agency of the Department of Science and Technology
(DST) of South Africa; the Onsala Space Observatory
(OSO) national infrastructure, for the provisioning
of its facilities/observational support (OSO receives
funding through the Swedish Research Council under
grant 2017-00648) the Perimeter Institute for Theoretical
Physics (research at Perimeter Institute is supported
by the Government of Canada through the Department
of Innovation, Science and Economic Development
and by the Province of Ontario through the
Ministry of Research, Innovation and Science); 
the Spanish Ministerio de Ciencia e Innovación (grants PGC2018-098915-B-C21, AYA2016-80889-P,
PID2019-108995GB-C21, PID2020-117404GB-C21); 
the State
Agency for Research of the Spanish MCIU through
the "Center of Excellence Severo Ochoa" award for
the Instituto de Astrofísica de Andalucía (SEV-2017-
0709); the Toray Science Foundation; the Consejería de Economía, Conocimiento, 
Empresas y Universidad 
of the Junta de Andalucía (grant P18-FR-1769), the Consejo Superior de Investigaciones 
Científicas (grant 2019AEP112);
the M2FINDERS project which has received funding by the European Research Council (ERC) under 
the European Union’s Horizon 2020 Research and Innovation Programme (grant agreement No 101018682);
the US Department
of Energy (USDOE) through the Los Alamos National
Laboratory (operated by Triad National Security,
LLC, for the National Nuclear Security Administration
of the USDOE (Contract 89233218CNA000001);
 the European Union’s Horizon 2020
research and innovation programme under grant agreement
No 730562 RadioNet;
Shanghai Pilot Program for Basic Research, Chinese Academy of Science, 
Shanghai Branch (JCYJ-SHFY-2021-013);
ALMA North America Development
Fund; the Academia Sinica; Chandra DD7-18089X and TM6-
17006X; the GenT Program (Generalitat Valenciana)
Project CIDEGENT/2018/021. This work used the
Extreme Science and Engineering Discovery Environment
(XSEDE), supported by NSF grant ACI-1548562,
and CyVerse, supported by NSF grants DBI-0735191,
DBI-1265383, and DBI-1743442. XSEDE Stampede2 resource
at TACC was allocated through TG-AST170024
and TG-AST080026N. XSEDE JetStream resource at
PTI and TACC was allocated through AST170028.
The simulations were performed in part on the SuperMUC
cluster at the LRZ in Garching, on the
LOEWE cluster in CSC in Frankfurt, and on the
HazelHen cluster at the HLRS in Stuttgart. This
research was enabled in part by support provided
by Compute Ontario (http://computeontario.ca), Calcul
Quebec (http://www.calculquebec.ca) and Compute
Canada (http://www.computecanada.ca). 
CC acknowledges support from the Swedish Research Council (VR).
We thank
the staff at the participating observatories, correlation
centers, and institutions for their enthusiastic support.
This paper makes use of the following ALMA data:
ADS/JAO.ALMA\#2016.1.01154.V. ALMA is a partnership
of the European Southern Observatory (ESO;
Europe, representing its member states), NSF, and
National Institutes of Natural Sciences of Japan, together
with National Research Council (Canada), Ministry
of Science and Technology (MOST; Taiwan),
Academia Sinica Institute of Astronomy and Astrophysics
(ASIAA; Taiwan), and Korea Astronomy and
Space Science Institute (KASI; Republic of Korea), in
cooperation with the Republic of Chile. The Joint
ALMA Observatory is operated by ESO, Associated
Universities, Inc. (AUI)/NRAO, and the National Astronomical
Observatory of Japan (NAOJ). The NRAO
is a facility of the NSF operated under cooperative agreement
by AUI.
Support for this work was also provided by the NASA Hubble Fellowship grant HST-HF2-51431.001-A awarded 
by the Space Telescope Science Institute, which is operated by the Association of Universities for 
Research in Astronomy, Inc., for NASA, under contract NAS5-26555.
Hector Olivares and  Gibwa Musoke
were supported by Virtual Institute of Accretion (VIA) postdoctoral fellowships 
from the Netherlands Research School for Astronomy (NOVA).
APEX is a collaboration between the
Max-Planck-Institut f{\"u}r Radioastronomie (Germany),
ESO, and the Onsala Space Observatory (Sweden). The
SMA is a joint project between the SAO and ASIAA
and is funded by the Smithsonian Institution and the
Academia Sinica. The JCMT is operated by the East
Asian Observatory on behalf of the NAOJ, ASIAA, and
KASI, as well as the Ministry of Finance of China, Chinese
Academy of Sciences, and the National Key R\&D
Program (No. 2017YFA0402700) of China. Additional
funding support for the JCMT is provided by the Science
and Technologies Facility Council (UK) and participating
universities in the UK and Canada. 
Simulations were performed in part on the SuperMUC cluster at the LRZ in Garching, on the 
LOEWE cluster in CSC in Frankfurt, on the HazelHen cluster at the HLRS in Stuttgart, 
and on the Pi2.0 and Siyuan Mark-I at Shanghai Jiao Tong University.
The computer resources of the Finnish IT Center for Science (CSC) and the Finnish Computing 
Competence Infrastructure (FCCI) project are acknowledged.
Junghwan Oh was supported by Basic Science Research Program through the National Research
Foundation of Korea(NRF) funded by the Ministry of Education(NRF-2021R1A6A3A01086420).
We thank Martin Shepherd for the addition of extra features in the Difmap software 
that were used for the CLEAN imaging results presented in this paper.
The computing cluster of Shanghai VLBI correlator supported by the Special Fund 
for Astronomy from the Ministry of Finance in China is acknowledged.
The LMT is a project operated by the Instituto Nacional
de Astrófisica, Óptica, y Electrónica (Mexico) and the
University of Massachusetts at Amherst (USA). The
IRAM 30-m telescope on Pico Veleta, Spain is operated
by IRAM and supported by CNRS (Centre National de
la Recherche Scientifique, France), MPG (Max-Planck-
Gesellschaft, Germany) and IGN (Instituto Geográfico
Nacional, Spain). The SMT is operated by the Arizona
Radio Observatory, a part of the Steward Observatory
of the University of Arizona, with financial support of
operations from the State of Arizona and financial support
for instrumentation development from the NSF.
Support for SPT participation in the EHT is provided by 
the National Science Foundation through award OPP-1852617 
to the University of Chicago. Partial support is also 
provided by the Kavli Institute of Cosmological Physics 
at the University of Chicago. The SPT hydrogen maser was 
provided on loan from the GLT, courtesy of ASIAA.
Support for this work was provided by NASA through the NASA Hubble Fellowship grant
\#HST--HF2--51494.001 awarded by the Space Telescope Science Institute, which is operated 
by the Association of Universities for Research in Astronomy, Inc., for NASA, 
under contract NAS5--26555.
The EHTC has
received generous donations of FPGA chips from Xilinx
Inc., under the Xilinx University Program. The EHTC
has benefited from technology shared under open-source
license by the Collaboration for Astronomy Signal Processing
and Electronics Research (CASPER). The EHT
project is grateful to T4Science and Microsemi for their
assistance with Hydrogen Masers. This research has
made use of NASA’s Astrophysics Data System. We
gratefully acknowledge the support provided by the extended
staff of the ALMA, both from the inception of
the ALMA Phasing Project through the observational
campaigns of 2017 and 2018. We would like to thank
A. Deller and W. Brisken for EHT-specific support with
the use of DiFX. We acknowledge the significance that
Maunakea, where the SMA and JCMT EHT stations
are located, has for the indigenous Hawaiian people.

%% file: S8_Appendices.tex
\appendix

\section{Light curve feedback on the EHT VLBI data calibration}
\label{appendix:VLBIfeedback}

Rapid variability in \sgra light curves affects the VLBI observations of the EHT as a modulation of the total intensity of the compact source resolved on the VLBI baselines. A detailed measurement of the mm light curve can therefore help inform simultaneous VLBI observations. For a sparse network like the EHT, the additional a priori information provided by time-dependent total intensity light curves can be of paramount importance for a successful reconstruction of the compact source structure. We make use of the light curve results for the VLBI data calibration in several ways. ALMA gains ($G_t$) derived as a by-product of the A1 pipeline through intra-field calibration (Section \ref{IntraField:subsec}), were used to produce ALMA a priori amplitude calibration metadata \citepalias[ANTAB tables,][]{SgraP2}, updated with respect to the standard ALMA QA2 tables derived under the constant flux density assumption \citep{Goddi2019}.
Moreover, the EHT array contains pairs of nearby stations providing intra-site baselines that do not resolve \sgra, and effectively measure a total compact flux density equal to the light curve amplitude up to the VLBI calibration station-based gain errors.
Using light curve information as a prior, all stations with a co-observing intra-site companion (ALMA, SMA, the Atacama Pathfinder Experiment [APEX], and the James Clerk Maxwell Telescope [JCMT]) can be absolutely flux calibrated by way of ``network calibration'' constraints \citep{blackburn2019a} that do not depend on any a priori VLBI station calibration. To that end, the combined light curves spanning the entire duration of the VLBI observations were constructed by merging the A1 and SM pipeline results (Sections \ref{IntraField:subsec} and \ref{sec:calibration:SMA}), after removing constant offsets between the pipelines (Section \ref{sec:consistency}, Table \ref{tab:medians}). A smoothed continuous representation of each full-day light curve was then generated through a smoothing spline interpolation \citep[\texttt{SciPy},][]{scipy2020}, and employed for the time-dependent network calibration. Similarly, light curves provide a natural variable flux density scaling for simple a priori source models suitable for initial self-calibration of the shortest inter-site EHT baselines. 
For the EHT VLBI observations of \sgra, such an approach was used to mitigate poor amplitude calibration of the Large Millimeter Telescope (LMT; \citetalias{paperiii}), through modeling visibilities on the shortest baseline ($< 2$\,G$\lambda$) to a well-calibrated SMT station with a Gaussian \citepalias{SgraP2}. The size of the Gaussian was selected based on the previous VLBI measurements \citep{johnson2018} and the pre-imaging constraints derived for the 2017 data set \citepalias{SgraP2}. Finally, the effect of total compact flux density modulation can be mitigated in the calibrated VLBI data sets by uniformly renormalizing visibility amplitudes on all baselines by the time-dependent light curves. In this way, a significant contribution to the total source intrinsic variability is removed \citep{Georgiev2021,Broderick2022}, increasing the robustness of imaging and modeling observations of \sgra with a static source model \citepalias{SgraP3,SgraP4}. All calibration procedures described above were applied separately to data from the LO and HI frequency bands, in which EHT VLBI observations were performed in 2017 \citepalias{paperii}. 

\section{Full bandwidth SMA data}
\label{appendix:SMA}

\begin{figure}[h!]
    \centering
    \includegraphics[width=0.75\textwidth,trim=0.0cm 0.0cm -0.0cm 0cm,clip]{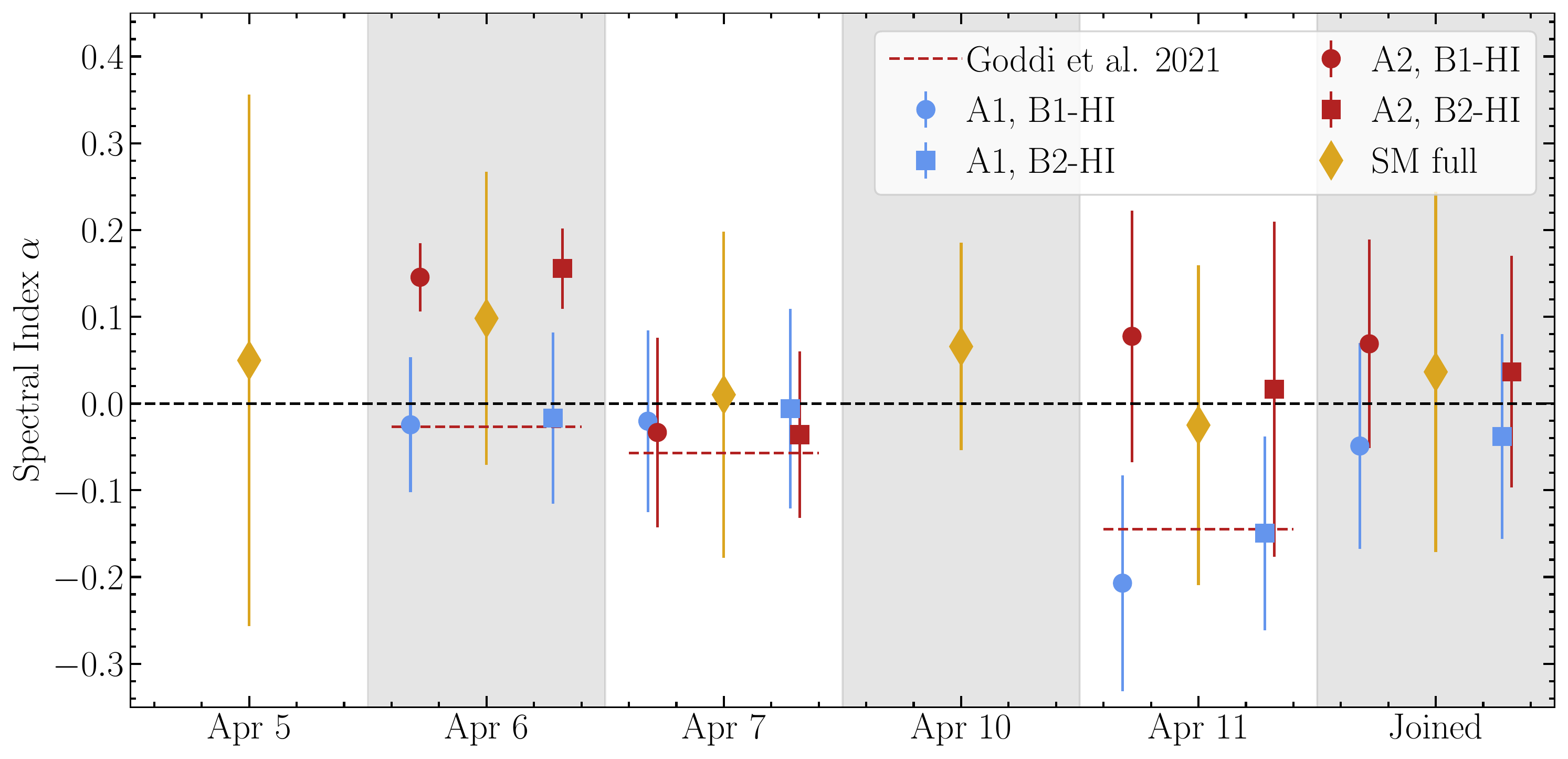}
    \caption{The spectral index measured from the 220\,GHz light curves of \sgra in 2017 April. The ALMA data points follow values reported in Figure \ref{fig:spectral_index}. The SMA results were obtained by fitting to all of the 8 sub-bands with frequencies ranging from 208.1 to 232.1\,GHz.}
    \label{fig:SI_SMA}
\end{figure}

In this paper, we presented SMA light curve results corresponding to the VLBI observing bands, LO at 227.1\,GHz and HI at 229.1\,GHz. However, the SMA observed \sgra with a particularly wide band, with 4 sub-bands, 2\,GHz wide in the 208.1--216.1 GHz\,range, and another 4 sub-bands in the 224.1--232.1\,GHz range. Since these data confirm the findings obtained for the SMA LO and HI bands, and the correlation between separate SMA sub-bands is overall very high, we only briefly comment on the entire SMA data set in this Appendix. A wide SMA bandwidth is particularly useful for measuring the spectral index. We estimate it using linear regression on amplitudes in all 8 sub-bands, for each separate timestamp. In Figure \ref{fig:SI_SMA}, we show the SMA results alongside the ALMA spectral index measurements, reported in Section \ref{sec:spectral_index}. The error bars correspond to the sample standard deviation in the spectral index distribution. Hence, they capture the intrinsic time-variability of the spectral index on top of the statistical uncertainties. We find the SMA spectral index to be consistent with zero, which corroborates the ALMA results (Section \ref{sec:spectral_index}).

\section{Model-fitting cornerplots}
\label{appendix:corns}

\begin{figure}[h!]
    \centering
    \includegraphics[width=0.495\textwidth,trim=0.0cm 0.0cm -0.0cm 0cm,clip]{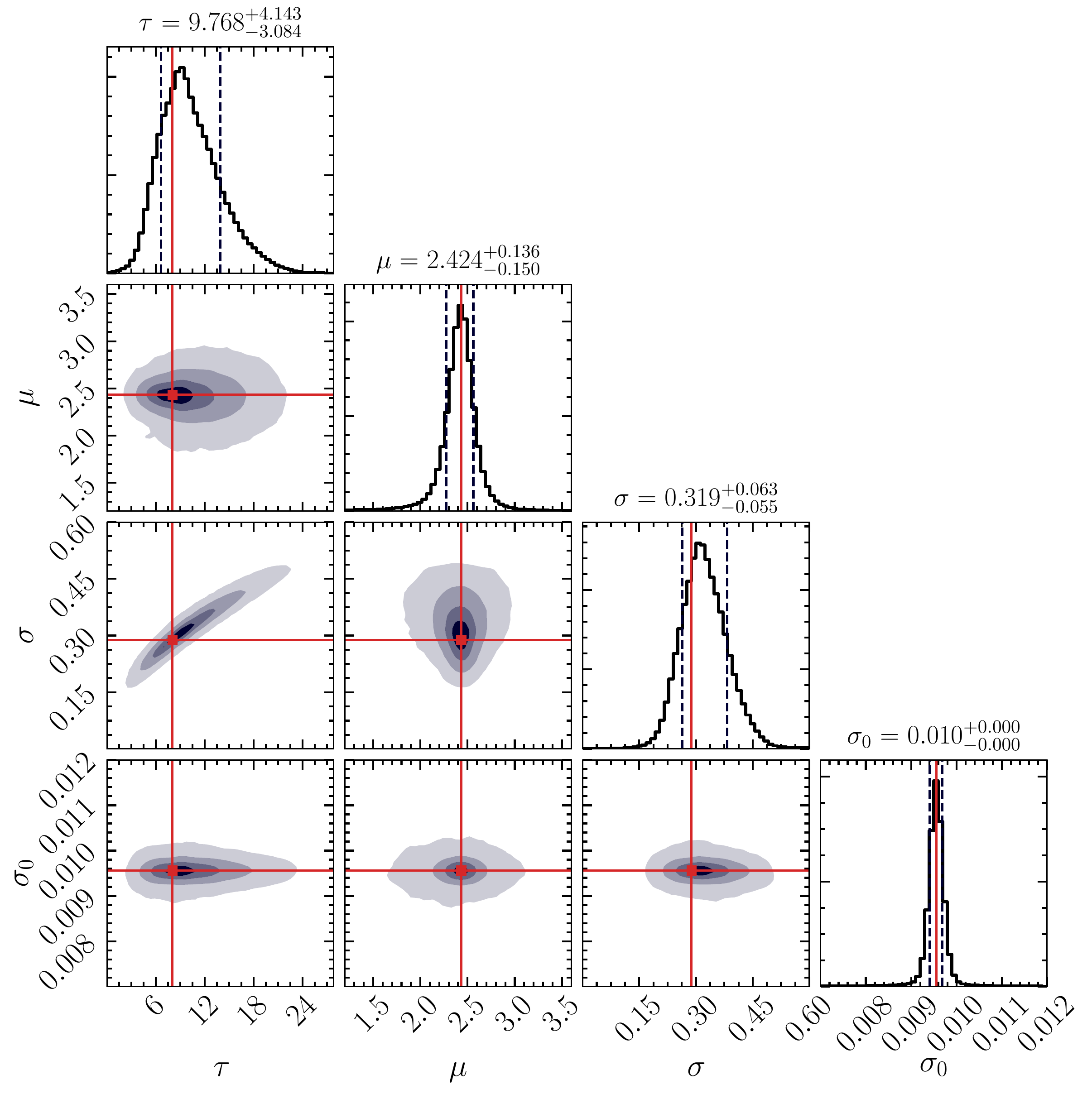}
     \includegraphics[width=0.495\textwidth,trim=0.0cm 0.0cm -0.0cm 0cm,clip]{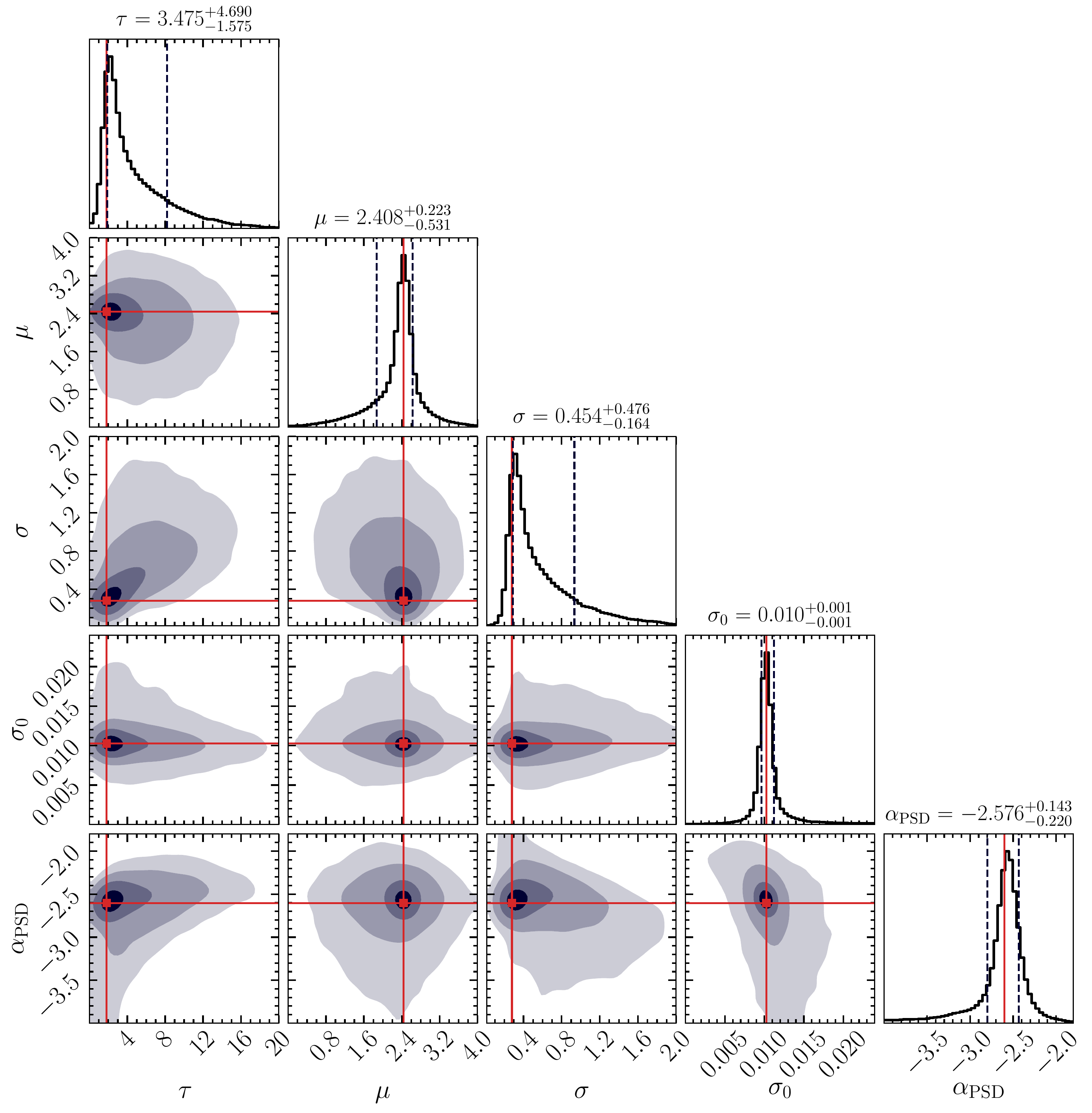}
    \caption{\textit{Left:} The DRW model best-fit to the EHT light curves of \sgra. Contours correspond to 0.2, 0.5, 0.8, and 0.95 of the posterior volume, values of median estimators on the marginalized posteriors are presented. The red line corresponds to the ML estimator, reported in Table \ref{tab:GPresults}. \textit{Right:} Same as above, but for the Mat\'{e}rn process model fit.} 
    \label{fig:cornplots}
\end{figure}

In Figure \ref{fig:cornplots}, we present the posterior distribution corner plots corresponding to the fiducial fits to the entire EHT 2017 \sgra light curves data set, FULL HI in Table \ref{tab:GPresults}. These cornerplots correspond to the two different GP models discussed in Section \ref{sec:modeling}, fitted to the observational data with a nested sampling posterior exploration algorithm.

